\RequirePackage{lineno} 

\documentclass[twocolumn,preprintnumbers,amsmath,amssymb,nofootinbib]{revtex4}

\usepackage{graphicx}
\usepackage{dcolumn}
\usepackage{bm}
 
\usepackage{epstopdf}

\def\bea{\begin{eqnarray}}
\def\eea{\end{eqnarray}}

\def\pp{\mbox{$p$-$p$}}

\def\pa{\mbox{$p$-A}}
\def\da{\mbox{$d$-A}}
\def\auau{\mbox{Au-Au}}

\def\pbpb{\mbox{Pb-Pb}}
\def\ppb{\mbox{$p$-Pb}}

\def\pn{\mbox{$p$-N}}
\def\aa{\mbox{A-A}}
\def\nn{\mbox{N-N}}

\def\ee{\mbox{$e^+$-$e^-$}}

\def\pt{$p_t$}
\def\mt{$m_t$}
\def\yt{$y_t$}

\def\nch{$n_{ch}$}
\def\mmpt{$\bar p_t$}

\begin{document} 

\setlength{\pdfpagewidth}{8.5in}
\setlength{\pdfpageheight}{11in}

\setpagewiselinenumbers
\modulolinenumbers[5]

\addtolength{\footnotesep}{-10mm}\

\preprint{version 1.7\textsl{}}

\title{Nuclear modification factors for identified hadrons from 5 TeV $\bf p$-Pb \\ collisions  and their relation to the Cronin effect
}

\author{Thomas A.\ Trainor}\affiliation{University of Washington, Seattle, WA 98195}


\date{\today}

\begin{abstract}

Nuclear modification factors (NMFs) are spectrum ratios rescaled by an estimate of the number of binary \nn\ collisions $N_{bin}$ within an A-B collision.  NMFs from more-central \aa\ collisions have been interpreted to indicate formation of a quark-gluon plasma (QGP) when compared with results from control \pa\ or \da\ collisions. However, subsequent analyses of such control systems are now also interpreted to indicate QGP formation, calling into question proper interpretation of NMFs. An additional complication is the nature of the so-called ``Cronin effect'' contribution to NMF structure that is not well understood. In the present study a two-component model of hadron production (TCM) is applied to identified-hadron (PID) \pt\ spectra from 5 GeV \ppb\ collisions extending up to 20 GeV/c. Hard components (jet fragment distributions)  are accurately isolated and their evolution with collision centrality parametrized. The TCM is then applied to NMFs {\em without} rescaling by $N_{bin}$, allowing direct comparisons between NMF evolution and hard-component  evolution with centrality. To address the Cronin effect the TCM is applied to fixed-target \pa\ spectra from the Chicago-Princeton (C-P) collaboration, the origin of the Cronin effect.  Inferred C-P spectrum hard components are {\em quantitatively} consistent with extrapolation of jet-related structure from higher energies. As a general conclusion spectrum ratios such as NMFs are difficult  to interpret, whereas direct differential analysis of isolated spectra may be interpreted simply and accurately.
\end{abstract}


\maketitle

\section{Introduction}

This article presents analysis of identified-hadron (PID) spectra from 5 TeV \ppb\ collisions reported by Ref.~\cite{alicenucmod}. These more-recent data  extend up to 20 GeV/c some previously published \pt\ spectra for light hadrons from the same collision system reported by Ref.~\cite{aliceppbpid} within more-limited \pt\ intervals. In the present study PID spectra are described by a two-component (soft+hard) model (TCM) of hadron production in high-energy nuclear collisions that provides accurate and complete isolation of jet-related hard components. The TCM has been applied previously to PID spectra from 5 TeV \ppb\ collisions~\cite{ppbpid,pidpart1,pidpart2} and 13 TeV \pp\ collisions~\cite{pppid}.

Specific motivations for the present study include claims that (a) certain data features (e.g.\ spectrum shape evolution with centrality) emerging from small collision systems are consistent with radial flow, (b) that nuclear modification factor (NMF) $R_\text{pPb}$ (for \ppb\ collisions) can provide information about jet modification within \ppb\ collisions and (c) that PID spectrum {\em ratios} (i.e.\ hadron/pion ratios) also provide evidence for radial flow.

Summarizing conventional opinion Ref.~\cite{alicenucmod} states that ``Recent measurements in high multiplicity pp, p-A and d-A collisions at different energies have revealed strong flow-like effects even in these small systems.'' Commenting on its own results Ref.~\cite{alicenucmod} states that ``...the spectra [for lighter hadrons] behave like in \pbpb\ collisions, i.e., the $p_T$ distributions become harder as the multiplicity increases and the change is most pronounced for [more-massive baryons]. In heavy-ion collisions this effect is commonly attributed to radial flow.'' In fact, PID spectrum evolution for \ppb\ collisions has been demonstrated to arise from jet production {\em increasing dramatically} with \ppb\ centrality~\cite{ppbpid,pidpart1,pidpart2} compared to nonjet production, a trend  first reported for 200 GeV \pp\ collisions in Ref.~\cite{ppprd}.

The statement continues...``For larger momenta the spectra follow a power-law shape as expected from perturbative QCD [pQCD].'' In fact, a hadron spectrum trend at higher \pt\ is described (factorization theorem) by the convolution of a {\em measured} parton (jet) energy spectrum for a given collision energy $\sqrt{s_\text{NN}}$~\cite{jetspec2} with an ensemble of {\em measured} fragmentation functions corresponding to parton (jet) energies~\cite{eeprd} as reported in Ref.~\cite{fragevo}. Thus, a spectrum shape at higher \pt\ is mainly related to combinations of {\em measured nonperturbative} phenomena.

Implementation of NMF $R_\text{pPb}$, a spectrum ratio rescaled by an estimate of the number of binary proton-nucleon \pn\ collisions $N_{bin}$ within a \ppb\ collision, requires estimation of $N_{bin}$ which is conventionally provided by a classical (not quantum mechanical) Glauber Monte Carlo. That procedure is based on a dramatic assumption that \pn\ collisions within \ppb\ collisions (for any centrality) are equivalent to inelastic or minimum-bias (MB) \pp\ collisions. The same assumption is the basis for the Glauber model wherein \pn\ interactions within \ppb\ collisions depend only on an inelastic cross section, and {\em multiple} \pn\ collisions may occur {\em simultaneously}~\cite{tomexclude}.

Regarding PID NMF trends Ref.~\cite{alicenucmod} observes that ``An enhancement of protons in the same \pt\ range [$2 < p_t < 5$ GeV/c] is also observed in heavy-ion collisions, {\em where it commonly is interpreted as} [indicating the presence of] {\em radial-flow} [emphasis added] and has a strong centrality dependence.''
Concerning PID spectrum ratios Ref.~\cite{alicenucmod} comments that in contrast to the apparent kaon/pion ratio invariance across collision systems ``... the proton-to-pion ratios exhibit {\em similar flow-like features} [emphasis added] for the \ppb\ and \pbpb\ systems, namely, the ratios are below the \pp\ baseline for $p_t < 1$ GeV/c and above for $p_t > 1.5$ GeV/c.'' The centrality trend of the proton/pion spectrum ratio is conjectured to ``play a dominant role'' in Cronin enhancement.

There are several fundamental problems with such data treatments: Most information from {\em costly} particle data is obscured or discarded by several mechanisms including (a) plotting formats that minimize visual access to critical structure, (b) combining basic data structures into ratios and (c) describing resulting data trends with simplistic {\em qualitative} language. The consequence is a projection screen upon which desired outcomes are perceived. Compounding those issues is a tendency to invoke {\em argument from analogy} -- because systems A and B share {\em one} common feature other aspects of the systems may be equivalent -- {\em and  stop there}. False analogy is countered by demonstrating critical {\em differences} between A and B.

The present study has the following goals: 
(a) demonstrate that the TCM describes PID hadron \pt\ spectra from Ref.~\cite{alicenucmod} within their point-to-point uncertainties over the interval $p_t \in [0.1,20]$ GeV/c, 
(b) demonstrate that jet production (spectrum hard component) dominates 
spectrum {\em evolution} with charge multiplicity \nch\ or \ppb\ centrality, 
(c) demonstrate that NMF (ratio) $R_\text{pPb}$ is {\em deceptive} as regards estimation of its rescale factor $N_{bin}$ via Glauber Monte Carlo and its relevance to modification of jet production, 
(d) demonstrate that hadron1/hadron2 spectrum ratios again abandon critical information and present a misleading picture of hadron production in small collision systems (among others), and
(e) analyze fixed-target spectra that motivated the Cronin effect (``enhancement'') within a TCM context.

What has emerged since the commencement of LHC heavy ion operations is a fundamental conflict between two mechanisms for transport of longitudinal projectile momentum to transverse momentum of produced hadrons  near midrapidity: (a) thermalization of longitudinal momentum via a conjectured energy-loss mechanism, developing pressure gradients and responding particle-source flows and (b) transport of longitudinal momentum via (i) Gribov diffusion within parton splitting cascades~\cite{gribov,gribov2} and (ii) large-angle scattering of projectile partons (mainly gluons) with fragmentation to minimum-bias jets near mid-rapidity~\cite{mbdijets}. To re-establish small collision systems as a reference for claims of QGP formation, jet production in small systems and its several manifestations must be established {\em quantitatively and completely} before invocation of fluid dynamics.
 
This article is arranged as follows:
Section~\ref{spectrumtcm} defines a TCM for PID spectra from seven event classes of 5 TeV \ppb\ collisions.
Section~\ref{smallspec} presents data spectra and their hard components with parametrizations of the latter.
Section~\ref{geometry} reviews collision geometry estimation for \ppb\ collisions from two methods that produce dramatically different results.
Section~\ref{nucmod} presents nuclear modification factors, including centrality dependence, derived from the \ppb\ spectra.
Section~\ref{specrat} presents PID spectrum {\em ratios} derived from \ppb\ spectra.
Section~\ref{cronineff} presents a TCM analysis of fixed-target $p$-X spectra (where X = $p$, Be, Ti and W) that were the origin for the so-called Cronin effect. The resulting TCM representation demonstrates factorization of soft and hard components each into collision energy $\sqrt{s}$ and nucleus size A dependences. 
Sections~\ref{disc} and~\ref{summ} present discussion and summary. 

\section{PID TCM for $\bf p$-P$\bf b$ spectra}  \label{spectrumtcm}

In this section a PID TCM for  5 TeV \ppb\ collisions is introduced, referring to previous analyses of PID spectra from \ppb\ collisions for lower-mass hadrons~\cite{ppbpid,pidpart1,pidpart2}. 

\subsection{PID spectrum TCM definition}   \label{pidspec}

Given a \pt\ spectrum TCM for unidentified-hadron spectra~\cite{ppprd,newpptcm} a corresponding TCM for identified hadrons can be generated by assuming that each hadron species $i$ comprises certain {\em fractions} of soft and hard TCM components denoted by $z_{si}(n_s)$ and $z_{hi}(n_s)$. The PID spectrum TCM can then be written as
\bea \label{pidspectcm}
\bar \rho_{0i}(p_t,n_s) &\approx&  z_{si}(n_s) \bar \rho_{s} \hat S_{0i}(p_t) +   z_{hi}(n_s) \bar \rho_{h} \hat H_{0i}(p_t),
\eea
where for \pp\ collisions $\bar \rho_h \approx \alpha(\sqrt{s}) \bar \rho_s^2$~\cite{newpptcm} and $\bar \rho_0 = \bar \rho_{s} + \bar \rho_h$ is the measured event-class charge density. $\bar \rho_s$ is then obtained from  $\bar \rho_0$ as the root of $\bar \rho_0 = \bar \rho_{s} + \alpha \bar \rho_s^2$. \ppb\ data spectra are plotted here as densities on \pt\ (as published) vs transverse rapidity $y_{t} = \ln[(p_t + m_{t\pi})/m_\pi]$.  Unit-normal model functions $\hat S_{0i}(m_t)$ (soft) and $\hat H_{0i}(y_t)$ (hard) are defined on those respective variables where they have simple forms and then transformed to \pt\ via appropriate Jacobians as necessary. Detailed definitions are provided in Refs.~\cite{ppbpid,pidpart1,pidpart2}. For convenience in what follows note that \yt\ = 2 corresponds to $p_t \approx 0.5$ GeV/c, \yt\ = 2.7 to 1 GeV/c, \yt\ = 4 to 3.8 GeV/c and \yt\ = 5 to 10 GeV/c.

The PID spectrum TCM for \ppb\ collisions includes $\bar \rho_s \rightarrow (N_{part}/2) \bar \rho_{sNN}$ and $\bar \rho_h \rightarrow N_{bin} \bar \rho_{hNN}$, where $N_{part}$ is the number of participant nucleons N, $N_{bin}$ ($= N_{part}-1$ for \ppb) is the number of \nn\ (in this case \pn) binary collisions, and densities $\bar \rho_{xNN}$ are averages over all participant pairs. If \pn\ ({\em not} NSD \pp) linear superposition within \ppb\ collisions is assumed then $\bar \rho_{hNN} \approx \alpha \bar \rho_{sNN}^2$ as for \pp\ collisions. Hard/soft ratio $\bar \rho_h / \bar \rho_s \equiv x(n_s) \nu(n_s)$, with $x \approx \alpha \rho_{sNN}$ and $\nu = 2 N_{bin} / N_{part}$, is of central importance. Parameter estimation requires accurate determination of \ppb\ centrality or collision geometry vs measured charge density $\bar \rho_0$ as reported in Ref.~\cite{tomglauber}. 

Table~\ref{rppbdata} shows centrality parameters for seven event classes of 5 TeV \ppb\ collisions. The primed numbers were reported in Ref.~\cite{aliceglauber}. The unprimed numbers are derived in Ref.~\cite{tomglauber} based in part on a TCM analysis of ensemble-mean \mmpt\ data from the same collision system. Those numbers are used in this study (see Sec.~\ref{geometry}).

\begin{table}[h]
	\caption{TCM fractional cross section $\sigma / \sigma_0$ (bin centers) and charge density $\bar \rho_0$, \nn\ soft component $\bar \rho_{sNN}$ and TCM hard/soft ratio $x(n_s)$ used for 5 TeV \ppb\ PID spectrum analysis~\cite{ppbpid}. $\sigma' / \sigma_0$ and $N_{bin}'$ values are from Table~2 of Ref.~\cite{aliceglauber}. Other parameter values are from Ref.~\cite{tomglauber}.
	}
	\label{rppbdata}
	\begin{center}
		\begin{tabular}{|c|c|c|c|c|c|c|c|c|} \hline
			$n$ &   $\sigma' / \sigma_0$ & $N_{bin}'$  &  $\sigma / \sigma_0$     & $N_{bin}$  & $\nu$ & $\bar \rho_0$ & $\bar \rho_{sNN}$ & $x(n_s)$ \\ \hline
			1	   &      0.025  & 14.7  & 0.15   & 3.20   & 1.52 & 44.6 & 16.6  & 0.188 \\ \hline
			2	 &  0.075  & 13.0 & 0.24    & 2.59   & 1.43 & 35.9 &15.9  & 0.180 \\ \hline
			3	 &  0.15  & 11.7 & 0.37 & 2.16  &  1.37 & 30.0  & 15.2  & 0.172 \\ \hline
			4	 &  0.30 & 9.36 & 0.58  & 1.70   & 1.26  & 23.0  & 14.1  & 0.159  \\ \hline
			5	 &  0.50  & 6.42 &0.80    & 1.31   & 1.13 & 15.8 &   12.1 & 0.137  \\ \hline
			6	 &  0.70 & 3.81 & 0.95   & 1.07   & 1.03  & 9.7  &  8.7 & 0.098 \\ \hline
			7	 & 0.90  & 1.94 & 0.99  & 1.00  & 1.00  &  4.4  & 4.2 &0.047  \\ \hline
		\end{tabular}
	\end{center}
\end{table}

\subsection{$\bf \hat S_{0i}$ and $\bf \hat H_{0i}$ model-function parameters}

Table~\ref{pidparams} shows 5 TeV \ppb\ TCM model parameters for hard component $\hat H_0(y_t)$ (first three) and soft component $\hat S_0(y_t)$ (last two) reported in Ref.~\cite{pidpart1}. Hard-component parameters vary significantly with hadron species and centrality: Gaussian widths $\sigma_{y_t}$ are greater for mesons than for baryons while exponents $q$ are substantially greater for baryons than for mesons. Peak modes $\bar y_t$ for baryons shift to higher \yt\ with increasing centrality while widths above the mode for mesons decrease. Soft-component model parameter $T$ is independent of collision energy but increases substantially with hadron mass. L\'evy exponent $n$ and hard-component exponent $q$ have substantial systematic collision-energy dependence~\cite{alicetomspec}. 

\begin{table}[h]
	\caption{TCM model parameters for identified hadrons from 5 TeV \ppb\ collisions from Ref~\cite{pidpart1}: hard-component parameters $(\bar y_t,\sigma_{y_t},q)$ and soft-component parameters $(T,n)$. These values apply to the  mid-central \ppb\ event class. 
	}
	\label{pidparams}
	\begin{center}
		\begin{tabular}{|c|c|c|c|c|c|} \hline
			& $\bar y_t$ & $\sigma_{y_t}$ & $q$ & $T$ (MeV) &  $n$  \\ \hline
			$ \pi^\pm $     &  $2.43\pm0.01$ & $0.62\pm0.01$ & $4.2\pm0.2$ & $145\pm3$ & $8.5\pm0.5$ \\ \hline
			$K^\pm$    & $2.65\pm0.03$  & $0.57\pm0.01$ & $4.1\pm0.2$ & $200\pm10$ & $14\pm3$ \\ \hline
			$p$        & $2.97\pm0.03$  & $0.47\pm0.01$ & $4.9\pm0.2$ & $210\pm10$ & $14\pm3$ \\ \hline
		\end{tabular}
	\end{center}
\end{table}

The entries for \ppb\ collisions in Table~\ref{pidparams} define a {\em fixed} TCM reference independent of centrality that describes the {\em mid-central} event class (wherein $\bar y_t \approx 2.95$ for baryons)~\cite{pidpart1}. In Refs.~\cite{pidpart2,pppid} variation of some hard-component model parameters is determined so as to describe all event classes within statistical uncertainties (e.g.\ see Fig.~4 of Ref.~\cite{pidpart2}). Required variations are linear on hard/soft ratio $x(n_s) \nu(n_s)$: with increasing ratio hard-component modes $\bar y_t(n_s)$ shift to higher \yt\ for baryons while (for \ppb\ but not \pp) hard-component widths above the mode decrease for mesons. 

\subsection{PID hadron fractions $\bf z_{si}(n_s)$ and $\bf z_{hi}(n_s)$}

Table~\ref{otherparamsx} shows PID parameters $z_{0i}$ and $\tilde z_i = z_{hi}/ z_{si}$ for three hadron species, determined from PID spectrum data for 5 TeV \ppb\ collisions, as reported in Ref.~\cite{pidpart1}. While $z_0$ was found to be independent of \ppb\ centrality within uncertainties the \ppb\ $\tilde z_i(n_s)$ exhibit significant centrality dependence as shown in Fig.~8 of Ref.~\cite{pidpart1}. It is notable that the $\tilde z_i(n_s)$ depend only on hadron mass, not on strangeness or baryon identity. Measurements of individual centrality trends for $z_{si}(n_s)$ and $z_{hi}(n_s)$ based on spectrum analysis are presented in Sec.~IV of Ref.~\cite{pidpart1}.  Individual fractions $z_{si}(n_s)$ and $z_{hi}(n_s)$ may also be derived from model parameters $\tilde z_i(n_s)$ and $z_{0i}$ via relations
\bea \label{zsix}
z_{si}(n_s) &=& \frac{1 + x(n_s) \nu(n_s)}{1 + \tilde z_i(n_s)x(n_s) \nu(n_s)} z_{0i}
\\ \nonumber
z_{hi}(n_s) &=& \tilde z_i(n_s)z_{si}(n_s).
\eea
 5 TeV \ppb\ geometry (centrality) parameters $x(n_s)$ and $\nu(n_s)$ are determined in Refs.~\cite{alicetommpt,tommpt,tomglauber} based on ensemble-mean \mmpt\ data (see Table~\ref{rppbdata}). The $\tilde z_i = z_{hi}/ z_{si}$ values in Table~\ref{otherparamsx} represent averages over \ppb\ centrality. Parameters $\tilde z_i(n_s)$ are observed to be strictly proportional to hadron mass, and parameters $z_{0i}$ are approximately consistent with statistical-model predictions~\cite{pidpart1}. Note that the corrected value for pion $\tilde z_i$ is $0.60\pm0.05$. The biased value 0.85 reported in Ref.~\cite{pidpart1} is attributed to systematic error of the $dE/dx$ PID process in which a fraction of protons is misassigned as pions~\cite{pppid}.
 
 The TCM parameter values reported in this section arise from previous analyses in Refs.~\cite{ppbpid,pidpart1,pidpart2,pppid} based on spectrum data for lighter hadrons over more-limited \pt\ intervals. Parameter values are reexamined in Sec.~\ref{tcmparams}.

\begin{table}[h]
	\caption{\label{otherparamsx}
		TCM model parameters for identified hadrons from 5 TeV \ppb\ collisions in Ref.~\cite{pidpart1}. Values for $\tilde z_i = z_{hi} / z_{si}$ are averages over \ppb\ centrality. Parameters $ \bar p_{ts}$ and $\bar p_{th}$ are determined by model functions $\hat S_0(y_t)$ and $\hat H_0(y_t)$ defined by Table~\ref{pidparams}.  Values for $\bar p_{thi}$ are event-class averages from Ref.~\cite{pidpart2}.
	}
$^\dagger$Correct value satisfying charge conservation is $0.60\pm0.05$.
	\begin{center}
		\begin{tabular}{|c|c|c|c|c|} \hline
			&   $z_{0i}$    &  $\tilde z_i$ &   $ \bar p_{tsi}$ (GeV/c)  & $ \bar p_{thi}$ (GeV/c)  \\ \hline
			$ \pi^\pm$        &   $0.82\pm0.01$  & $0.85\pm0.05^\dagger$  & $0.40\pm0.02$ &    $1.15\pm0.03$  \\ \hline
			$K^\pm $   &  $ 0.128\pm0.002$   &  $2.7\pm0.2$ &  $0.60\pm0.02$&  $1.34\pm0.03$   \\ \hline
			$p $        & $ 0.065\pm0.002$    &  $5.6\pm0.2$ &  $0.73\pm0.02$&   $1.57\pm0.03$   \\ \hline
		\end{tabular}
	\end{center}
\end{table}


\section{5 $\bf TeV$ $\bf p$-P$\bf b$ PID Spectra}
 \label{smallspec}

The 5 TeV \ppb\ PID spectrum data used for this analysis were reported by Ref.~\cite{alicenucmod} as an extension to higher \pt\ of PID spectra from the same collision system reported in Ref.~\cite{aliceppbpid}.
Collision events (12.5 million  events for charged hadrons) are sorted into seven multiplicity classes based on charge accumulated within a V0A detector. Mean charge densities $\langle dN_{ch} / d\eta \rangle \rightarrow \bar \rho_0$  as integrated within $|\eta| < 0.5$ or angular acceptance $\Delta \eta = 1$ are 45, 36.2, 30.5, 23.2, 16.1, 9.8 and 4.4 for event classes $n \in [1,7]$. 

\subsection{Proton spectrum inefficiency correction} \label{ineff}

\begin{figure}[h]
	\includegraphics[width=3.3in]{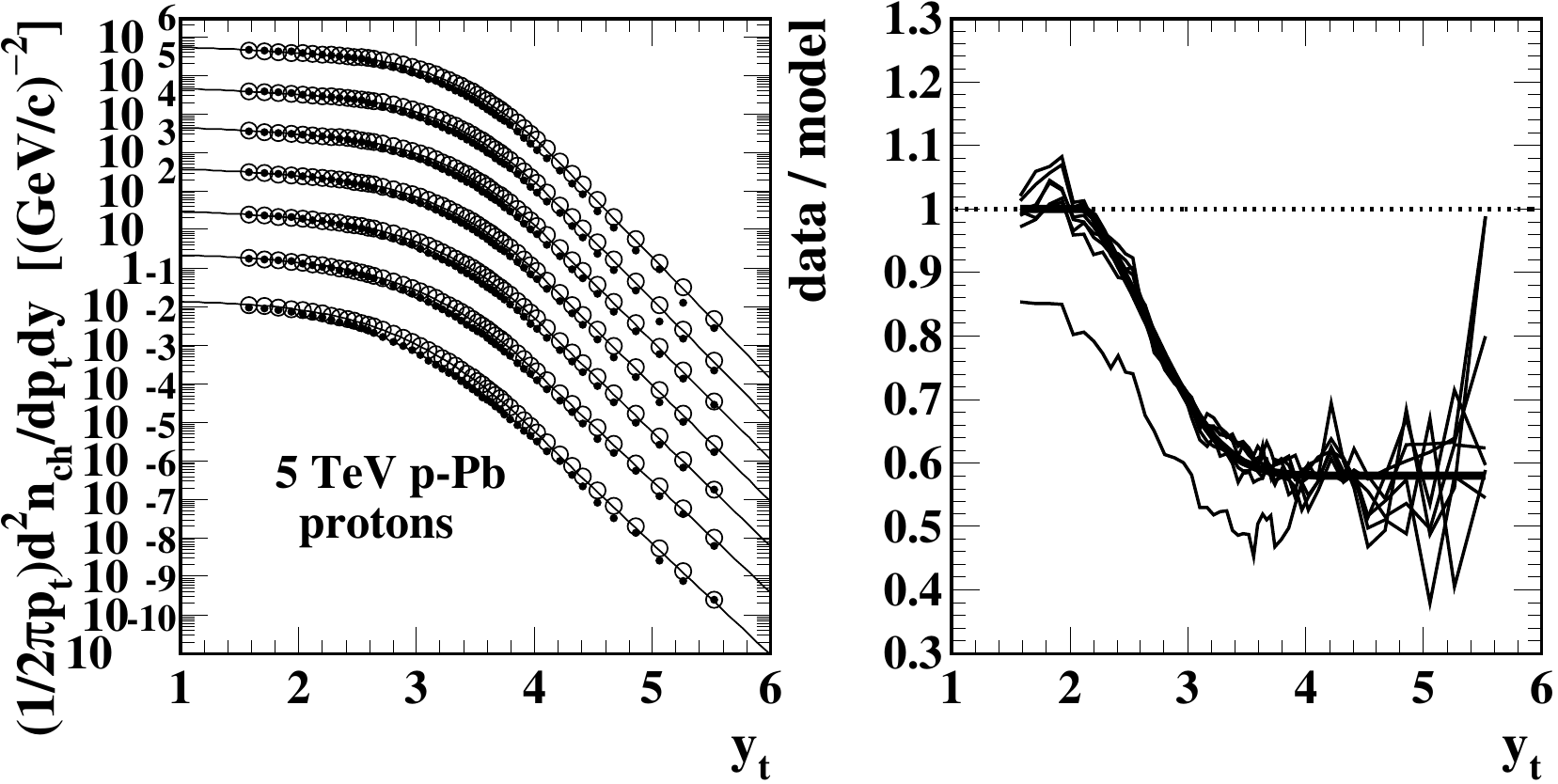}
	\caption{\label{eppsprotons}
		Left: Published proton spectra from 5 TeV \ppb\ collisions (solid dots) as reported in Ref.~\cite{alicenucmod}. The solid curves and open circles are TCM predictions based on low-\pt\ spectrum structure as summarized in Ref.~\cite{pppid}.
		Right: Ratios of proton data to TCM predictions from the left panel. The bold solid curve is Eq.~(\ref{effppb}) applied as a correction to published spectra.
	} 
\end{figure}

In previous analyses of PID spectra from 5 TeV \ppb\ collisions~\cite{ppbpid,pidpart1,pidpart2} and 13 TeV \pp\ collisions~\cite{pppid} it was observed that proton spectra above 0.5 GeV/c (\yt\ = 2) are strongly suppressed ($\approx 40$\%) relative to TCM predictions. The basis for that observation is reviewed in Sec.~IV A of Ref.~\cite{pppid}.
The \ppb\ data-model comparison in Fig.~\ref{eppsprotons} (left) for the present study indicates similar systematic suppression of protons. The solid dots are published data; the solid curve and open circles are TCM predictions based on low-\pt\ trends. The efficiency correction derived for 5 TeV \ppb\ protons in Ref.~\cite{pidpart1} is
\bea \label{effppb}
\epsilon_p(\ppb) \hspace{-.05in} &=& \hspace{-.05in} \left\{0.58+(1\hspace{-.02in} - \hspace{-.02in}\tanh[(p_t \hspace{-.02in}- \hspace{-.02in}0.45)/0.95])/2\right\}
\\ \nonumber &\times& (1+0.003 p_t^3).
\eea
However, the PID spectra from Ref.~\cite{alicenucmod} extended to 20 GeV/c demonstrate that the last factor is inconsistent with the new data. It is therefore omitted in what follows.

Equation~(\ref{effppb}) describes the bold solid curve in the right panel that is consistent with the data/TCM ratio within point-to-point uncertainties. That expression
is assumed to describe, {\em independent of \ppb\ event class}, a proton instrumental inefficiency arising from $dE/dx$ PID measurements reported in Ref.~\cite{alicenucmod}. The inefficiency appears significant only above 0.5 GeV/c ($y_t \approx 2$). The same correction is applied consistently to seven \ppb\ event classes.

\subsection{5 $\bf TeV$ $\bf p$-P$\bf b$ PID spectrum data} \label{fitqual}

Spectra for the present study are presented as densities on \pt\ (i.e.\ published \pt\ spectra) plotted vs pion rapidity $y_{t\pi}$ with pion mass assumed. Soft-component model $\hat S_{0i}(m_{ti})$ for species $i$ is defined by a L\'evy distribution on $m_{ti}$ with hadron mass $m_i$. Hard-component model $\hat H_{0}(y_{t\pi})$  is defined as a density on $y_{t\pi}$ with a simple form, a Gaussian on \yt\ with exponential high-\yt\ tail, then converted to $\hat H_{0}(p_t)$ via the Jacobian factor $y_{t\pi} / m_{t\pi} p_t $. 

\begin{figure}[h]
	\includegraphics[width=3.3in]{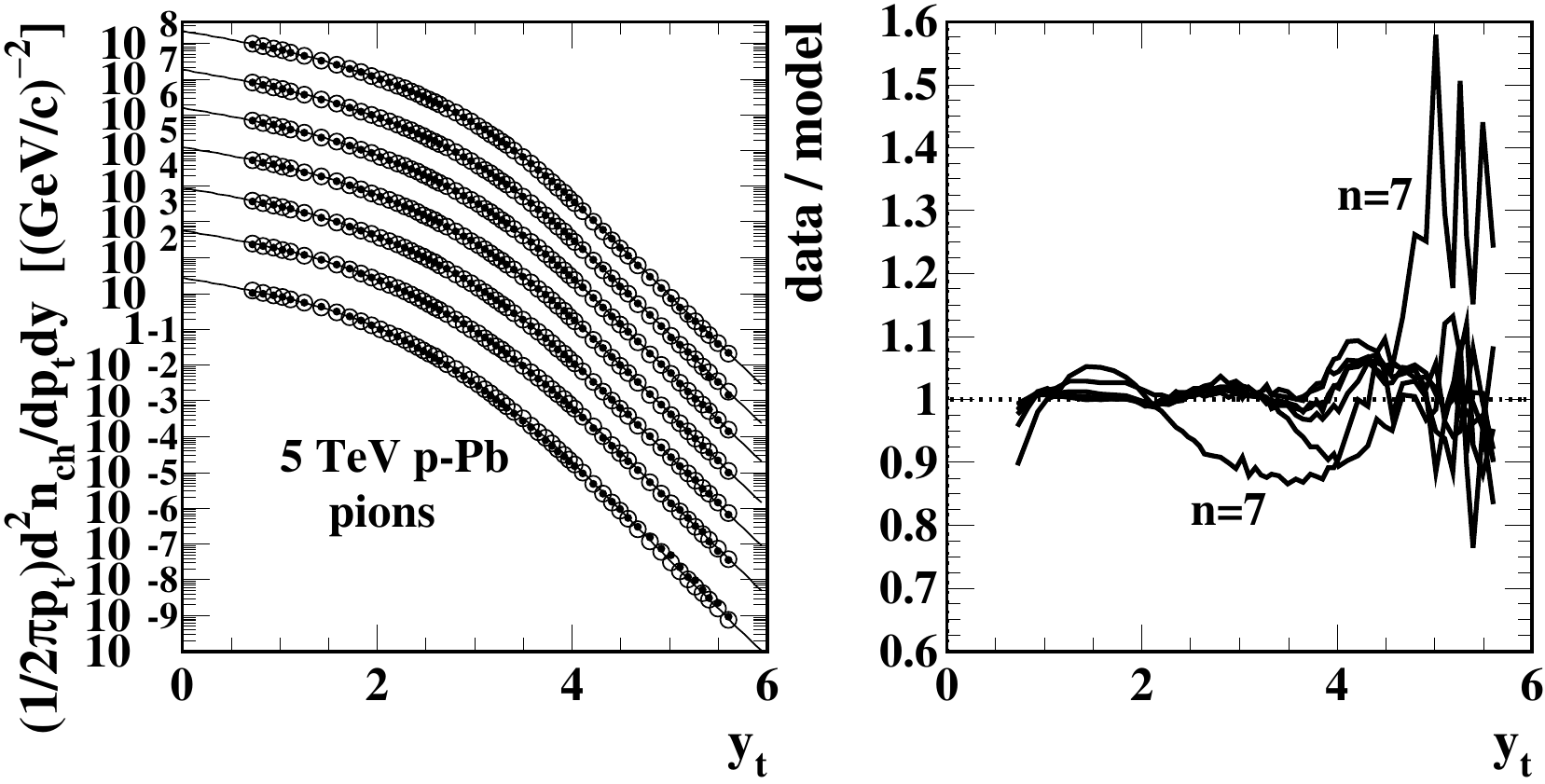}
	\caption{\label{pionspec}
		Left: Pion spectra (solid dots) reported in Ref.~\cite{aliceppbpid} from seven multiplicity classes of 5 TeV \ppb\ collisions. Solid curves are the corresponding TCM on a continuum (100 points uniformly distributed on \yt). The TCM defined on data \yt\  values appears as open circles.
		Right: Data/model ratios corresponding to solid dots and open circles in the left panel. The most-peripheral  ratio ($n = 7$) shows strong bias typically encountered for \pp\ and \ppb\ collision systems~\cite{newpptcm}.
	} 
\end{figure}
\begin{figure}[h]
	\includegraphics[width=3.3in]{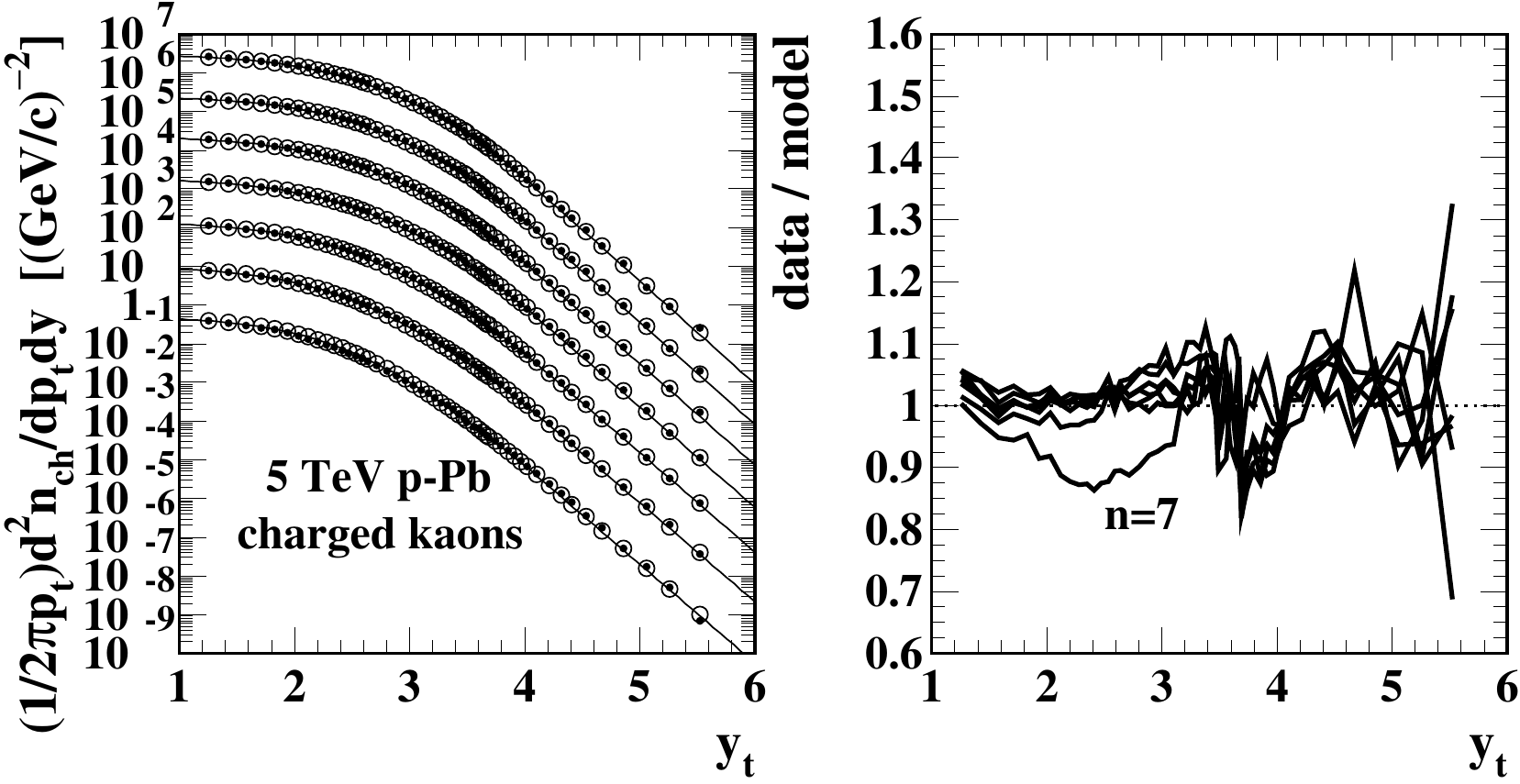}
	\caption{\label{kaonspec}
Same as Fig.~\ref{pionspec} but charged kaons. The sharp deviation near \yt\ = 3.5 appears to arise from a calibration issue when extending the PID spectra from Ref.~\cite{aliceppbpid} to higher \pt.	
} 
\end{figure}
\begin{figure}[h]
	\includegraphics[width=3.3in]{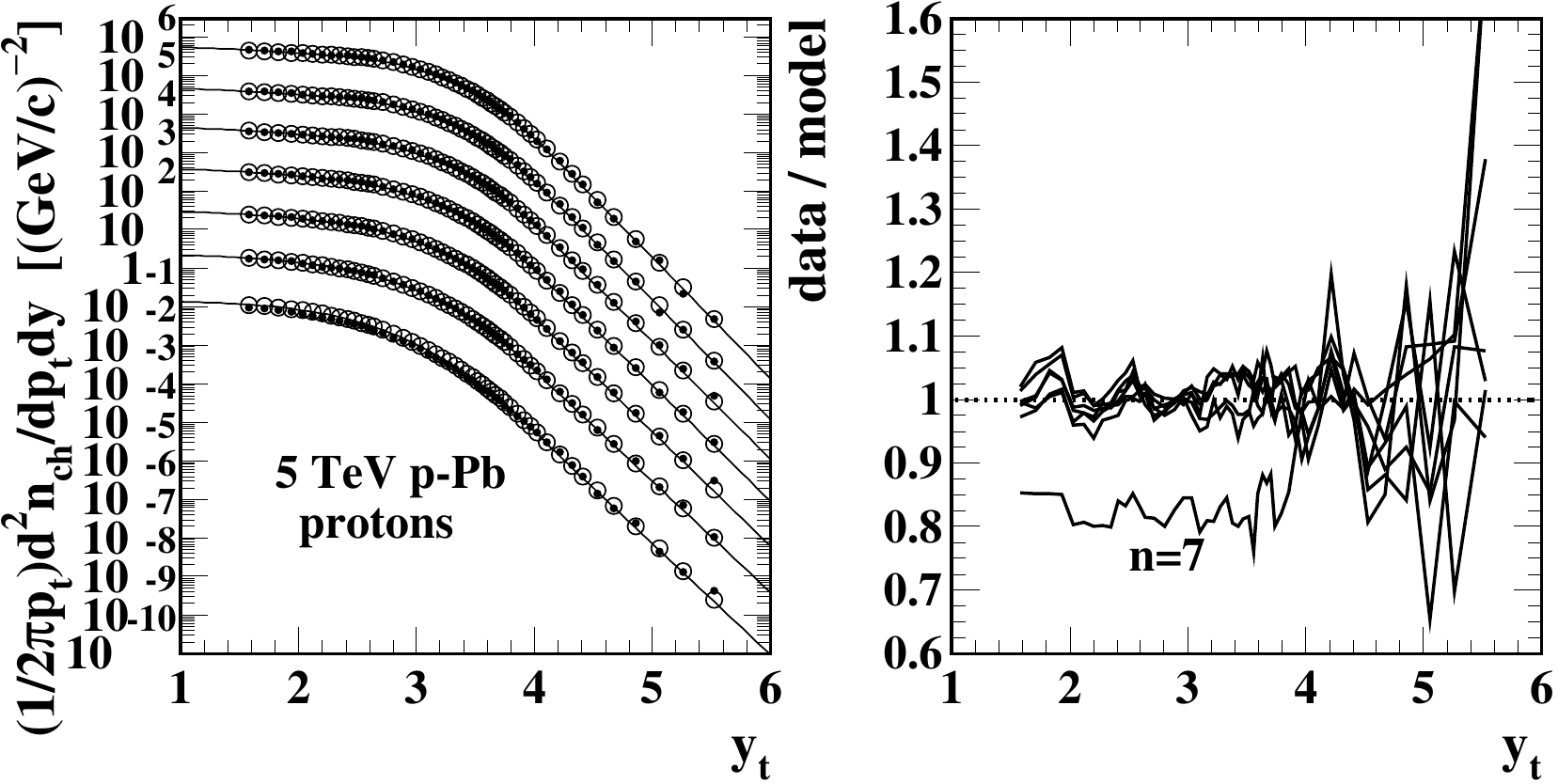}
	\caption{\label{protonspec}
Same as Fig.~\ref{pionspec} but for protons. The sharp deviation near \yt\ = 4 appears to arise from a calibration issue when extending the PID spectra from Ref.~\cite{aliceppbpid} to higher \pt.	
	} 
\end{figure}

Figures~\ref{pionspec}-\ref{protonspec} (left) show PID \pt\ spectra (solid dots) for charged pions, kaons and protons plotted vs pion rapidity \yt. The solid curves (continuum) and open circles (defined on data \yt\ values) are {\em variable}-TCM spectra determined as described below. 

Figures~\ref{pionspec}-\ref{protonspec} (right) show data/model ratios. The most-peripheral event class for $n=7$ is singled out in each case because of the substantial bias common to low-\nch\ data that has consequences for the present analysis.

The non-single-diffractive (NSD) \ppb\ spectra appearing in Ref.~\cite{alicenucmod} and in the present study are approximately consistent with n=5 \ppb\ spectra that have $\bar \rho_0 \approx 16$. Significant deviations arise from the broad average compared to specific centrality ($\bar \rho_0$) event classes. The MB \pp\ collision spectra reported in Ref.~\cite{alicenucmod} are approximately consistent with $n =7$ \ppb\ collisions, but again significant deviations appear for the MB \pp\ average compared to a specific \ppb\ event class.

\subsection{PID spectrum hard components} \label{spechard}

Experience derived from PID spectrum analysis of a broad array of A-B collision systems and collision energies~\cite{ppbpid,pbpbpid,pidpart1,pidpart2} indicates that spectrum soft components modeled by $\hat S_{0i}(y_t)$ are independent of event \nch\ or centrality, permitting accurate isolation of complementary jet-related spectrum hard components by a simple procedure. Given Eq.~(\ref{pidspectcm}), repeated in the first line below, spectrum {\em data} hard components described by model functions $\hat H_{0i}(y_t,n_s)$ are obtained in the second line by
\bea \label{pidspectcmx}
\bar \rho_{0i}(y_t,n_s) &\approx& z_{si}(n_s)  \bar \rho_{s} \hat S_{0i}(y_t) +  z_{hi}(n_s)  \bar \rho_{h} \hat H_{0i}(y_t,n_s)~~
\\ \nonumber
&& \hspace{-.8in} \frac{\bar \rho_{0i}(y_t,n_s) - z_{si}(n_s)\bar \rho_s \hat S_{0i}(y_t)}{ \hat H_{0i}(\bar y_t,n_s)\bar \rho_h} \approx z_{hi}(n_s) \frac{\hat H_{0i}(y_t,n_s)}{\hat H_{0i}(\bar y_t,n_s)},
\eea
where $\hat H_{0i}(\bar y_t,n_s)$ is the value of $\hat H_{0i}(y_t,n_s)$ at its mode. The difference in the numerator at left is first obtained from densities on \pt\ as in left panels above. The difference is then transformed to \yt\ suitable for the second line of Eq.~(\ref{pidspectcmx}). Accuracy of such differential analysis relies on precise determination of \ppb\ geometry, nonPID charge densities $ \bar \rho_{s}$ and $ \bar \rho_{h}$ and PID species fractions $z_{si}(n_s)$ and $z_{hi}(n_s)$. PID spectrum hard components are isolated and described precisely to well below 0.5 GeV/c ($y_t \approx 2$).

\begin{figure}[h]
	\includegraphics[width=3.3in]{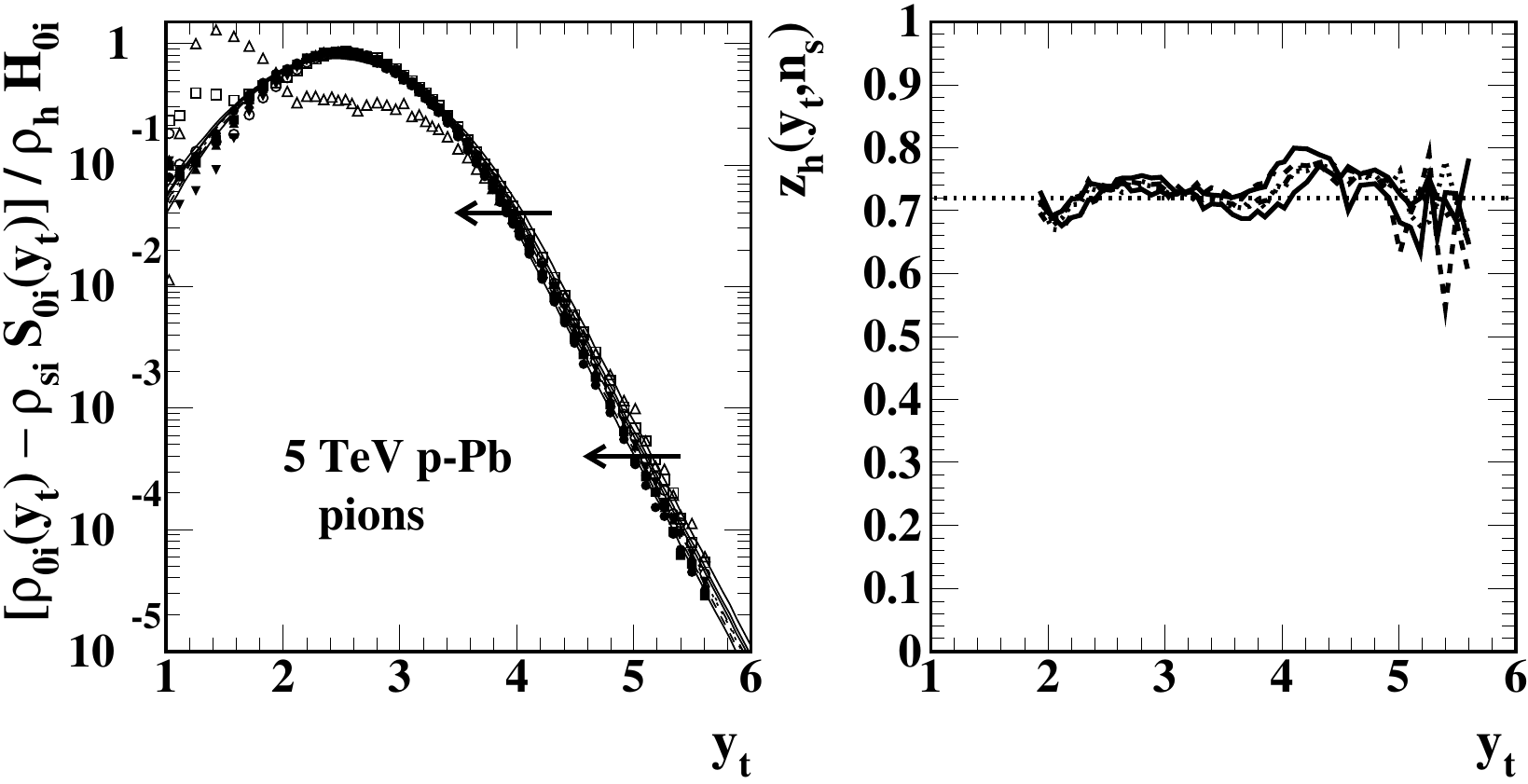}
	\caption{\label{pionhc}
		Left: Pion spectrum hard components (points) isolated according to Eq.~(\ref{pidspectcmx}) (second line, left). The solid curves are the pion variable TCM with hard-component parameters as shown in Tables~\ref{pidparam1} and \ref{pidparam2}. In that formulation the peak values correspond to $z_{hi}(n_s)$.
		Right: Data in the left panel divided by pion TCM hard components $\hat H_{0i}(y_t,n_s)/\hat H_{0i}(\bar y_t,n_s)$.
	} 
\end{figure}

Figure~\ref{pionhc} (left) shows pion data hard components (points) in  the form of Eq.~(\ref{pidspectcmx}) (second line) and variable TCM hard components (solid curves). The right panel shows hard-component data/model ratios. In  this and two other figures arrows in the left panels indicate evolution from peripheral to central \ppb\ collisions.

\begin{figure}[h]
	\includegraphics[width=3.3in]{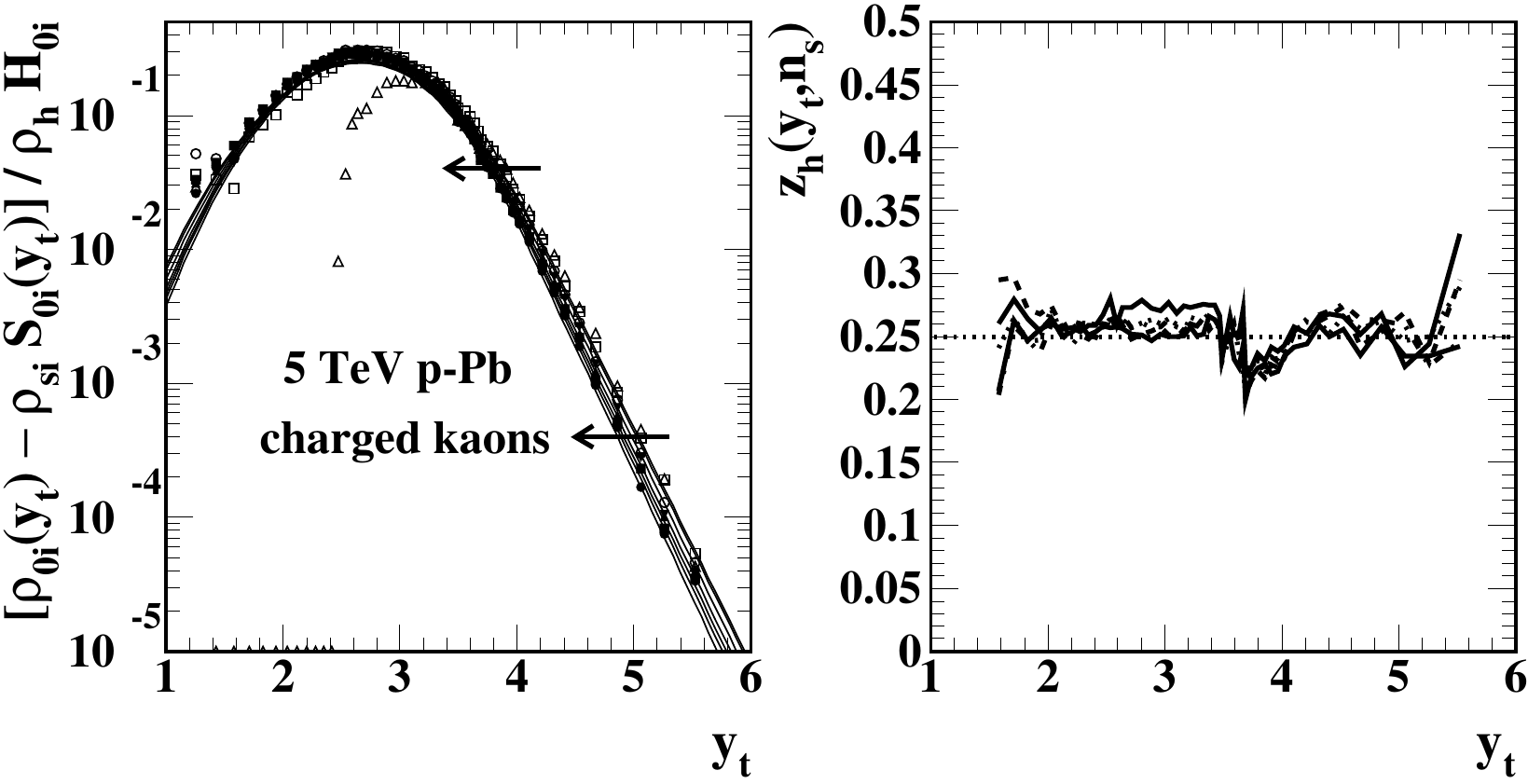}
	\caption{\label{kaonhc}
Same as Fig.~\ref{pionhc} but for charged kaons. Note the discontinuity near \yt\ = 3.5 corresponding to the endpoint of kaon spectra from Ref.~\cite{aliceppbpid} extended to higher \pt\ by Ref.~\cite{alicenucmod}. The right panels display only $n \in [1,5]$ for better clarity.
} 
\end{figure}

Figure~\ref{kaonhc} shows charged-kaon spectrum hard components and data/model ratios. Notable in the right panel is the step-like discontinuity near \yt\ = 3.5 that coincides with the interface between lower-\pt\ data from Ref.~\cite{aliceppbpid} and the extension to higher \pt\ reported in Ref.~\cite{alicenucmod}.

\begin{figure}[h]
	\includegraphics[width=3.3in]{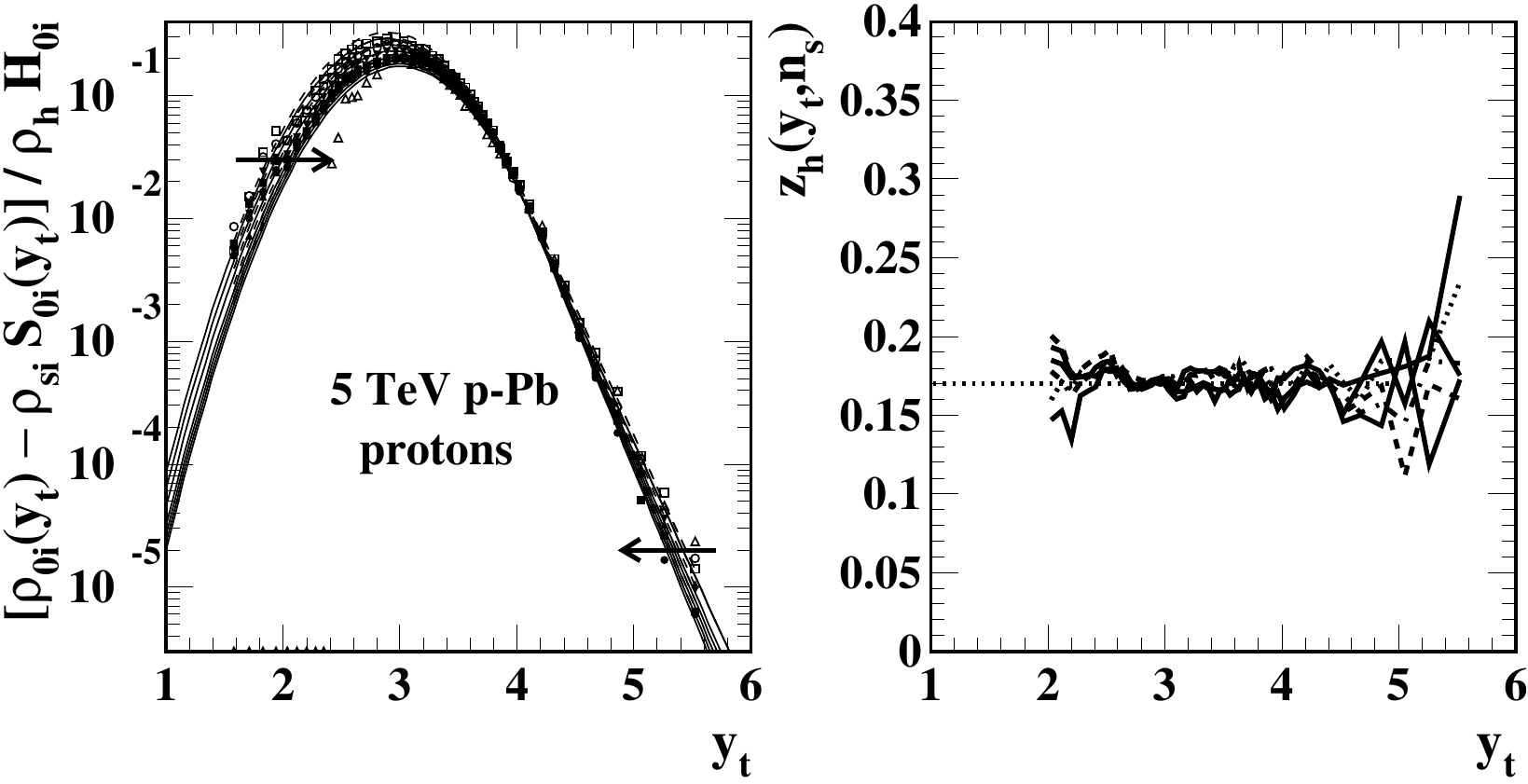}
	\caption{\label{protonhc}
Same as Fig.~\ref{pionhc} but for protons. The proton hard-component mode shifts to higher \yt\ with increasing \ppb\ centrality coordinated with decreasing amplitude $z_{hi}(n_s)$ so the exponential tail remains stationary near \yt\ = 4.2. At higher \yt\ exponent $q$ increases with softening of the exponential tail.
} 
\end{figure}

Figure~\ref{protonhc} shows similar proton hard-component results. Note that variation of hard-component structure relative to {\em fixed} TCM model functions as above represents all the new spectrum information and summarizes completely what could be called ``jet modification'' in \ppb\ collisions.

\subsection{Hard-component parameter trends} \label{tcmparams}

For hard-component peak mode $\bar y_t(n_s)$ and width $\sigma_{y_t}(n_s)$, exponent $q(n_s)$ and fractional-abundance ratio $\tilde z_i(n_s)$ variation with \ppb\ centrality or charge multiplicity soft-component $n_s$ is found to be linear on hard/soft ratio $x(n_s)\nu(n_s)$ to good approximation. TCM parameter $X(n_s)$ values are thus expressed as
\bea \label{star}
X(n_s) &=& X^* + \delta X^* x(n_s)\nu(n_s),
\eea
where the starred quantities appear in the tables below.

Table~\ref{pidparam1} shows hard-component parameter values for  mode position $\bar y_{ti}(n_s)$ and width $\sigma_{y_ti}$. Fixed soft-component parameter values are as presented in Table~\ref{pidparams}.
%
\begin{table}[h]
	\caption{TCM hard-component model parameters $\bar y_{ti}(n_s)$ and $\sigma_{y_ti}(n_s)$ in the form of Eq.~(\ref{star}) for charged hadrons from 5 TeV \ppb\ collisions.
	} \label{pidparam1}
	\begin{center}
		\begin{tabular}{|c|c|c|c|c|} \hline
			& $\bar y_t^*$ & $\delta \bar y_t^*$ & $\sigma_{y_t}^*$ & $\delta \sigma_{y_t}^*$    \\ \hline
			$\pi$          &  $2.46\pm0.01$ &0 & $0.67\pm0.01$ & $-0.25\pm0.05$   \\ \hline
			$K^\pm$       & $2.65\pm0.03$ & $0$ & $0.64\pm0.02$ & $-0.25\pm0.05$   \\ \hline	
			$ p$     &  $2.87\pm0.05$ & $0.47\pm0.05$ & $0.47\pm0.02$ & $0$   \\ \hline
		\end{tabular}
	\end{center}
\end{table}
%
For protons the mode positions are determined by measured hadron fraction hard components $z_{hi}(n_s)$
\bea
\bar y_{ti}(n_s) &=& 3.0 + 0.204 \ln[z_{hi}(n_s) / 0.17]
\eea
such that as the mode shifts the exponential tail maintains a fixed position as observed for data hard components (see Fig.~\ref{protonhc}, left, just above \yt\ = 4). The value $z_{hi}(n_s) = 0.17$ is that corresponding to $\bar y_{ti}(n_s) \approx 3.0$.

Table~\ref{pidparam2} shows values for hard-component exponent $q_i(n_s)$ and hadron species fraction hard/soft ratio $\tilde z_i(n_s) = z_{hi}(n_s)/z_{si}(n_s)$. Fractional abundance coefficients $z_{0i}(n_s)$  in Eqs.~(\ref{zsix}) vary significantly with \ppb\ centrality for Cascades and Omegas as noted in Ref.~\cite{pidsss}. However, for lighter hadrons the variation is not significant except for event classes falling below NSD $\bar \rho_0$ and are assumed independent of charge multiplicity for lighter hadrons as in Refs.~\cite{ppbpid,pidpart1,pidpart2,pppid}.

\begin{table}[h]
	\caption{TCM hard-component model parameter $q_i(n_s)$ and fractional abundance parameter $\tilde z_i(n_s)$ in the form of Eq.~(\ref{star}) for charged hadrons from 5 TeV \ppb\ collisions. The $\tilde z_i(n_s)$ parameters correspond to Fig.~1 (right) of Ref.~\cite{pidsss}.
		} \label{pidparam2}
	\begin{center}
		\begin{tabular}{|c|c|c|c|c|} \hline
			& $q^*$ & $\delta q^*$ & $\tilde z_i^*$ & $\delta \tilde z_i^*$    \\ \hline
			$\pi$          &  $4.25\pm0.1$ &$0$ & $0.60\pm0.05$ & $0.30\pm0.05$   \\ \hline
			$K^\pm$       & $3.97\pm0.02$ & $1.0\pm 0.1$ & $2.63\pm0.1$ & $1.20\pm0.2$   \\ \hline	
			$ p$     &  $4.55\pm0.05$ & $2.9\pm0.2$ & $5.30\pm0.1$ & $2.80\pm0.2$   \\ \hline
		\end{tabular}
	\end{center}
\end{table}

Note that hard component $\hat H_{0i}(y_t,n_s)$ is defined as a Gaussian on pion \yt\ with exponential tail where  the transition point from Gaussian to exponential (near \yt\ = 4) is determined by slope matching. Thus, if the width above the mode varies the point at which the exponential begins shifts accordingly but the exponential slope does not change. However, if exponent $q$ alone varies any shift of the point at which the exponential begins is small.

The evolution trends (arrows) in Figs.~\ref{pionhc}, \ref{kaonhc} and \ref{protonhc} are the {\em only deviations} from a fixed \ppb\ TCM reference, at the level of data statistical uncertainties. In effect, those deviations represent the entire novel information carried by data from Ref.~\cite{alicenucmod} relative to previous analyses.

\subsection{Data description quality}

In previous TCM analyses of PID spectra from 5 TeV \ppb\ collisions~\cite{ppbpid,pidpart1,pidpart2} and 13 TeV \pp\ collisions~\cite{pppid} data description quality was studied in detail using the Z-score statistic~\cite{zscore}. It was established that spectrum data are described by the TCM {\em within their statistical uncertainties}. The data/TCM ratios (right panels above) are generally consistent within statistical uncertainties with a few exceptions: a sharp deviation for charge kaons near \yt\ = 3.5 and a sharp deviation for protons near \yt\ = 4.1. It is notable that the upper bounds for PID spectra published in Ref.~\cite{aliceppbpid} are 3, 2.5 and 4 GeV/c for pions, kaons and protons respectively, corresponding to \yt\ = 3.75, 3.5 and 4.05. It is possible that the sharp deviations arise from mismatches in extending the spectra to higher \pt\ for Ref.~\cite{alicenucmod}. Jet-related hard components are accurately described within the large interval $p_t \in [0.5,20]$ GeV/c.

\section{$\bf p$-P$\bf b$ geometry determination} \label{geometry}

NMFs are motivated by an A-B collision model that assumes linear superposition of (in effect) the equivalent of MB \pp\ collisions. It is claimed that NMFs should return to unity within a relevant \pt\ interval if that assumption is correct: no modification of \nn\ collisions relative to MB \pp\ in the absence of new phenomena.

As conventionally defined, nuclear modification factors require estimation of the number of binary \nn\ collisions $N_{bin}$. \pa\ collision geometry may be addressed in at least two ways with quite different results. Reference~\cite{aliceppbgeom} asserts that ``The use of centrality estimators in p-A collisions based on multiplicity or summed energy in certain pseudo-rapidity intervals is motivated by the observation that they show a linear dependence with $N_{part}$ or $N_{coll}$. ... The {\em total rapidity integrated multiplicity} [emphasis added] of charged particles measured in hadron-nucleus collisions...at [various collision energies] is consistent with a linear dependence on $N_{part}$: $N^{h\text{-A}}_{ch} = N^{pp}_{ch} \cdot N_{part}/2$.'' An approximate proportionality for {\em total particles integrated over $4\pi$ acceptance} may be related to the transport phenomenon noted in Ref.~\cite{transport} wherein, for {\em each} hadron species, transport from soft (nonjet) to hard (jet-related) component via large-angle parton scattering preserves a total yield consistent with statistical-model predictions. What matters for {\em differential} spectra and {\em local} densities near midrapidity are the relative fractions of nonjet and jet-related hadrons that, for \ppb\ collisions especially, rely on accurate A-B geometry determination. 

\subsection{Classical Glauber model and $\bf p$-Pb centrality}
 
A joint distribution $P(n_{ch},N_{part})$ relating observable $n_{ch}$ to geometry parameter $N_{part}$ can in principle be factorized according to Bayes' theorem as $P(n_{ch}|N_{part}) P(N_{part})$. Projection of the joint distribution onto a measurable distribution $P(n_{ch})$ is written as
\bea \label{glaubeq}
P(n_{ch}) &=& \sum_{N_{part}} P(n_{ch}|N_{part}) P(N_{part}).
\eea
To summarize the approach of Ref.~\cite{aliceppbgeom} $P(N_{part})$ is estimated using a classical Glauber Monte Carlo (MC) with its complex of assumptions. It is further assumed that $N_{part} \propto n_{ch}$ as noted above and that $P(N_{part})$ is a relatively broad distribution on $N_{part}$ whereas conditional $P(n_{ch}|N_{part})$ is  relatively narrow on $n_{ch}$ to approximate the assumed relation $N_{part} \propto n_{ch}$. The conditional distribution $P(n_{ch}|N_{part})$ is then modeled with a negative binomial distribution or NBD. The resulting product $P(n_{ch},N_{part})$ is then projected onto \nch\ as $P(n_{ch})$ in Eq.~(\ref{glaubeq}) and fitted to an experimental distribution on some $n_x \sim n_{ch}$ (e.g.\ V0A) to determine NBD parameters. 

That approach fails catastrophically when confronted with ensemble-mean \mmpt\ data as reported in Ref.~\cite{tomglauber} because the assumption $n_{ch} \propto N_{part}$ implies that the mean charge density per \nn\ pair $\bar \rho_{0NN}$ is approximately independent of centrality. 
Primed quantities in Table~\ref{rppbdata} are from Table 2 of Ref.~\cite{aliceppbgeom} and are consistent with $\bar \rho_0 \approx (N_{part}'/2) \bar \rho_{0NN}$ with $\bar \rho_{0NN} \approx 4.5$ vs $\approx 5$ for 5 TeV NSD \pp\ collisions. 
But that in turn implies that jet production per \nn\ binary collision (approximately $\propto \bar \rho_{0NN}^2$) remains the same for all centralities. 
The number of binary collisions per \nn\ pair $\nu = 2 N_{bin} / N_{part}$, with $N_{part} = N_{bin}+1$ for \pa\ collisions, must be less than 2. Ensemble-mean \mmpt\ in the form $\bar P_t / n_s$ ({\em total} \mmpt\ over \nch\ soft component $n_s$) is well described by $\bar p_{ts} + x\nu \bar p_{th}$, $\nu$ is described above and $x \equiv \bar \rho_{hNN} / \bar \rho_{sNN} \approx \alpha \bar \rho_{sNN} \approx 0.05$ for NSD 5 TeV \pn\ collisions. For protons or Lambdas $\bar p_{ts} \approx 0.75$ GeV/c and $\bar p_{th} \approx 1.6$ GeV/c~\cite{pidpart2}, and from above $x\nu \leq 0.10$. Thus, predicted $\bar P_t / n_s$ given assumptions has upper limit $0.75 + 0.1 \times 1.6 \approx 0.9$ GeV/c, but measured $\bar P_t / n_s$ for baryons is as high as $4$ GeV/c~\cite{pidpart2}. 

Section~VII D of Ref.~\cite{tomglauber} explains the relation of the incorrect Glauber strategy above to a proper procedure based on the TCM: Equation~(\ref{glaubeq}) represents an {\em inverse problem} wherein frequency distribution $P(n_{ch})$ is the Data, cross section distribution $P(N_{part}) \rightarrow (1/\sigma_0) d\sigma / dN_{part}$ is the Model, and conditional distribution $P(n_{ch}|N_{part})$ relating \nch\ to $N_{part}$ is the Kernel. The inverse problem is solved if given Data and Kernel the Model can be inferred. For the analysis of Ref.~\cite{aliceppbgeom}  \ppb\ geometry Model $P(N_{part})$ is adopted from a classical Glauber MC based on certain assumptions. The relation of $n_{ch}$ to $N_{part}$ (hadron production model) is also assumed as an NBD distribution combined with the assumption $N_{part} \propto n_{ch}$ as above. The resulting convolution integral is compared to Data in the form of V0A frequency distribution $P(n_{x})$ to ``validate'' Model $P(N_{part})$. 

Several issues are  apparent: (a) Model $P(N_{part})$ -- differential cross section on $N_{part}$ -- is determined by a classical Glauber MC assuming that projectile protons may interact with several target nucleons simultaneously which is contradicted by data. (b) Data $P(n_{ch}) \rightarrow P(n_x)$ -- frequency distribution on V0A $n_x$ -- is derived from a different $\eta$ acceptance than that used to determine the actual $N_{part}$ vs $\bar \rho_0$ relation via \mmpt\ data. (c) Frequency distribution $P(n_x)$ at left in Eq.~(\ref{glaubeq}) is not commensurate with cross section distribution $P(N_{part}) \rightarrow (1/\sigma_0) d\sigma/dN_{part}$ at right. (d) Kernel $P(n_{ch}|N_{part})$ -- conditional distribution -- is modeled by an NBD with free parameters fitted to (a) and (b) via Eq.~(\ref{glaubeq}). The result is qualitatively inconsistent with other measured quantities: ensemble-mean \mmpt\ data as noted and the relevant NSD \pp\ \nch\ distribution (the NBD from Glauber fit is incorrect)~\cite{tomglauber}.

\subsection{TCM and $\bf p$-Pb centrality determination}

As noted above, the unprimed numbers in Table~\ref{rppbdata} are derived from a TCM analysis of ensemble-mean \mmpt\ data from 5 TeV \ppb\ collisions as reported in Ref.~\cite{tommpt}. The large differences between those results and what is reported from the ALICE collaboration in Ref.~\cite{aliceglauber} (primed numbers in Table~\ref{rppbdata}) are explained in Refs.~\cite{tomglauber,tomexclude}. What follows is a brief summary of the results of Ref.~\cite{tomglauber}.

\begin{figure}[h]
\includegraphics[width=1.65in]{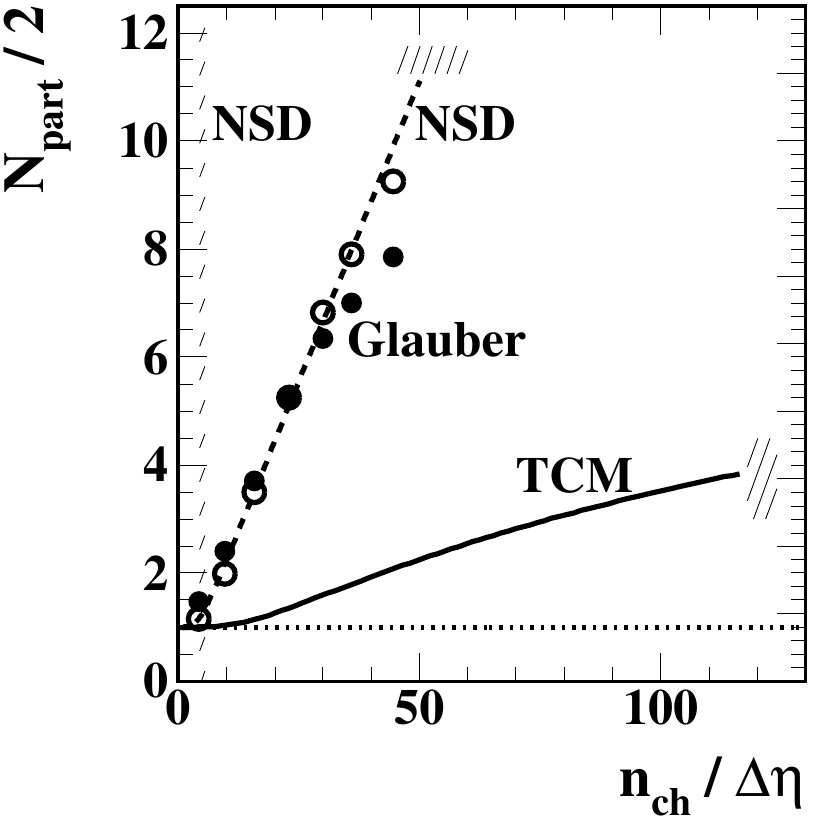}
\includegraphics[width=1.65in]{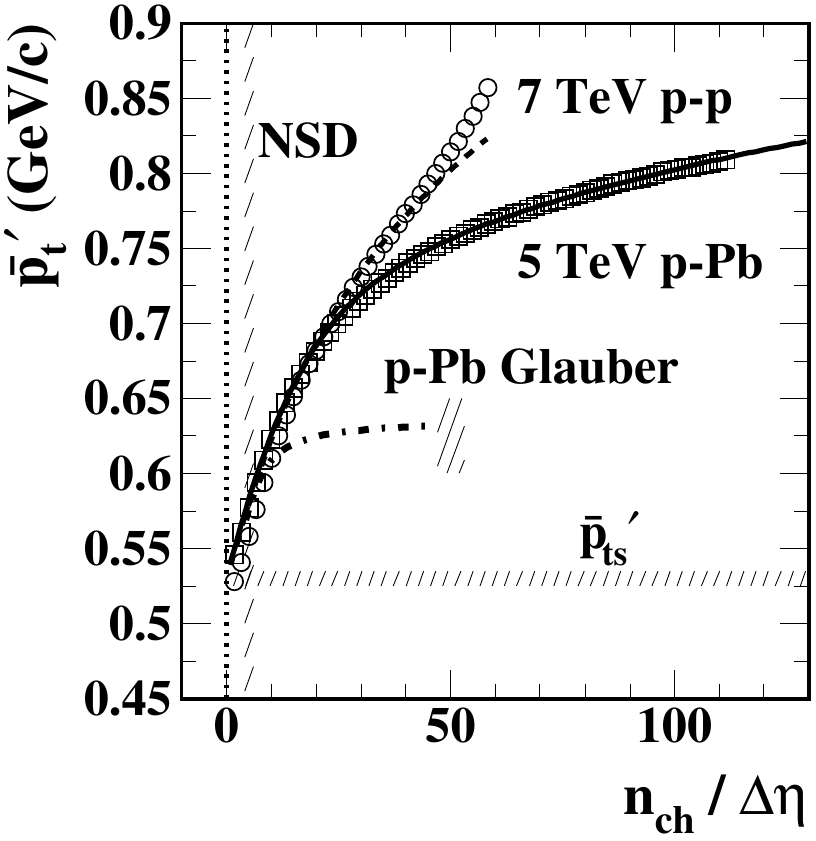}
	\caption{\label{mmpt}
		Left:  5 TeV \ppb\  $N_{part}/2$ vs charge density $\bar \rho_0$ for classical Glauber Monte Carlo (points) and TCM analysis of \mmpt\ data (solid curve). The dashed line represents the assumption that $\bar \rho_0 \approx (N_{part}/2) \bar \rho_{0\text{NSD}}$. The hatched band indicates the NSD value $\bar \rho_0 \approx 5$.
		Right: Ensemble-mean \mmpt\ data for 7 TeV \pp\ collisions and 5 TeV \ppb\ collisions (open circles, open squares), corresponding TCM descriptions (dashed curve, solid curve) compared to a prediction derived from the classical Glauber Monte Carlo (dash-dotted curve).
	} 
\end{figure}

Figure~\ref{mmpt} (left) shows $N_{part}$ vs $\bar \rho_0$ trends for the Glauber analysis in Ref.~\cite{aliceppbgeom} (dashed line and points) and as inferred in Ref.~\cite{tommpt} by inverting the TCM description of ensemble-mean \mmpt\ data (solid curve). The dashed line corresponds to assumed $\bar \rho_0 \approx (N_{part}/2)\bar \rho_{0\text{NSD}}$. The solid dots correspond to primed entries in Table~\ref{rppbdata}.

Figure~\ref{mmpt} (right) shows uncorrected 5 TeV \ppb\ \mmpt\ data from Ref.~\cite{alicempt} (open squares) vs the TCM description from Ref.~\cite{tommpt} (solid curve) corresponding to the solid curve in the left panel. Those \mmpt\ data are derived from spectra that were not extrapolated to zero \pt, hence ``uncorrected.'' The open circles are \mmpt\ data from 7 TeV \pp\ collisions reported in Ref.~\cite{alicempt}. The dashed curve is a \mmpt\ TCM for 5 TeV \pp\ collisions also from Ref.~\cite{tommpt}. This comparison clearly indicates that up to $\bar \rho_0 \approx 15$ \ppb\ collisions are indistinguishable from {\em single peripheral} \pn\ collisions. The dash-dotted curve is a prediction based on primed values from Table~\ref{rppbdata} with $\bar p_{ts}' \approx 0.53$ GeV/c and $\bar p_{th} \approx 1.3$ GeV/c where 0-5\% central occurs at $\bar \rho_0 \approx 45$.

Failure of the Glauber-model approach to describe \mmpt\ data hints at a solution to the \ppb\ geometry problem: As a first step the critical $N_{part}$ vs $\bar \rho_0$ relation may be determined via TCM analysis of \mmpt\ data that are particularly sensitive to {\em jet-related contributions} to spectra and yields. That relation ({Kernel}) as derived from nonPID \mmpt\ data is established in Ref.~\cite{tommpt} wherein \ppb\ \mmpt\ data are accurately represented over a range of charge densities $\bar \rho_0$ up to 115.  Centrality determination in Ref.~\cite{aliceppbgeom} extends only up to $\bar \rho_0 \approx 45$ that it then associates with 0-5\% central \ppb\ collisions.  Results from Ref.~\cite{tommpt} and their large differences from Ref.~\cite{aliceppbgeom} are interpreted in the context of the Glauber model and TCM by Refs.~\cite{tomglauber,tomexclude}.

\begin{figure}[h]
	\includegraphics[width=1.63in]{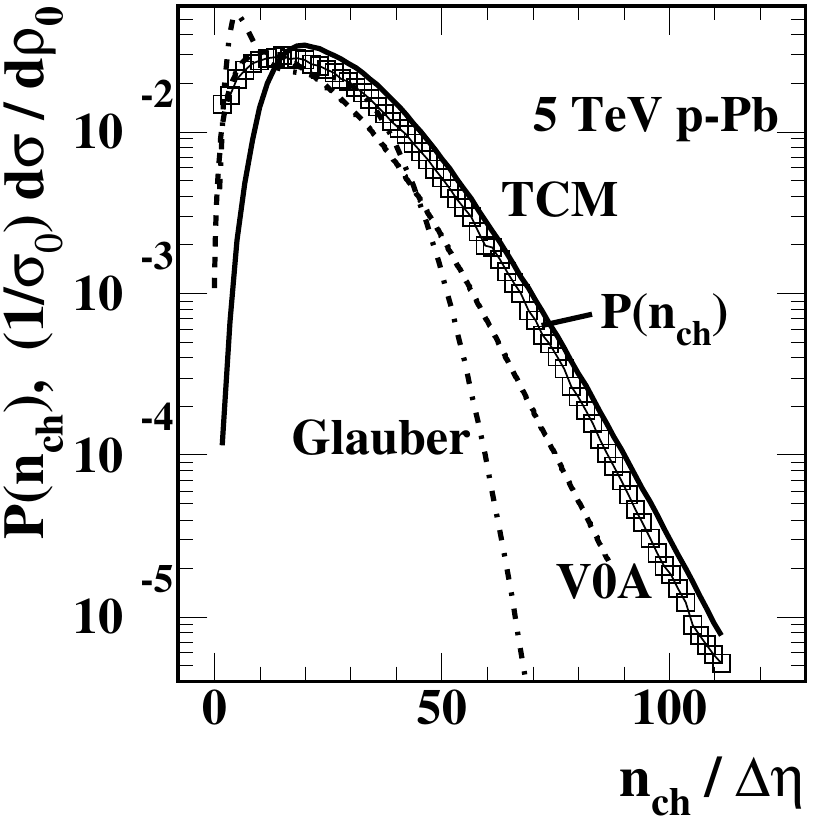}
	\includegraphics[width=1.67in]{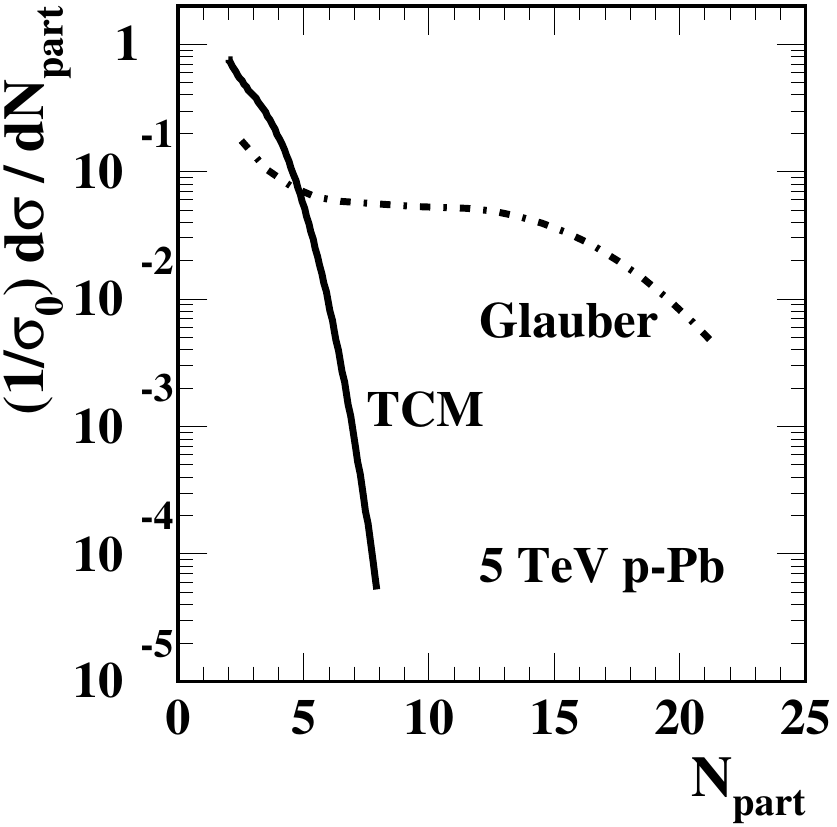}
	\caption{\label{xsects}
		Left: 5 TeV \ppb\ cross-section $(1/\sigma_0)d\sigma /d\bar \rho_0$ and event-frequency $P(n_{ch})$ distributions on charge density $\bar \rho_0$. See text for details.
		Right: Differential cross-section distributions on participant number $N_{part}$ derived from a classical Glauber Monte Carlo (dash-dotted)~\cite{aliceglauber} and from TCM (solid) via analysis of 5 TeV \ppb\ ensemble-mean \mmpt\ data~\cite{tomglauber}.
	} 
\end{figure}

Figure~\ref{xsects} (left) shows an event frequency distribution $P(n_{ch})$ on $\bar \rho_0$ ({Data}, open boxes) derived from statistical errors published with the \mmpt\ data from Ref.~\cite{alicempt}. Utilization of published statistical errors was required to infer a {\em directly-measured} event frequency distribution $P(n_{ch})$ for the relevant $\eta$ acceptance. The dashed curve is the corresponding distribution for V0A data on $n_x$ transformed to \nch\ by an appropriate Jacobian.
The dash-dotted curve is the Glauber cross-section distribution (right) transformed to \nch\ by an appropriate Jacobian. 

The solid curve is derived from Eq.~(\ref{glaubeq}) as follows: The Glauber cross section is modified in product by an additional function. The result is introduced to Eq.~(\ref{glaubeq}) to generate a cross section $(1/\sigma_0) d\sigma/dn_{ch}$ hypothesis on \nch. The modifying function is then adjusted until the cross-section shape (solid curve) best matches the $P(n_{ch})$ data (open boxes) for higher \nch. That procedure solves the inverse problem by {\em trial and error}. The TCM curve is high vs data because both distributions are unit normal. Disagreement between cross section and frequency distribution $P(n_{ch})$ for lower \nch\ is expected. In Fig.~\ref{mmpt} (left) $N_{part}$ is approximately constant ($\approx 2$) at lower \nch, in which case the cross section on $N_{part}$ is constant ($\sigma \approx \sigma_0$) and $d\sigma/dn_{ch} \rightarrow 0$. Details are provided in Ref.~\cite{tomglauber}.

Figure~\ref{xsects} (right) shows the Glauber cross section derived in Ref.~\cite{aliceppbgeom} (dash-dotted) and TCM cross section derived via trial and error through Eq.~(\ref{glaubeq}) (solid). The two shapes are assumed to match for small $N_{part}$ (where {\em exclusivity} should not play a role~\cite{tomexclude}). The TCM version  (sought-after {Model}) then deviates strongly downward to accommodate frequency data $P(n_{ch})$ ({Data}) (via the {Kernel} established from \mmpt\ data) as shown in the left panel (open boxes). The TCM cross section is above Glauber at low $N_{part}$ because both distributions are unit normal.
The result indicates that $N_{part}$ does not exceed 8 for the most central \ppb\ collisions. A Pb nucleus is approximately eight tangent nucleons in diameter.

\section{Nuclear modification factors} \label{nucmod}

The nuclear modification factor (NMF), a rescaled spectrum ratio, is intended to reveal changes in jet production due to possible formation of a dense medium in nucleus-nucleus collisions. The measure is based on the assumption that in the absence of such effects \nn\ collisions within A-B collisions should be statistically equivalent to isolated MB \pp\ collisions.  ``In the absence of nuclear effects the $R_\text{pPb}$ [at higher \pt] is expected to be one''~\cite{alicenucmod}. Given detailed spectrum structure in a TCM context and alternative determinations of \ppb\ geometry as described above it is of interest to revisit the details of NMFs for PID spectra from 5 TeV \ppb\ collisions.

\subsection{Nuclear modification factor definition}

The nuclear modification factor was introduced as a possible method to detect modification of jet production in high-energy nuclear collisions in the event that a deconfined QGP is produced, e.g.\ to reveal significant parton energy loss within a dense QCD medium. The basic concept is comparison of a \pt\ spectrum from more-central \aa\ collisions with a reference spectrum from minimum-bias \pp\ collisions in which QGP effects should not be present. The NMF is a spectrum ratio wherein the jet-related contribution (how defined?) is of primary interest. In that case, {\em based on several assumptions}, rescaling the spectrum ratio by an estimated number of \nn\ binary collisions $N_{bin}$ should allow direct evaluation of jet contributions for individual \nn\ binary collisions. That reasoning is related to certain ``Cronin effect'' manifestations appearing in spectrum ratios as discussed in Sec.~\ref{cronineff}. In what follows rescaling is omitted from NMFs to provide clearer correspondence with TCM hard components.

\subsection{Circumventing the rescale problem}

Equation~(\ref{rpbp}) defines spectrum ratio $R_{p\text{Pb}}'$ expressed in terms of a PID TCM for \ppb\ collisions. The prime indicates that estimated $N_{bin}$ has been omitted from this ratio definition. The first line incorporates TCM algebra that predicts individual PID spectra within their point-to-point data uncertainties. In the numerator $\bar \rho_s = (N_{part}/2) \bar \rho_{sNN}$ and $\bar \rho_h = N_{bin} \bar \rho_{hNN}$ for \ppb\ collisions.
\bea \label{rpbp}
R_{p\text{Pb}}' &=& \frac{ z_{si}(n_s) \bar \rho_{s} \hat S_{0i}(p_t) +   z_{hi}(n_s) \bar \rho_{h} \hat H_{0ip\text{Pb}}(p_t,n_s)}{z_{sipp} \bar \rho_{spp} \hat S_{0i}(p_t) + z_{hipp} \bar \rho_{hpp} \hat H_{0ipp}(p_t)}~~
\\ \nonumber 
&\rightarrow& \frac{ z_{si}(n_s) (N_{part}/2) \bar \rho_{sNN} \hat S_{0ipPb}(p_t)}{z_{sipp} \bar \rho_{spp} \hat S_{0ipp}(p_t)} ~~\text{for low \pt}
\\ \nonumber 
&\rightarrow& \frac{z_{hi}(n_s) N_{bin} \bar \rho_{hNN} \hat H_{0ip\text{Pb}}(p_t,n_s)}{z_{hipp} \bar \rho_{hpp} \hat H_{0ipp}(p_t)}~~\text{for high \pt},
\eea
with conventional unprimed NMF $R_{p\text{Pb}}=(1/N_{bin})R_{p\text{Pb}}'$.

For the limiting case at low \pt\ the soft-component model $\hat S_{0i}(p_t)$ is observed to be independent of multiplicity or centrality for any collision system. Refer to results in Sec.~\ref{smallspec} where spectra maintain the same form within statistical uncertainties below \yt\ = 2 ($p_t \approx 0.5$ GeV/c) and where assumption of a fixed $\hat S_{0i}(p_t)$ model function leads to isolation of spectrum hard components with intact shapes down to or even below 0.5 GeV/c. If $\hat S_{0i}(p_t)$ is then canceled in ratio what remain are factors $z_{si}(n_s)\bar \rho_{sNN}$ for \ppb\ and \pp\ collisions and geometry parameter $N_{part}/2$ {\em that should be common to all hadron species}. Given the hypothesis that \ppb\ collisions are linear superpositions of MB \pn\ collisions, cancellation of the $z_{si}(n_s)\bar \rho_{sNN}$ factors leaves $N_{part}/2$ isolated, implying that $R_{p\text{Pb}}'$ at low \pt\ provides an estimate {\em from spectrum data} of the required $N_{bin} = N_{part} - 1$ for \ppb\ collisions.

For the limiting case at high \pt, and based on the same linear superposition assumption, factors $z_{hi}(n_s)\bar \rho_{hNN}$ from \ppb\ and \pp\ collisions should cancel leaving the expression $N_{bin}\hat H_{0ip\text{Pb}}(y_t,n_s)/\hat H_{0ipp}(y_t)$. In  the absence of jet modification that expression simplifies to $N_{bin}$ alone (in principle determined from the low-\pt\ limiting case) justifying the definition of conventional {\em unprimed} $R_{p\text{Pb}}$.

\subsection{Nuclear modification factor data} \label{nmfdata}

\begin{figure}[h]
	\includegraphics[width=3.3in]{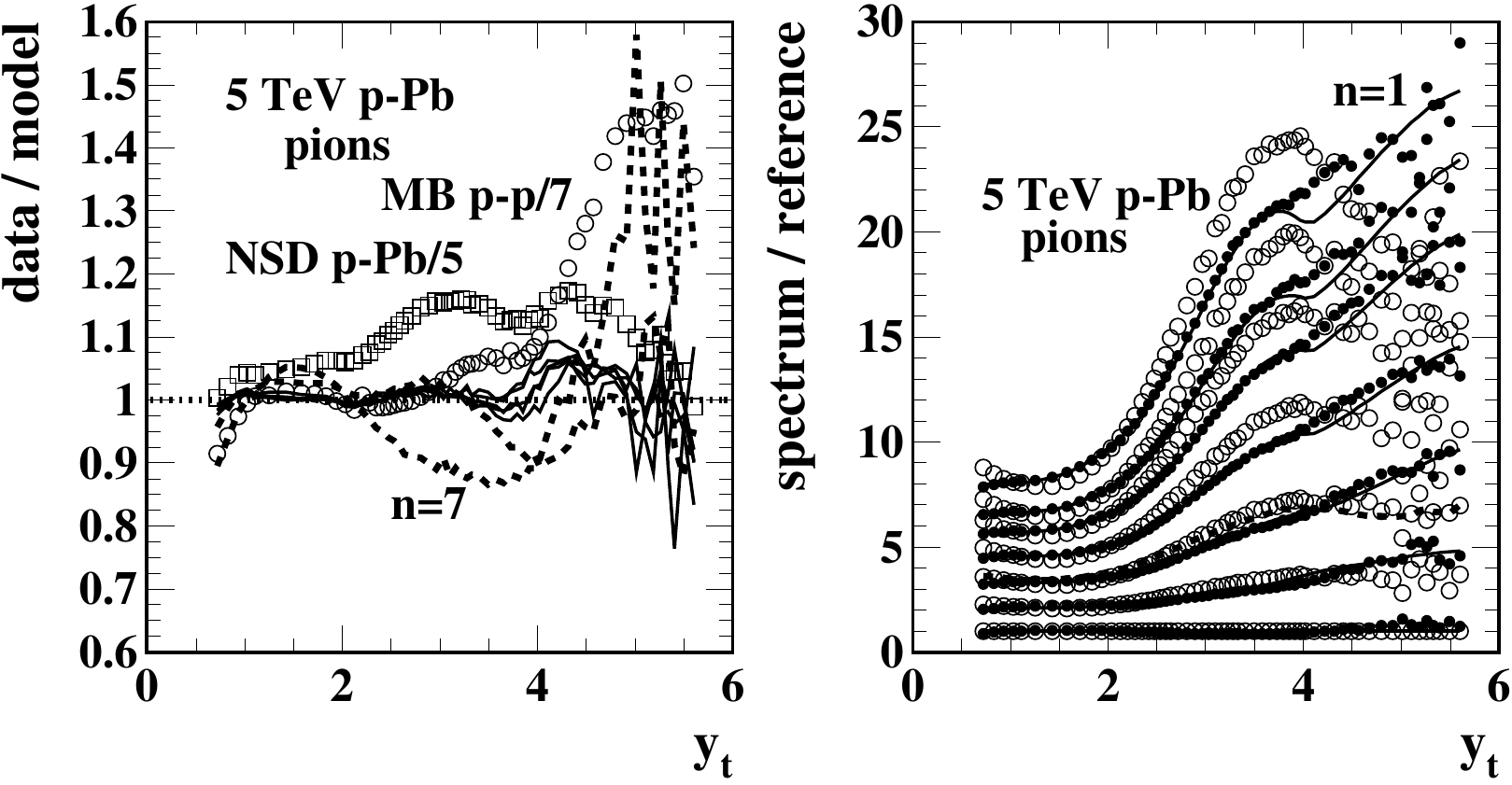}
	\caption{\label{10a}
		Left: Data/model ratios for PID spectra vs TCM (solid and dashed), NSD \ppb\ vs TCM $n=5$ (open boxes) and MB \pp\ vs TCM $n=7$ (open circles).
		Right: Spectrum/reference ratios defined by Eq.~(\ref{rpbp}) for spectrum data vs data reference $n = 7$ (open circles), for spectrum data vs TCM reference $n = 7$ (solid dots) and for TCM vs TCM reference (solid curves). The spectrum ratio NSD \ppb\ vs MB \pp\ is shown as the dashed curve near other $n=5$ ratios.
	}   alippbhi10a
\\ redo
\end{figure}

Figure~\ref{10a} (left) shows data/model ratios for 5 TeV \ppb\ pion spectra vs TCM repeated from Fig.~\ref{pionspec} (right) (solid and dashed), the spectrum from NSD \ppb\ collisions in ratio to TCM model spectrum for $n=5$ (open squares) and the spectrum from MB \pp\ collisions in ratio to TCM spectrum for $n = 7$ (open circles). The $n=7$ (peripheral) data/model ratios (dashed) are emphasized because they are strongly biased, as is typical for spectra for low-\nch\ \pp\ or peripheral \ppb\ collisions~\cite{alicetomspec}.

Figure~\ref{10a} (right) shows spectrum/reference ratios $R_{p\text{Pb}}' $ where the reference is the $n = 7$ data spectrum (open circles), the $n=7$ TCM spectrum (solid dots) or the ratio is NSD \ppb\ spectrum to MB \pp\ spectrum (dashed curve close to $n=5$ \ppb\ ratios). The solid curves are TCM spectra in ratio to the TCM reference for $n=7$. If the biased $n=7$ data spectrum is used as reference the NMF data ratios in the right panel (open circles) are in turn biased, whereas with the TCM $n=7$ spectrum as the reference the data ratios (solid dots) are well represented by  TCM model ratios (solid curves).

\begin{figure}[h]
	\includegraphics[width=3.3in]{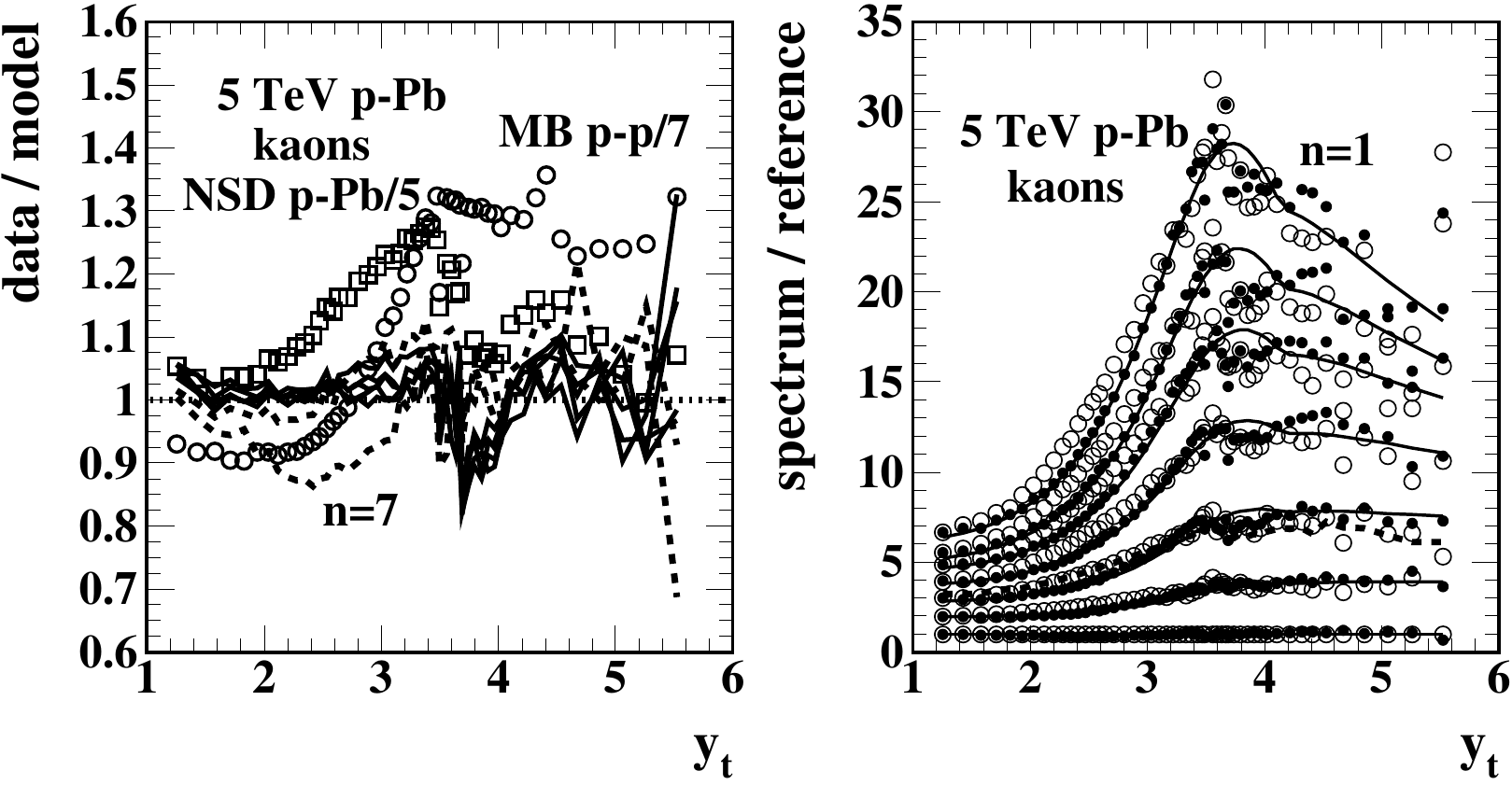}
	\caption{\label{10b}
Same as Fig.~\ref{10a} except for charged kaons.
	}   alippbhi10b
\\ redo
\end{figure}

Figure~\ref{10b}  shows similar results for charged kaons.

\begin{figure}[h]
	\includegraphics[width=3.3in]{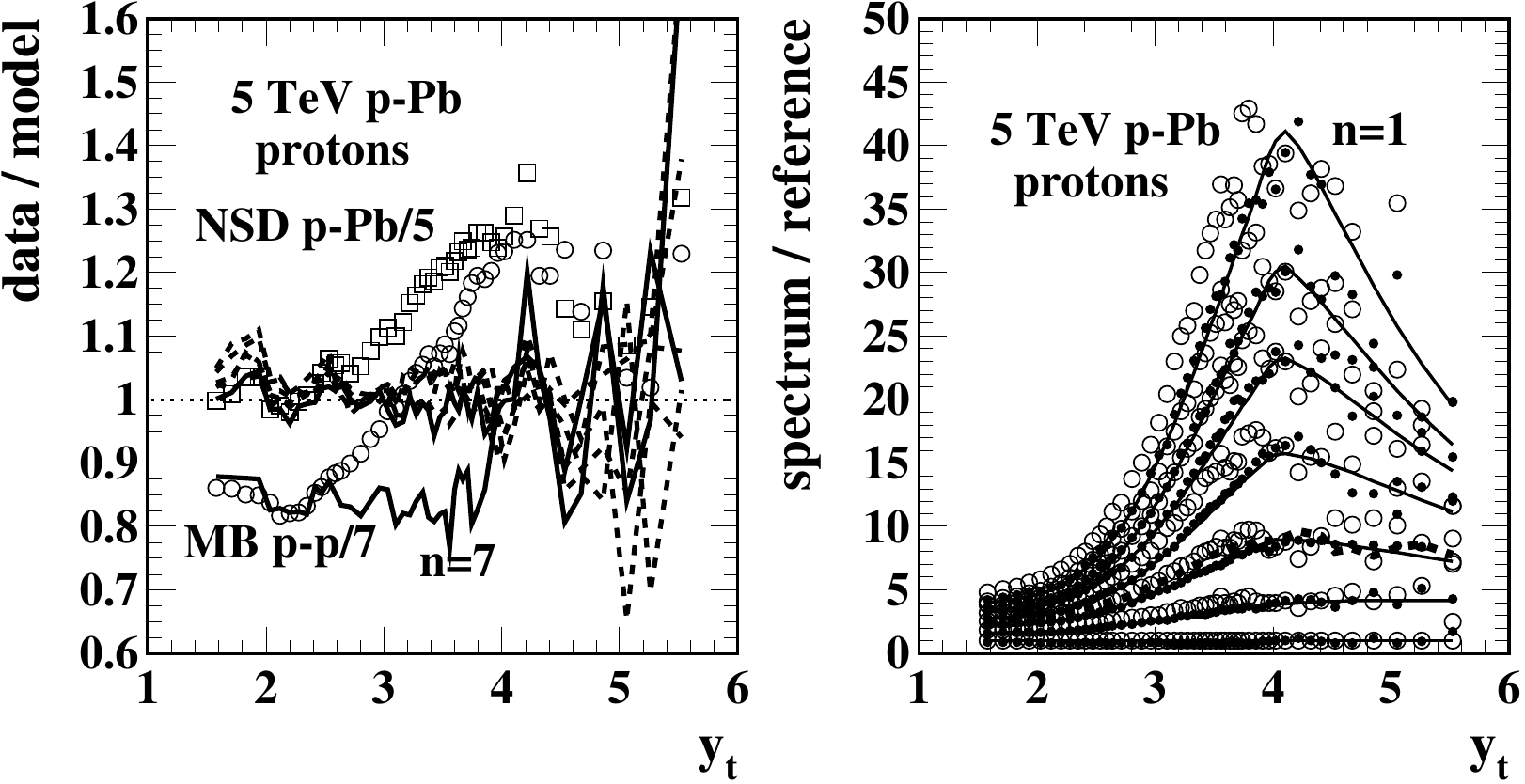}
	\caption{\label{10c}
Same as Fig.~\ref{10a} except for protons.
	}   alippbhi10c
\\ redo
\end{figure}

Figure~\ref{10c} shows similar results for protons.

As noted above, if the assumption of linear superposition of MB \pn\ collision within \ppb\ collisions were correct values for $N_{part}/2$ could be read off the low-\pt\ parts of the NMFs in the form $R_{p\text{Pb}}' $, and they should have the same common value for three hadron species. That is obviously not the case, the values for event class $n = 1$ (most central) being approximately 8, 6 and 4 for pions, kaons and protons. In turn, those values should predict $N_{bin} = N_{part} - 1$ leading to values 15, 11 and 7 for pions, kaons and protons. If there were no nuclear modification those values should correspond in some sense to ratios $R_{p\text{Pb}}' $ at high \pt, but that is also not the case. Those discrepancies indicate {\em in part} the substantial problems for interpretation of nuclear modification factors.


The low-\pt\ trend for $R_{p\text{Pb}}'$ can be evaluated as an example for event class $n = 1$ (central) vs reference $n = 7$ (peripheral). As noted, based on an array of specific TCM spectrum analyses it is established that model function $\hat S_{0i}(p_t)$ can be canceled as invariant. According to Table~\ref{rppbdata} $N_{part}/2$ is 2.1 and 1 for the two cases. The charge-density ratio for $\bar \rho_{sNN}$ is 16.6 / 4.2 = 3.95. The combination is thus $2.1 \times 3.95 = 8.3$ that should be common to all hadron species. What remains is the species fraction ratios $ z_{si}(n_s) / z_{sipp}$ that {\em are} unique to each hadron species. Measured values are obtained from Table~IV of Ref.~\cite{pidpart1}. For pions the $z_{si}$ ratio is consistent with unity; the resulting 8.3 is consistent with Fig.~\ref{10a} (right). For kaons the ratio is 0.091 / 0.113 = 0.80 with the total ratio being $0.80 \times 8.3 = 6.6$ consistent with Fig.~\ref{10b} (right). For protons the ratio is 0.028 / 0.053 = 0.53 with the total ratio being $0.53 \times 8.3 = 4.4$  consistent with Fig.~\ref{10c} (right). Thus, previous analysis of \ppb\ geometry and PID hadron fractions {\em predicts} the low-\pt\ trends of ratio $R_{p\text{Pb}}'$ but falsifies the assumption that \pn\ collisions within \ppb\ collisions are independent of \ppb\ centrality.


In the high-\pt\ limit the expression $ N_{bin}\, \bar \rho_{hNN}/ \bar \rho_{hpp} \rightarrow 3.2 \times (3.95)^2 \approx 50$ where the relation $\bar \rho_{hNN} \propto \bar \rho_{sNN}^2$ has been invoked and the numbers come from Table~\ref{rppbdata}. That serves as a reference value for further discussion. The $z_{hi}$ ratios can be obtained from Table~V of Ref.~\cite{pidpart1} as $\approx 1$ for pions, 0.25 / 0.27 = 0.93 for kaons and 0.17 / 0.29 = 0.60 for protons. What remains is the data hard-component ratios approximated by TCM models as $\hat H_{0ip\text{Pb}}(y_t,n_s)/\hat H_{0ipp}(y_t)$. Given the assumptions invoked for $R_{p\text{Pb}}$, for {\em no jet modification} hard-component shapes (model functions) should cancel and ratios for central collisions should saturate at the constant values 50, $0.93 \times 50 \approx 46$ and $0.60 \times 50 \approx 30$ for pions, kaons and protons.

That reveals a major problem for interpretation of spectrum {\em ratios}. The jet-related hard-component modes for all hadron species appear near \yt\ = 2.7 ($p_t \approx 1$ GeV/c) (see Sec.~\ref{spechard}). Variation of hard-component structure below the mode is as meaningful as that above the mode, but within ratios of {\em complete} spectra hard components are strongly suppressed by soft components below \yt\ = 4 ($p_t \approx 3.8$ GeV/c) where  the {\em great majority} of jet fragments resides. Ratio structure at higher \pt\ does not distinguish between biases for more-peripheral event classes (common) and relevant {\em jet fragmentation} evolution for more-central event classes. In short, spectrum ratios $R_{p\text{Pb}}$ {\em alone} are {\em not physically interpretable}.

\subsection{NMF physical interpretation via TCM} \label{interpret}

Accurate isolation of PID spectrum hard components as in Sec.~\ref{spechard} and parametrization of their evolution with \ppb\ centrality as in Tables~\ref{pidparam1} and \ref{pidparam2} does permit further interpretation of NMFs at higher \pt\ in terms of parameters $\bar y_t$, $\sigma_{y_t}$ and $q$. For all hadron species the NMF increases rapidly above \yt\ = 2 as the spectrum hard/soft (jet/nonjet) ratio increases with \yt.  The following comments apply most directly to central \ppb\ event class $n = 1$ in ratio to peripheral $n = 7$.

Since jet production increases with \nch\ proportional to hadron mass~\cite{transport} the low-mass pion soft component still influences the pion NMF above 4 GeV/c. The hard-component mode is stationary, but the width above the mode decreases with increasing \ppb\ centrality while $q$ remains constant. Due to the width decrease, above 4 GeV/c the pion NMF decreases, {\em uniformly} across \yt, by approximately factor 2 from reference value 50 to approximately 23. The pion NMF continues to increase {\em with} \yt\ due to decreasing soft component vs the jet contribution.

For kaons as for pions the mode is approximately stationary, and width above the mode decreases at a similar rate. The resulting NMF values above \yt\ =4 are thus also approximately 23.  However, kaon exponent $q$ {\em increases} substantially (the exponential tail becomes softer) leading to the negative slope of the kaon NMF above \yt\ = 4, increasingly so with increasing \ppb\ centrality.

For protons the mode shifts substantially to higher \yt\ with increasing centrality while the width above the mode is stationary. Average values near \yt\ = 4 are thus {\em above} the proton reference value $0.6 \times 50 \approx 30$ (taking into account species fraction ratio 0.6). Exponent $q$ increases very substantially for protons leading to increasingly negative slope of the NMF on \yt\ above \yt\ = 4, more so than for kaons. The combined mode shift and exponent increase lead to pronounced narrowing with centrality of the proton NMF peak near \yt\ = 4 relative to kaons. Those trends relate to observation of varying degrees of ``Cronin enhancement'' for different hadron species.

While the established TCM context permits {\em qualitative} interpretation of NMF evolution with \yt, centrality and hadron species there is no quantitative information accessible. Spectrum {\em ratios} provide some sensitivity to changes in hard-component peak shapes but not the absolute shapes that might be physically interpretable. For the latter see Sec.~\ref{spechard}. The log of a ratio is a difference, also a relative measure that does not convey absolute quantities. In either case essential information carried by particle spectra is discarded, reducing the ability to challenge theory -- a basic element of the scientific method.

\begin{figure}[h]
	\includegraphics[width=1.65in]{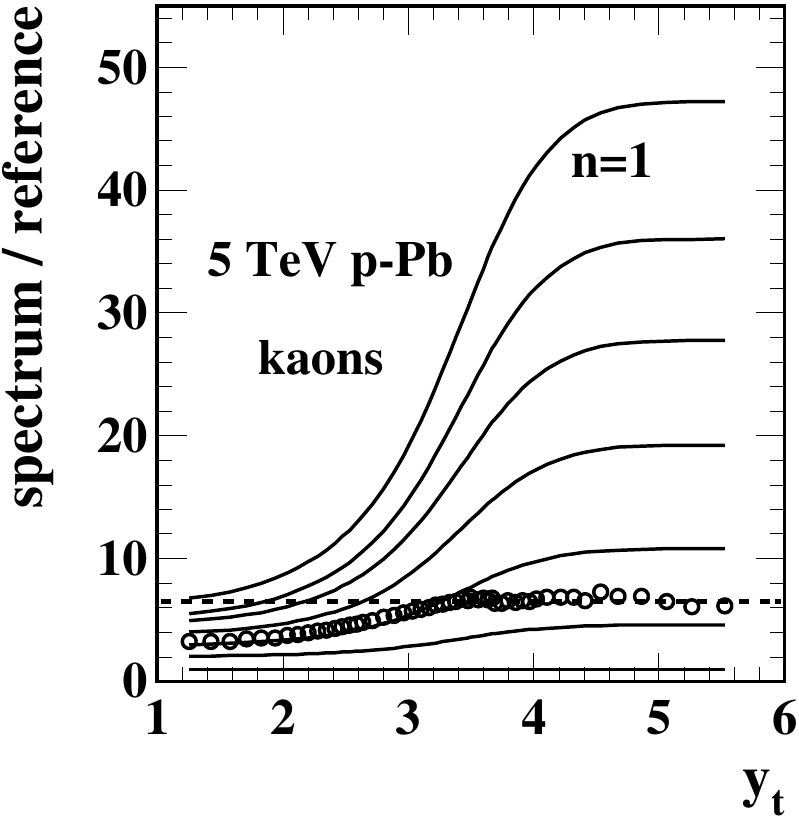}
	\includegraphics[width=1.65in]{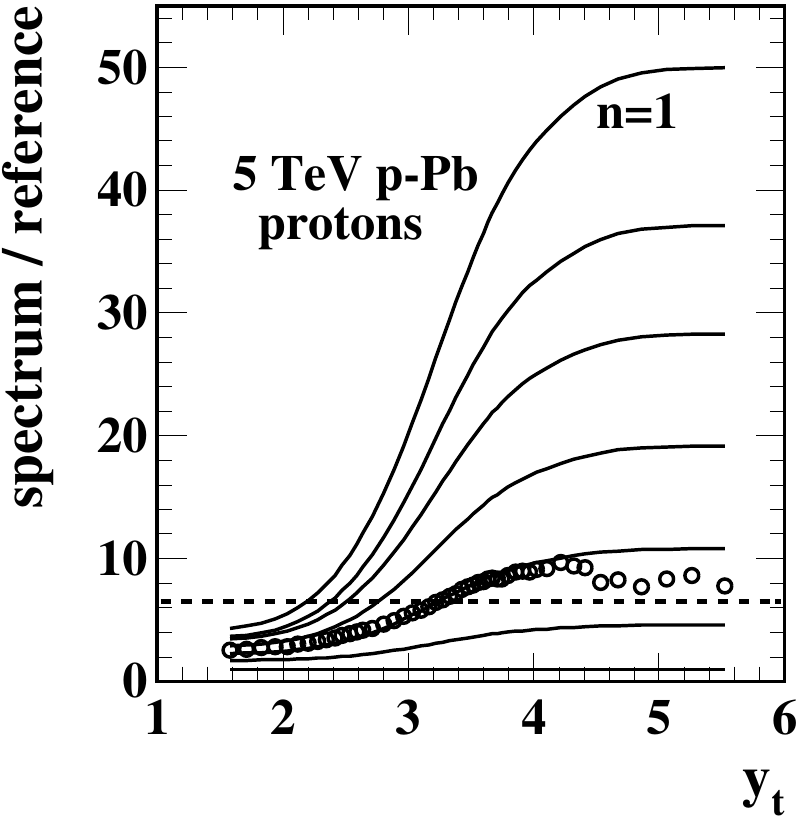}
\put(-145,92) {\bf (a)}
\put(-23,92) {\bf (b)}
\\
	\includegraphics[width=1.65in]{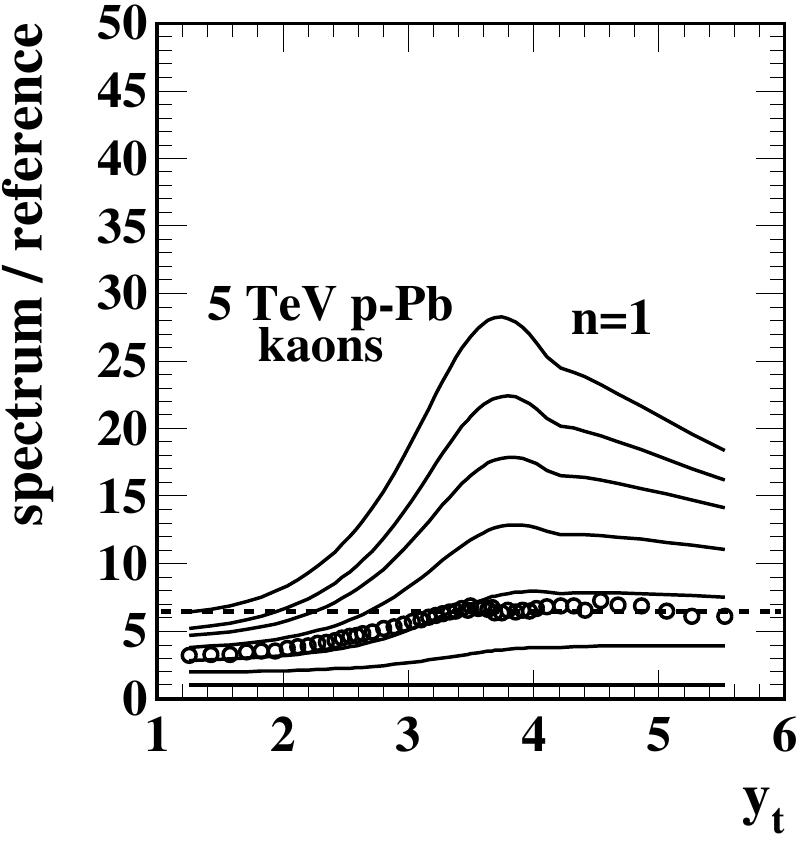}
\includegraphics[width=1.65in]{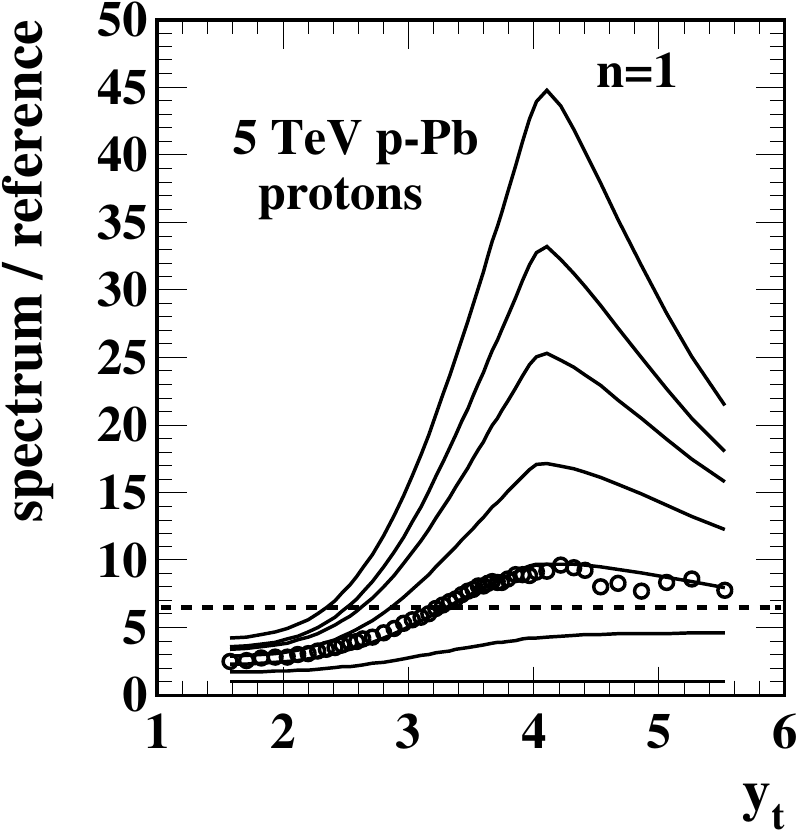}
\put(-145,90) {\bf (c)}
\put(-23,90) {\bf (d)}
	\caption{\label{off} 
		(a,b) Figures~\ref{10b} and \ref{10c} (right) with the TCM held {\em fixed} at parameter values for \ppb\ event class $n=4$. The open circles are NSD \ppb/MB \pp\ spectrum ratios from Ref.~\cite{alicenucmod}. The other points are omitted for clarity.
		(c,d) These solid curves represent the {\em variable} TCM appearing in Figs.~\ref{10b} and \ref{10c} (right) that describes spectrum data within uncertainties. This comparison demonstrates what parts of NMF spectrum ratios are actually sensitive to the jet fragment contribution.
		} 
\end{figure}

Figure~\ref{off} (a,b) shows spectrum/reference ratios with a {\em fixed} TCM: the hard-component models are fixed at parameter values for event class $n = 4$, i.e.\ no ``nuclear modification'' by construction. For a fixed TCM the models $\hat H_{0ixx}(y_t,n_s)$ cancel in the third line of Eq.~(\ref{rpbp}) leaving factor $z_{hi}(n_s) / z_{hipp}$ times the ratio of factors $N_{bin}(n_s) \bar \rho_{hNN}(n_s)$ that are {\em common to all hadron species} for event class $n$ vs 7. For protons $z_{hi}(n_s)$ is also held fixed because it determines the mode position. Those $z_{hi}$ factors then cancel leaving only the common ratio $\approx 50$. For kaons the  $z_{hi}(n_s)$ factors are retained leading to somewhat smaller high-\pt\ amplitudes for the kaon trends. The solid points in Figs.~\ref{10b} and \ref{10c} have the wrong reference in this limiting case ($n = 4$, not $n = 7$) and are therefore omitted along with open circles to improve clarity.
What remains are TCM curves that transition from a low-\pt\ reference to a high-\pt\ reference with shape depending only on {\em relative} shapes of fixed soft- and hard-component model functions $\hat S_0(y_t)$ and $\hat H_0(y_t)$. 

Figure~\ref{off} (c,d) shows the solid curves in Figs.~\ref{10b} and \ref{10c} (right) with {\em variable} TCM for comparison. The point of this exercise is a demonstration that ratio structure below \yt\ = 3.5 ($p_t \approx 2.5$ GeV/c) is determined only by the general shapes of $\hat S_0(y_t)$ and $\hat H_0(y_t)$ and is otherwise not sensitive to hard-component details. The ratio shapes above \yt\ = 3.5 are largely controlled by hard-component width $\sigma_{y_t}$ above the mode and exponential slope $q$. As noted there is no quantitative information forthcoming. Substantial variation of NMFs  with centrality in Figs.~\ref{10a}, \ref{10b} and \ref{10c} lies above \yt\ = 4 ($p_t \approx 3.8$ GeV/c). Integration of hard-component model functions above 3.8 GeV/c indicates that {\em only about 2\% of the total fragment  distribution lies within that interval}, implying that 98\% of jet fragments go unrepresented by NMF systematics.

Returning to the issue of NMFs in Ref.~\cite{alicenucmod}, there is the assertion that ``At high \pt\ all nuclear modification factors are consistent with unity [within errors].''  That should apply to any centrality according to the assumption of linear superposition of \pn\ collisions in the sense that all are equivalent to MB \pp\ collisions. The required rescale parameter $N_{bin}$ is estimated as follows: The mean nuclear overlap function is given as $\langle T_\text{pPb}\rangle \approx 0.10$ mb$^{-1}$ and $\sigma_\text{inel} \approx 65$ mb for $\sqrt{s_\text{NN}} = 5$ TeV. The combination leads to $N_{bin} \approx 6.5$ which is marked by the dashed lines in Fig.~\ref{off}. The restriction to NSD \ppb\ collisions as compared to alternative event classes may simply follow the precedent, associated with the ``Cronin effect,'' set by the C-P collaboration as described in Sec.~\ref{cronineff} below.

In Fig.~9 of Ref.~\cite{alicenucmod} ``The total normalization uncertainty is indicated by a vertical scale of the empty box....'' The amplitude of the box appears to be about 5\% and presumably includes the uncertainty in the $N_{bin}$ estimate. It is true that the estimate 6.5 leads to the apparent agreement with unity in that figure, but there is good reason to discount the agreement as anything other than accidental. Event class $n = 5$ for \ppb\ collisions corresponds to $\bar \rho_0 \approx 16$ (see Table~\ref{rppbdata}). Referring to Fig.~\ref{mmpt} (right) that charge density corresponds to a point where \pp\ and \ppb\ \mmpt\ trends remain indistinguishable, implying that \ppb\ collisions at that charge density are almost all single peripheral \pn\ collisions. The corresponding $N_{bin}$ estimate is 1.3, not 6.5. The disconnect arises because \pn\ collisions within \ppb\ collisions have a mean event multiplicity that increases rapidly with \ppb\ centrality, is not constant as is conventionally assumed.

In Fig.~\ref{off} the open circles are NSD \ppb/MB \pp\ spectrum ratios for charged kaons (a,c) and protons (b,d) as they appear in Fig.~9 of Ref.~\cite{alicenucmod}. NSD \ppb\ spectra are approximately consistent with \ppb\ event class $n = 5$ whereas MB \pp\ spectra are approximately consistent with \ppb\ event class $n = 7$ (see Figs.~\ref{10b} and \ref{10c},  left) albeit there are quite substantial deviations from unity. Thus, the NSD \ppb/MB \pp\ ratios (open circles) in Fig.~\ref{off} are approximated by the $n=5$ TCM ratios. In fact, the TCM describes proton data within point-to-point uncertainties because the deviations from a TCM reference in Fig.~\ref{10c} (left) are nearly the same for NSD \ppb\ and MB \pp\ and thus cancel in the $R_{p\text{Pb}}'$ ratio.

\section{PID spectrum ratios} \label{specrat}

Reference~\cite{alicenucmod} applies PID spectra from 5 TeV \ppb\ collisions, extended to higher \pt, to kaon/pion and proton/pion spectrum ratios. It asserts that the kaon/pion ratio does not show any multiplicity dependence whereas the proton/pion ratio ``...shows a clear multiplicity evolution at low and intermediate \pt\ [$< 10$ GeV/c or \yt\ $<$ 5]. This multiplicity evolution is {\em qualitatively similar} [emphasis added] to the centrality evolution observed in \pbpb\ collisions.'' It is further asserted that within data uncertainties ``...the proton-to-pion ratios exhibit {\em similar flow-like features} [emphasis added] for the \ppb\ and \pbpb\ systems, namely the ratios are below the pp baseline for $p_t < 1$ GeV/c and above for $p_t > 1.5$ GeV/c.'' That description appears to confuse the Cronin effect in A-dependent spectrum ratios for a given hadron species with peak structure in ratios for different hadron species

In its introduction Ref.~\cite{alicenucmod} states that ``In heavy-ion collisions at ultra-relativistic energies, it is well established that a...[quark-gluon plasma or QGP] is formed. Some of the characteristic features...are strong collective flow and opacity to jets.'' Upon considering its \ppb\ spectrum ratios in relation to previous \pbpb\ results Ref.~\cite{alicenucmod} comments that ``...the results for \ppb\ collisions appear to raise questions about  the long standing ideas of specific [i.e.\ different] physics models for small and large systems.''  Taken together  those several statements appear to argue that PID spectrum ratios support inference of flows in \ppb\ collisions based on argument from analogy.

For spectrum ratios relating hadron species $i$ to species $j$ Eq.~(\ref{rpbp}) can be reformulated as
\bea \label{ijrat}
R_{ij} &=& \frac{z_{si}(n_s) \bar \rho_{s} \hat S_{0i}(p_t) +   z_{hi}(n_s) \bar \rho_{h} \hat H_{0ip\text{Pb}}(p_t,n_s)}{z_{sj}(n_s) \bar \rho_{s} \hat S_{0j}(p_t) +   z_{hj}(n_s) \bar \rho_{h} \hat H_{0jp\text{Pb}}(p_t,n_s)}~~
\\ \nonumber 
&\rightarrow& \frac{z_{si}(n_s) \hat S_{0i}(p_t)}{z_{sj}(n_s)  \hat S_{0j}(p_t)} ~~\text{for low \pt}
\\ \nonumber 
&\rightarrow& \frac{z_{hi}(n_s) \hat H_{0ip\text{Pb}}(p_t,n_s)}{z_{hj}(n_s)  \hat H_{0jp\text{Pb}}(p_t,n_s)}~~\text{for high \pt}.
\eea

\begin{figure}[h]
	\includegraphics[width=3.3in]{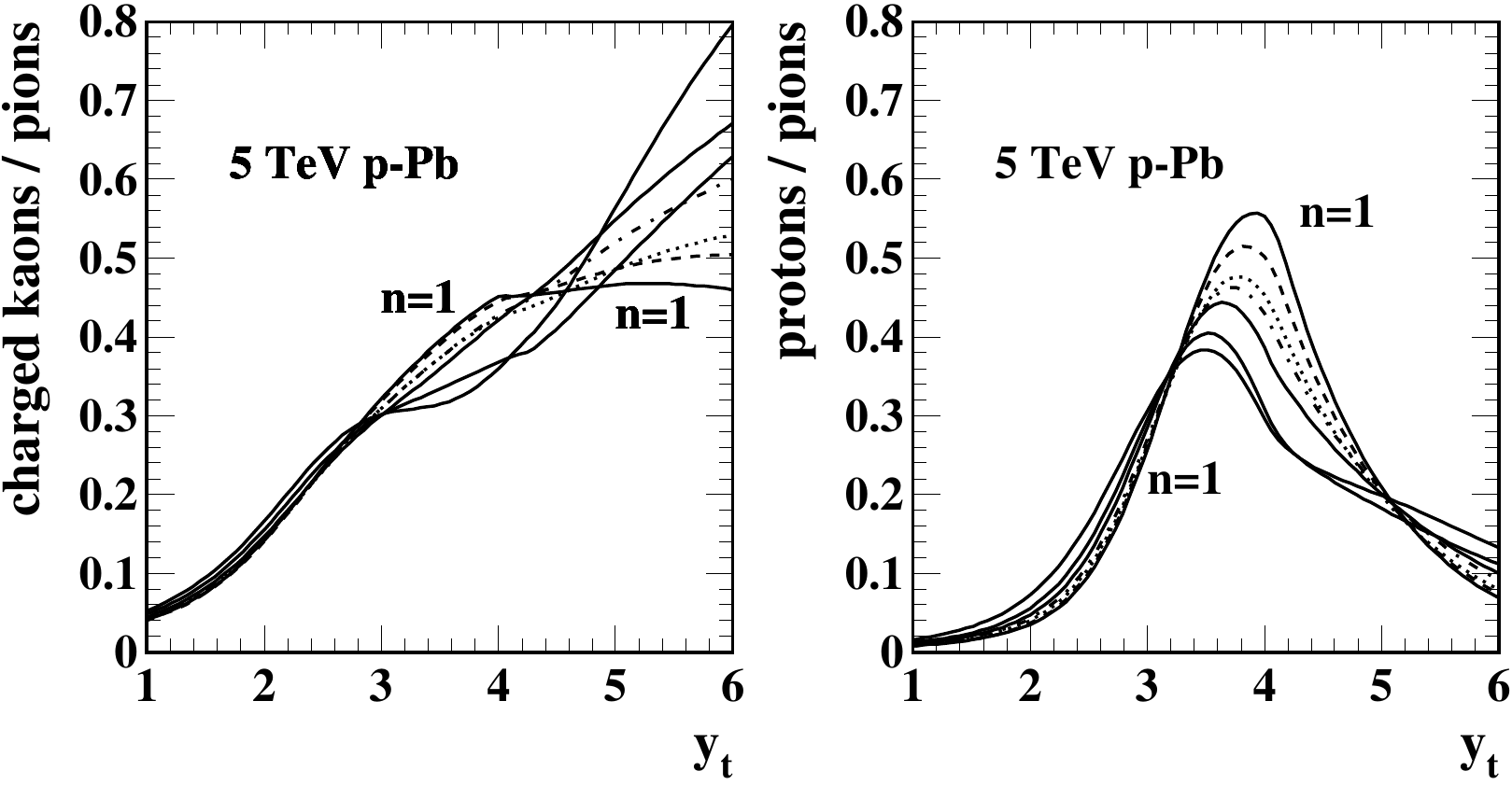}
	\caption{\label{pidrat}
		Left: Kaon/pion spectrum ratios for seven event classes of 5 TeV \ppb\ collisions. The most-central event class is $n = 1$.
		Right: Same as left panel except proton/pion ratios.
	}  
\end{figure}

Figure~\ref{pidrat} (left) shows TCM kaon/pion ratios. The PID TCM is demonstrated to represent charged meson spectra within point-to-point uncertainties and is therefore {\em statistically equivalent}. Given Eq.~(\ref{ijrat}) as representing PID spectrum ratios, at low \yt\ factor $N_{part}/2$ cancels as well as $\bar \rho_{sNN}$. Ratios of species fractions $z_{si}$ are $\approx 0.1 / 0.8 = 0.125$. However, unit-normal soft components $\hat S_{0i}(y_t)$ in this case do not cancel. Their ratio includes two aspects: (a) for higher-mass hadrons unit-normal $\hat S_0(y_t)$ extends to higher \yt\ and is therefore lower at low \yt, and (b) the pion soft component includes a substantial resonance contribution below 0.5 GeV/c ($y_t < 2$)~\cite{pidpart1}. Thus the ratio reduction to $\approx 0.05$.

If the hard components were equivalent at higher \yt\ (and so canceling) the spectrum ratio would transition to a $z_{hi}(n_s)$ ratio leading to saturation at 0.26/0.6 $\approx 0.4$.  Close examination of Figs.~\ref{pionhc} and \ref{kaonhc} (left) shows that the exponential tails approximately coincide as an apparent accident of displaced centroids and differing widths. Prominent rise of the kaon/pion ratio with \yt\ up to \yt\ = 4 followed by saturation near 0.4 (at least for more-central data)  is thus easily explained by the transition from smaller soft-component ratios to larger hard-component ratios per the hard-to-soft ratio of the model functions.

At higher \yt\ the ratio increases for peripheral events but remains flat for central events. That is consistent with the trend for exponent $q$ in Table~\ref{pidparam2} remaining constant for pions but {\em increasing} with centrality for kaons, decreasing the hardness of the kaon exponential tail. Peripheral kaon hard components are thus harder than central. Detailed structure of $K/\pi$ spectrum ratios is a delicate interplay of mode positions, widths and exponents.

Figure~\ref{pidrat} (right) shows TCM proton/pion ratios. The low-\yt\ $z_{si}$ ratio limit should be $\approx$ 0.035/0.8 = 0.044 but one observes $\approx$ 0.02. See above comments on soft-component issues. The contrast between kaon/pion and proton/pion ratios is explained as follows: As for kaons the proton/pion ratio increase for lower \yt\ is explained by the interaction between soft and hard model functions. The nominal species fraction $z_{hi}(n_s)$ ratio at higher \yt\ is approximately 0.2/0.6 = 0.33. However, with increasing \yt\ the difference between pion and proton hard components comes into play as explained below. 

\begin{figure}[h]
	\includegraphics[width=3.3in]{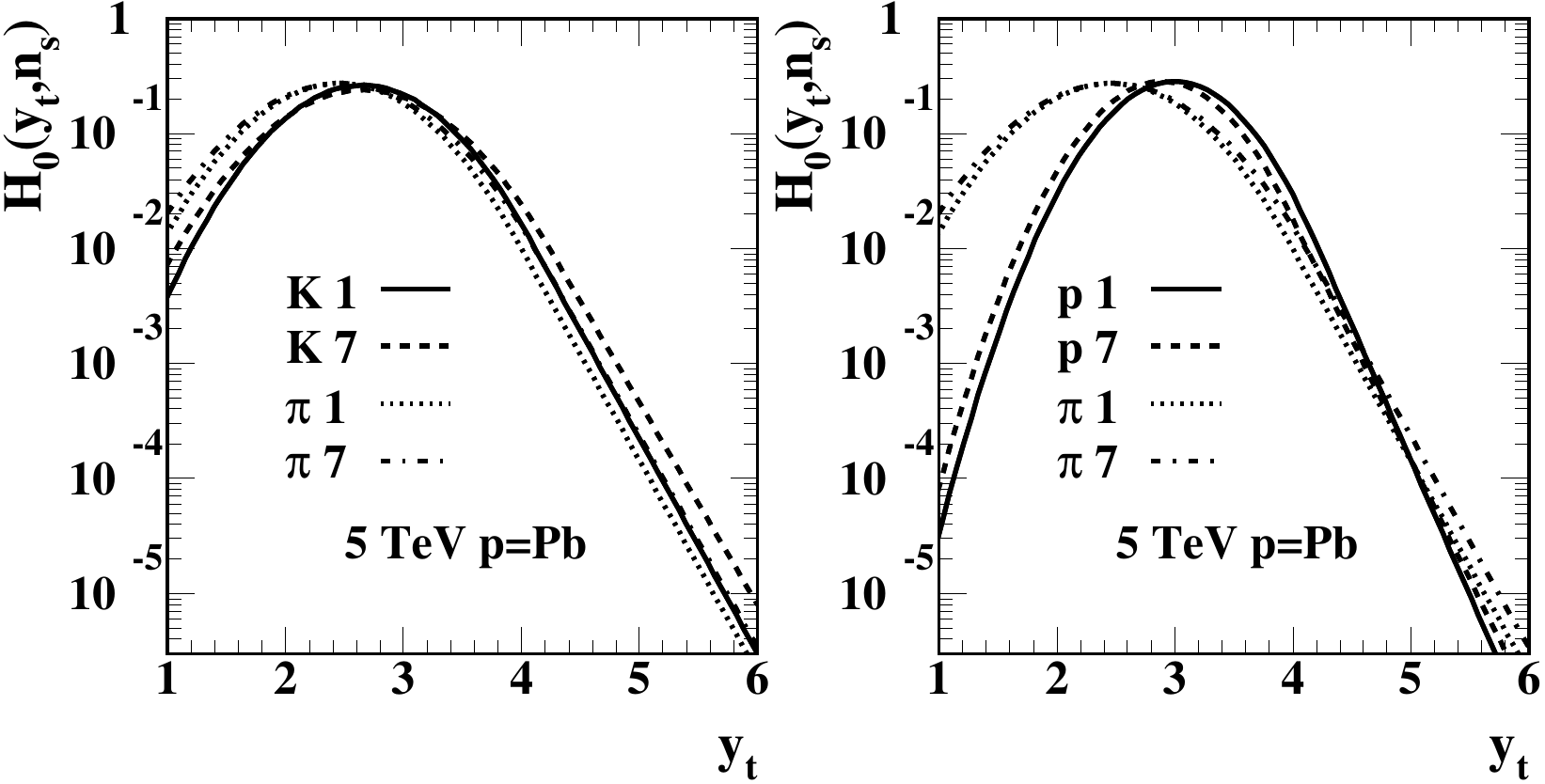}
	\caption{\label{h0rat}
		Left: Kaon/pion spectrum ratios for seven event classes of 5 TeV \ppb\ collisions. The most-central event class is $n = 1$.
		Right: Same as left panel except proton/pion ratios.
	}  
\end{figure}

Figure~\ref{h0rat} shows

the mode displacement on \yt\ -- 2.45 for pions vs nearly 3 for protons (both near \yt\ = 2.7 or 1 GeV/c) -- and the width difference 0.65 for pions vs 0.47 for protons. 

The substantial mode displacement implies that the ratio should increase rapidly near the proton mode. However, at higher \yt\ the broader pion hard component prevails in the denominator and the ratio falls. Evolution of the proton/pion ratio with centrality corresponds  to detailed evolution of proton hard-component mode positions as in Table~\ref{pidparam1}. The mode shifts to higher \yt\ but the exponential tail becomes softer.

Note that the peak values for the proton/pion spectrum ratio in Fig.~\ref{pidrat} (right) are much higher than what appears in Fig.~10 of Ref.~\cite{alicenucmod} due to the correction for misidentification discussed in Sec.~\ref{ineff} above. In PID ratios misidentification both reduces the proton numerator and increases the pion denominator since some fraction of protons are apparently misidentified as pions. See the discussion in Sec.~VII B of Ref.~\cite{pppid} on PID spectrum ratios and the effects of hadron species misidentification.

Reference~\cite{alicenucmod} observes that proton $R_{p\text{Pb}}$ is ``...significantly larger than those for pions and kaons, in particular in the [\pt] region where the Cronin peak was observed...at lower energies.'' It concludes that ``An enhancement of protons in the same \pt\ region [in \aa\ collisions]...commonly is interpreted as [indicating] radial-flow....'' Commenting on $p/\pi$ spectrum ratios and a centrality-dependent peak near 3 GeV/c it states that they ``exhibit similar flow-like features.''
The implication is that the peak in the proton/pion spectrum ratio is equivalent to the Cronin peak in $R_{p\text{Pb}}$ relating to ``...modification of the proton spectral shape going from pp to p-Pb collisions.''  There are actually  two different mechanisms at work. For the proton/{\em pion} ratio the different hard-component centroids and widths lead to the peak in Fig.~\ref{pidrat} (right). For {\em proton} $R_{p\text{Pb}}$ the hard-component {\em shape} changes very little from MB \pp\ to NSD \ppb\ collisions (i.e.\ from \ppb\ event class 7 to event class 5; see Fig.~\ref{protonhc}, left). The peak in $R_{p\text{Pb}}$ arises for any hadron species because {\em jet production} is greater in more-central \ppb\ collisions or for larger nucleus A (see Fig.~\ref{off}, d).

\section{Fixed-target spectrum data} \label{cronineff}

In the mid-seventies a Chicago-Princeton (C-P) collaboration including J. Cronin performed a series of fixed-target experiments at Fermilab with proton beams of 200, 300 and 400 GeV incident on several targets, including H, Be, Ti and W. The experiments revealed hardening of hadron spectra at higher \pt\ relative to a then-expected exponential decrease on \pt, increasingly so with increasing collision energy and target size A~\cite{cronin10,cronin1,cronin0}. Manifestations of those trends in spectrum {\em ratios} have come to be known as the ``Cronin effect.'' Spectrum ratios have been variously defined in this connection (even within C-P publications), but a common example is~\cite{accardi}
\bea \label{cronratio}
R_{AB} &=&\frac{B}{A}\frac{Ed\sigma_{pA}/d^3p}{Ed\sigma_{pB}/d^3p},
\eea
where cross sections $Ed\sigma_{pX}/d^3p$ are defined in terms of some $p$-X minimum-bias cross section $\sigma_{pX}$. According to Ref.~\cite{accardi} ``In absence of nuclear effects one would expect $R_{AB} = 1$, but for $A>B$ a suppression is observed experimentally at small \pt, and an enhancement at moderate \pt\ with $R_{AB} \rightarrow 1$ as $p_t \rightarrow \infty$.'' Assumptions regarding ``nuclear effects'' are challenged in the material below.

Spectra published in Refs.~\cite{cronin10,cronin1,cronin0} are reported as invariant cross sections in units  cm$^2$/GeV and include a proton-target cross section $\sigma_{pA}$ that may be the inelastic cross section for \pn\ collisions or the absorption cross section for \pa\ collisions. In this study published spectra are converted to {\em measured particle densities} on transverse mass $m_t$ and {\em longitudinal} rapidity $y_z$, distinguished from {\em transverse}  $y_t$ that is essential for describing jet-related spectrum hard components. The conversion factor is $2\pi 10^{27}/\sigma_{pA}$ using the cross section $\sigma_{pA}$ in mb applied to published data. Resulting spectra have the form $d^2n_i/m_t dm_t dy_z$ for hadron species $i$. That conversion is essential in order to better understand Cronin effects as usually presented.
There are two main issues related to the Cronin effect: collision-energy dependence and target size A (effective nuclear thickness) dependence.

\subsection{$\bf p$-A spectrum -- collision-energy $\bf \sqrt{s}$ dependence} \label{cronedep}

\begin{figure}[h]
	\includegraphics[width=1.65in]{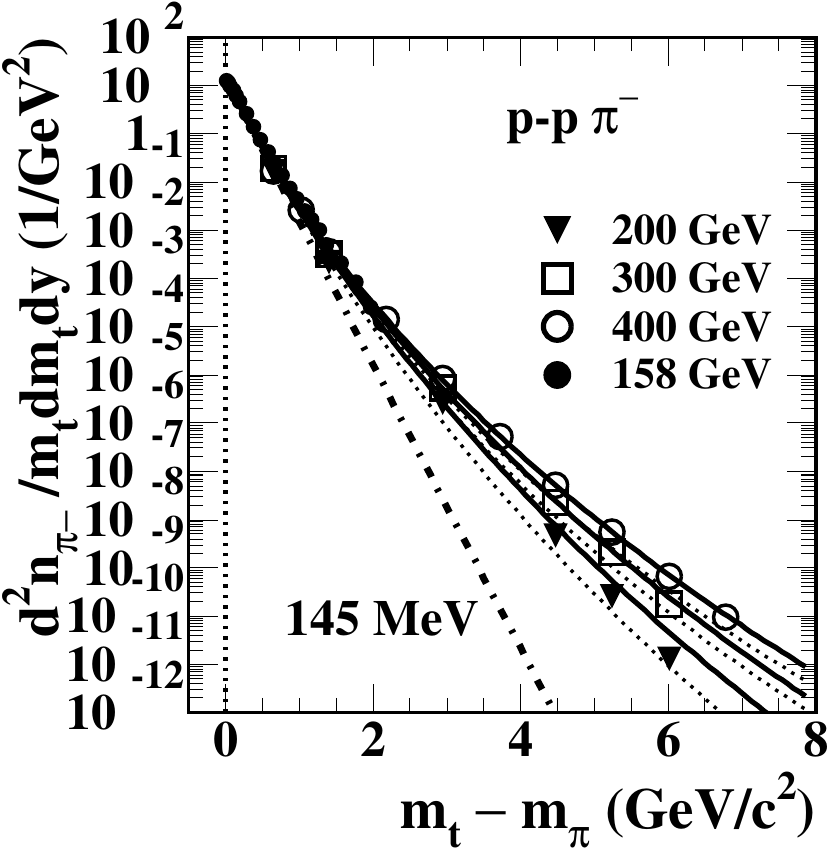}
	\includegraphics[width=1.6in]{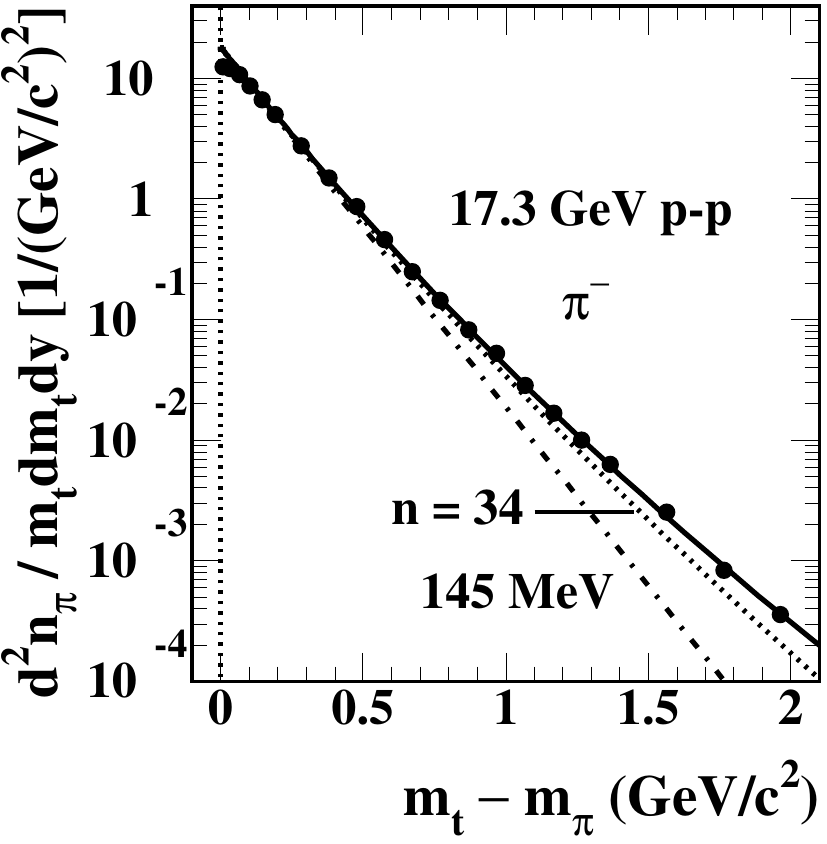}
	\caption{\label{softn}
		Left: $\pi^-$ spectra (points) extending to \pt\ $\approx 7$ GeV/c from \pp\ collisions at three fixed-target energies corresponding to $\sqrt{s} = 19.4$, 23.8 and 27.4~\cite{cronin0}. Also shown are data for 17.3 GeV from the right panel (solid dots). The dotted and solid curves are the TCM for those collision systems.
		Right:  $\pi^-$ spectra (points) from 17.3 GeV \pp\ collisions extending to \pt\ $\approx 2$ GeV/c~\cite{na49pions}. The dotted curve is a soft-component model with exponent $n = 34$ and temperature $T = 145$ MeV. The solid curve includes a jet contribution as discussed in the text.
	} 
\end{figure}

Figure~\ref{softn} (left) shows \pp\ fixed-target $\pi^-$ spectra for four beam energies corresponding to $\sqrt{s}= 17.3$, 19.4, 23.8 and 27.4 GeV. The 158 GeV data (solid dots, $p_t \in [0,2]$ GeV/c) are reported in Ref.~\cite{na49pions}. Data for the higher energies are from Ref.~\cite{cronin0}. The dash-dotted line is a Boltzmann exponential with $T = 145$ MeV that describes {\em low}-\pt\ pion data for all collision energies significantly above $\sqrt{s} \approx 10$ GeV. The soft-component model
\bea \label{s000}
\hat S_0(m_t;T,n) &=& \frac{C}{[1 + (m_t - m_i)/nT]^n}
\eea
with $T = 145$ MeV and exponents $n(\sqrt{s}) = 34$, 26.5 and 23.5 is shown as dotted curves for three increasing collision energies. Constant $C$ is adjusted for unit-normal model function and the dotted curves are defined by $ S(m_t;T,n) = \bar \rho_{s\pi^-} \hat S_0(m_t;T,n)$, where $\bar \rho_{s\pi^-} \approx 0.55\pm0.05$ with no significant energy dependence apparent from these \pp\ data.  
The functional form of $\hat S_0(m_t;T,n)$ describes an incompletely equilibrated thermal particle source where exponent $n$ is a measure of heterogeneity or resulting \pt\ variance~\cite{wilk}. The criteria for choosing soft-component $n$ values and composition of the solid curves through data are discussed below. This plot shows one manifestation of the Cronin effect: variation of the {\em soft} component (its power-law tail) with increasing collision energy. Hard components for solid curves are model functions from Fig.~\ref{softnx} (right) scaled down to $\bar \rho_{h\pi^-} \approx 0.002$ given $\bar \rho_{h\pi^-} \approx 0.027$ for $p$-W and $183^{1/2} = 13.5$.


\begin{figure}[h]
	\includegraphics[width=1.65in]{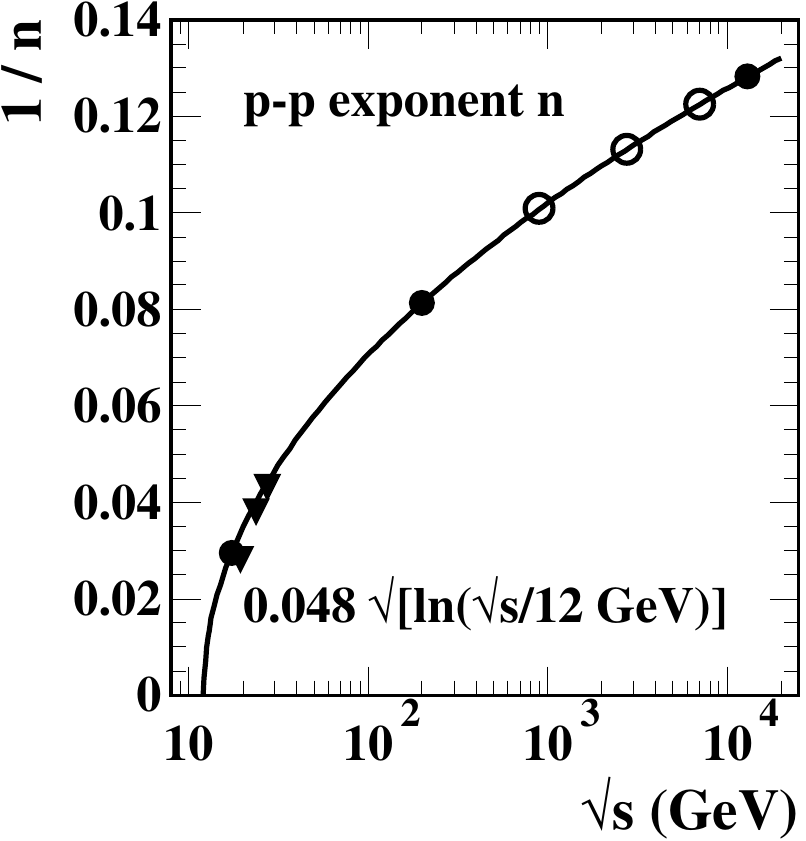}
	\includegraphics[width=1.65in]{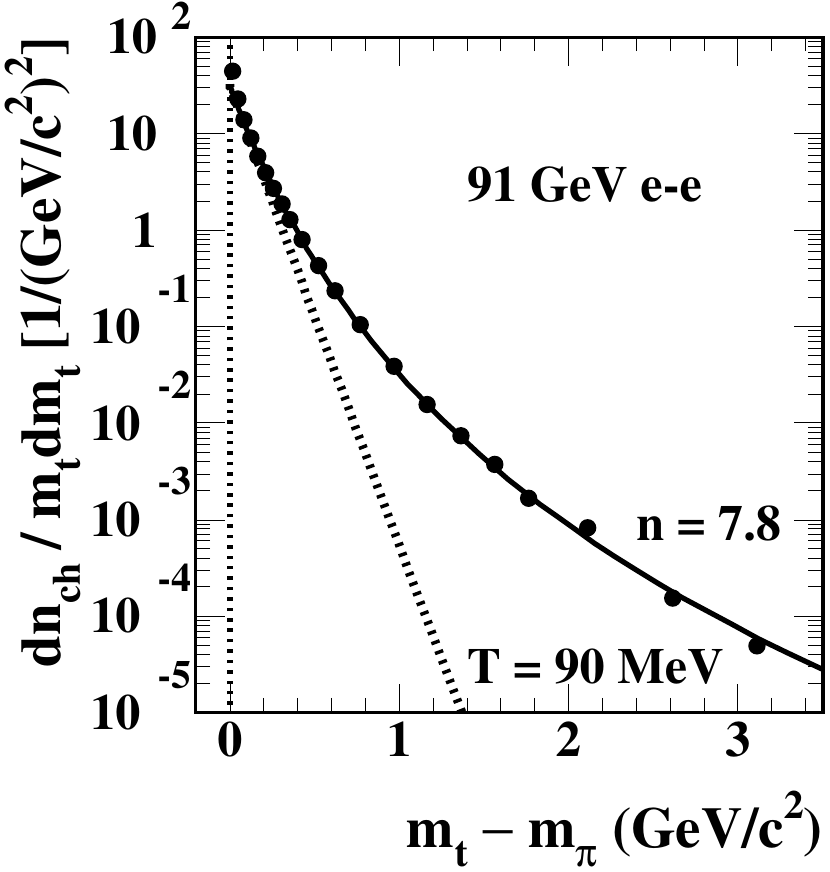}
	\caption{\label{spsdata}
		Left: Soft-component exponents $n$ for three collision systems (solid dots) from Ref.~\cite{alicetomspec}. The open circles are interpolations for relevant collision systems. The triangles are exponent values inferred from data in Figs.~\ref{softn} and \ref{softnx}. The solid curve is explained in the text.
		Right: \pt\ spectrum (momentum perpendicular to dijet axis) for dijets from 91 GeV \ee\ collisions~\cite{alephff}. Exponent $n = 7.8 \rightarrow 1/0.13$ is equivalent to the value for 13 TeV \pp\ collisions in the left panel.
		}  
\end{figure}

\begin{figure}[h]
	\includegraphics[width=1.65in]{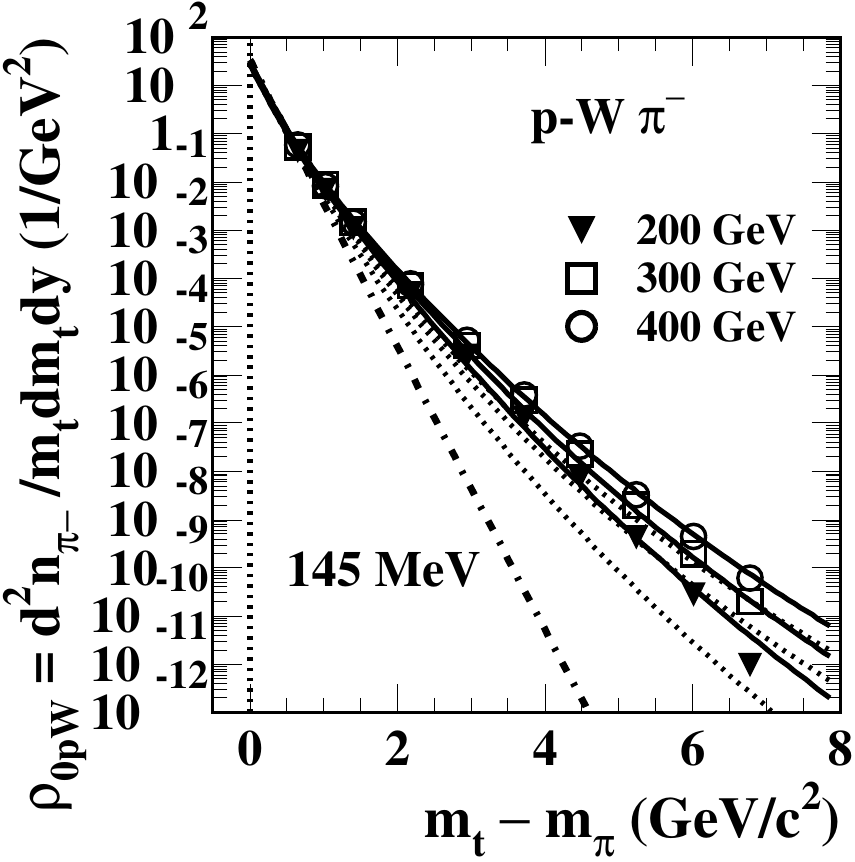}	\includegraphics[width=1.65in]{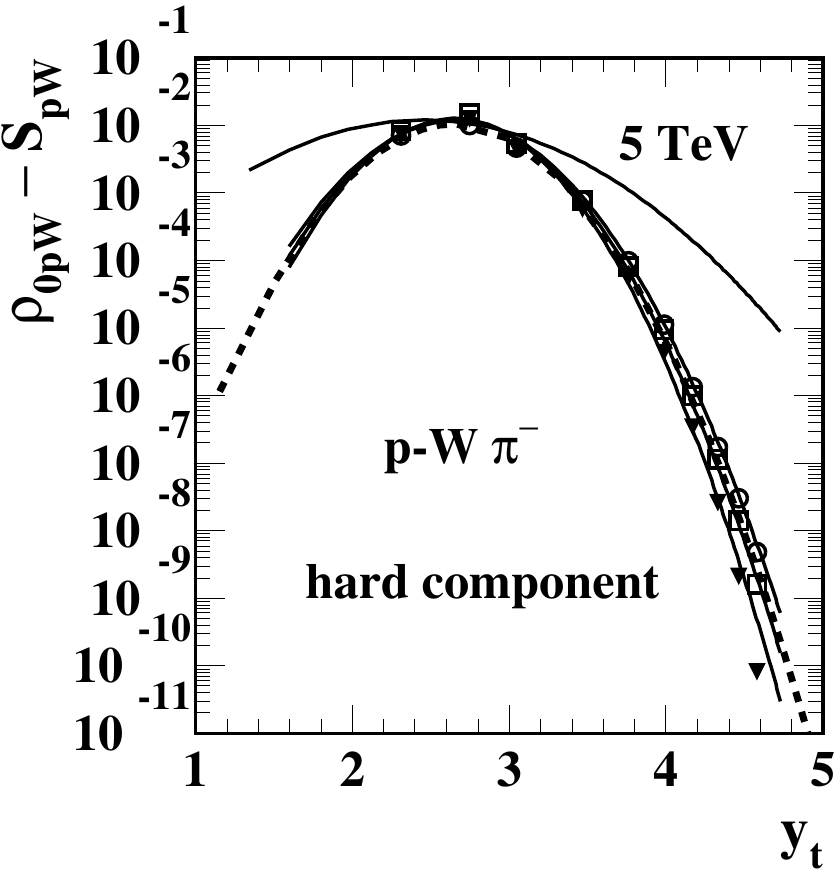}
	\caption{\label{softnx}
		Left:  $\pi^-$ spectra (points) extending to \pt\ $\approx 7$ GeV/c from $p$-W collisions at three fixed-target energies corresponding to $\sqrt{s} = 19.4$, 23.8 and 27.4~\cite{cronin10}.  The dotted and solid curves are the TCM for those collision systems.	
		Right: Data (points) and TCM (curves) spectrum hard components (jet fragments) inferred as described in the text. The limited energy evolution of data and TCM at these low collision energies may be compared with the TCM model for the pion spectrum hard component from 5 TeV \ppb\ collisions so labeled. The dashed curve is explained below Fig.~\ref{hardcom}.
	}  
\end{figure}

Fig.~\ref{softn} (right) shows a $\pi^-$ spectrum (solid dots) from 17.3 GeV \pp\ collisions reported in Ref.~\cite{na49pions}. The dotted curve is $\hat S_0(m_t;T,n) $ with parameter values shown in the plot. The dash-dotted line is a Boltzmann exponential on \mt\ with $T = 145$ MeV corresponding to the asymptotic limit $\hat S_0(m_t;T,n) $ with $1/n \rightarrow 0$. In a previous version of that plot the {\em solid} curve is Eq.~(\ref{s000}) with $n  =27$ assuming no jet contribution to data. From the present analysis the jet contribution for 19.3 GeV is determined precisely and appears here as the difference between dotted and solid curves. The dotted curve is now Eq.~(\ref{s000}) with $n = 34$. The solid dot at 17.3 GeV in Fig.~\ref{spsdata} (left, previously published) has been updated accordingly. These 17.3 GeV data are also shown as the solid dots in the left panel.

Deviations from an exponential for spectra as in Fig.~\ref{softn} derived from fixed-target experiments with $\sqrt{s_\text{NN}} \approx 20$ - 30 GeV have been claimed as evidence both for jet production~\cite{kuhn,straub,wangcronin}  and for radial flow~\cite{ssflow}. It has become clear that deviations from a Boltzmann exponential on \mt\ arise from a {\em combination} of splitting cascades associated with projectile nucleon dissociation (soft)~\cite{gribov,gribov2} and minimum-bias dijet production (hard)~\cite{fragevo,jetspec2,mbdijets,alicetomspec}.

Figure~\ref{spsdata} (left) shows Fig.~12 (right) from Ref.~\cite{alicetomspec} that describes the trend for $1/n$ as a fluctuation variance measure associated with Gribov diffusion within a parton splitting cascade. The solid dots are obtained from previous TCM analyses of \pp\ spectra for $\sqrt{s} =$ 17.3 GeV, 200 GeV and 13 TeV. As noted above the value for 17.3 GeV has been updated in light of this analysis. The open circles are interpolations for other relevant energies. The inverted triangles are obtained from the three higher-\pt\ spectra in Fig.~\ref{softn} (left). The solid curve is an algebraic hypothesis based on variation of the soft component related to Gribov diffusion~\cite{gribov}. Low-$x$ gluons result from a virtual parton splitting cascade within projectile nucleons whose mean depth on $x$ is determined by the collision energy. Each step of the cascade adds transverse-momentum components in a random-walk process. The depth of the cascade (``duration'' of the random walk) is proportional to $\ln(s/s_0)$, and $\sqrt{s_0} \approx 10$ GeV is inferred from dijet systematics~\cite{alicetomspec,jetspec2}. Note that the soft component of charged-hadron density $\bar \rho_0$ for NSD \pp\ collisions increases as $\bar \rho_s(\sqrt{s}) \approx 0.8 \ln(\sqrt{s} / \text{10 GeV})$ as reported in Ref.~\cite{alicetomspec} App.~C. Given the properties of a random walk, and with exponent $n$ in the form $1/n \sim \sigma^2_x / \bar x^2$ as a measure of transverse-momentum fluctuations~\cite{wilk}, its trend is estimated as $ 1/n\propto \sqrt{\ln(\sqrt{s} / \text{12 GeV})}$ (solid curve). 

The diffusion mechanism and consequences are clarified by Ref.~\cite{gribov2} that considers hadron transverse size for soft hadronic collisions at high energies. It notes that Gribov's emphasis was on parton diffusion in transverse {\em configuration} space: Reference~\cite{gribov} refers to ``...Gribov diffusion in the impact parameter space, giving rise to energy increase of the interaction radius....'' But Ref.~\cite{gribov2} notes that diffusion in transverse {\em momentum} space is also important. That mechanism is solely relevant for Fig.~\ref{spsdata} (left). In either case, parton rapidity (or number of steps $N$ in the splitting cascade) plays the role of time for the diffusion process as represented by $\log(s/s_0)$ for some $s_0$.

Fig.~\ref{spsdata} (right) shows an $m_t$ spectrum (points) from LEP \ee collisions at $\sqrt{s} \approx 91$ GeV~\cite{alephff}. \ee\ hadronic events from Z$_0$ decays are dominated by two-jet (dijet) events (see Fig.~13 of Ref.~\cite{alephff}).  The spectrum reveals the jet fragment momentum distribution transverse to the dijet axis. The {\em shape} of the $m_t$ spectrum from 91 GeV Z$_0$ decays  is precisely consistent with soft components $\hat S_0(m_t)$ of \pp\ spectra over a broad range of collision energies~\cite{alicetomspec}.  Exponent $n = 7.8$ for 91 GeV \ee\ data corresponds in magnitude to the point at 13 TeV in the left panel.
The parametrization in the left panel and $T \approx 145$ MeV predict the soft ({\em nonjet}) component for pion or unidentified-hadron spectra from any collision system in which \nn\ collisions are linearly superposed. The equivalence of soft-component shapes across energies and collision systems suggests that the $\hat S_0(m_t;T,n)$ model represents a universal feature of QCD fragmentation via splitting cascade for any leading particle. 

Figure~\ref{softnx} (left) shows $\pi^-$ spectra from $p$-W collisions (points) at three fixed-target energies~\cite{cronin10} that may be compared with the \pp\ results in Fig.~\ref{softn} (left). Soft-component models $\hat S_0(m_t)$ for $p$-W (dotted) are maintained identical to those for \pp\ spectrum data (dotted) in Fig.~\ref{softn}. Soft charge densities $\bar \rho_{s\pi^-}(\sqrt{s})$ are shown in Table~\ref{rootsparams}. Solid curves include spectrum hard components inferred in the right panel. Energy dependence in the left panel is dominated by soft-component evolution.

Figure~\ref{softnx} (right) shows spectrum hard components inferred as $H_\text{pW}(m_t) = \bar \rho_{0\text{pW}}(m_t) - S_\text{pW}(m_t)$ transformed with an appropriate Jacobian to transverse rapidity \yt\ where the spectrum hard component has a simple form (approximate Gaussian). As noted, the $S_\text{pW}(m_t) = \bar \rho_s \hat S_0(m_t)$ for three energies in the left panel are identical {\em in form} with those in Fig.~\ref{softn} (left). The data hard components in the right panel are represented by TCM models $\bar \rho_h \hat H_0(y_t,\sqrt{s})$ where hats indicate unit-normal functions. It is notable that collision-energy variation does not change charge density $\bar \rho_{h\pi^-} \approx 0.027\pm0.002$ significantly but does change the model shape at higher \pt. The lower three solid curves for $p$-W correspond to model widths $\sigma_{y_t} = 0.330$, 0.345 and 0.360 for increasing collision energy. The dashed curve is explained in the text below Fig.~\ref{hardcom}. Evolution of the soft component may be attributed to Gribov diffusion within scattered nucleons whereas evolution of the hard component may be attributed to hardening of the underlying parton spectrum leading to jets. Those forms for lower energies contrast with data from 5 TeV \ppb\ collisions (upper curve) where $\bar y_t = 2.46$ (vs 2.65 at the lower energies) and $\sigma_{y_t} = 0.60$.

Table~\ref{rootsparams} shows soft-component L\'evy exponent $n$ (and reciprocal $1/n$ for Fig.~\ref{spsdata}, left) consistent for H and W targets, hard-component Gaussian width $\sigma_{y_t}$ and soft-component charge density $\bar \rho_{s\pi^-}$ for $p$-W collisions.
Over this limited energy interval $\bar \rho_s$ increases approximately linearly with (but not proportional to) the trend $\ln(\sqrt{s/\text{10 GeV}})$, about 60\% as fast as at higher energies where  $\bar \rho_s \approx 0.8\ln(\sqrt{s/\text{10 GeV}})$~\cite{alicetomspec}. The trend is consistent with ISR observations at similar energies~\cite{isrnch}.

\begin{table}[h]
	\caption{Soft-component L\'evy exponent $n$ consistent with $p$ and W targets, hard-component Gaussian width $\sigma_{y_t}$ and soft charge density $\bar \rho_{s\pi^-}$ for W target vs $p$-W CM energy $\sqrt{s}$.
	} \label{rootsparams}
	\begin{center}
		\begin{tabular}{|c|c|c|c|c|} \hline
			$\sqrt{s}$ (GeV)	& $n$ & $1/n$ & $100\sigma_{y_t}$ & $\bar \rho_{s\pi^-}$     \\ \hline
			19.4          &  $34.0\pm2$ & $0.029\pm0.002$ & $33.0\pm0.5$ & $1.30\pm0.05$  \\ \hline
			23.8       & $26.5\pm1$ & $0.038\pm0.002$ & $34.5\pm0.5$ & $1.50\pm0.05$   \\ \hline	
			27.4     &  $23.5\pm1$ & $0.043\pm0.002$ & $36.0\pm0.5$  & $1.58\pm0.05$  \\ \hline
		\end{tabular}
	\end{center}
\end{table}

Establishment of $n$ values for $\hat S_0(m_t)$ proceeds as follows: The soft-component models in Fig.~\ref{softn} (left) are matched to data at low \pt\ by adjusting $\bar \rho_s$ in Eq.~(\ref{crontcm}). The respective $n$ values are then adjusted so that model curves fall {\em just below} the highest-\pt\ data points. Those same model functions are then introduced into Fig.~\ref{softnx} (left) where they are first matched to data at low \pt\ and then used to obtain the data hard components (points) appearing in the right panel by subtraction. Model curves $\hat H_0(y_t,\sqrt{s})$ are defined as the solid curves in the right panel that accommodate data at and above the mode. Soft-component densities $\bar \rho_s$ for $p$-W are then readjusted so that the lowest data points in Fig.~\ref{softnx} (right) fall on the model curves  (a sensitive requirement). Optimized hard-component models $\bar \rho_h \hat H_0(y_t,\sqrt{s})$ are then combined with $\bar \rho_s(\sqrt{s})\hat S_0(m_t,\sqrt{s})$ to form complete spectrum models appearing as solid curves in Fig.~\ref{softnx} (left). 

\subsection{$\bf p$-A spectrum -- target size A dependence}

In what follows symbol \pa\ refers to proton-nucleus systems whereas $p$-X refers to \pa\ or \pp\ systems. Equation~(\ref{pidspectcm}) can be reformulated in this context as
\bea \label{crontcm}
\bar \rho_\text{0pX}(m_t,\sqrt{s},A) &\approx&   S_\text{pX}(m_t) +   H_\text{pX}(m_t)
\\ \nonumber
&& \hspace{-.6in} \approx \bar \rho_s(\sqrt{s},A) \hat S_0(m_t,\sqrt{s}) + \bar \rho_h(A) \hat H_0(m_t,\sqrt{s}),
\eea
which assumes for simplicity no A dependence for the shapes of soft and hard components. That assumption is revisited below. Spectrum data for varying target size A are analyzed to obtain estimates of $\bar \rho_s(A)$ and $\bar \rho_h(A)$.

\begin{figure}[h]
	\includegraphics[width=3.3in]{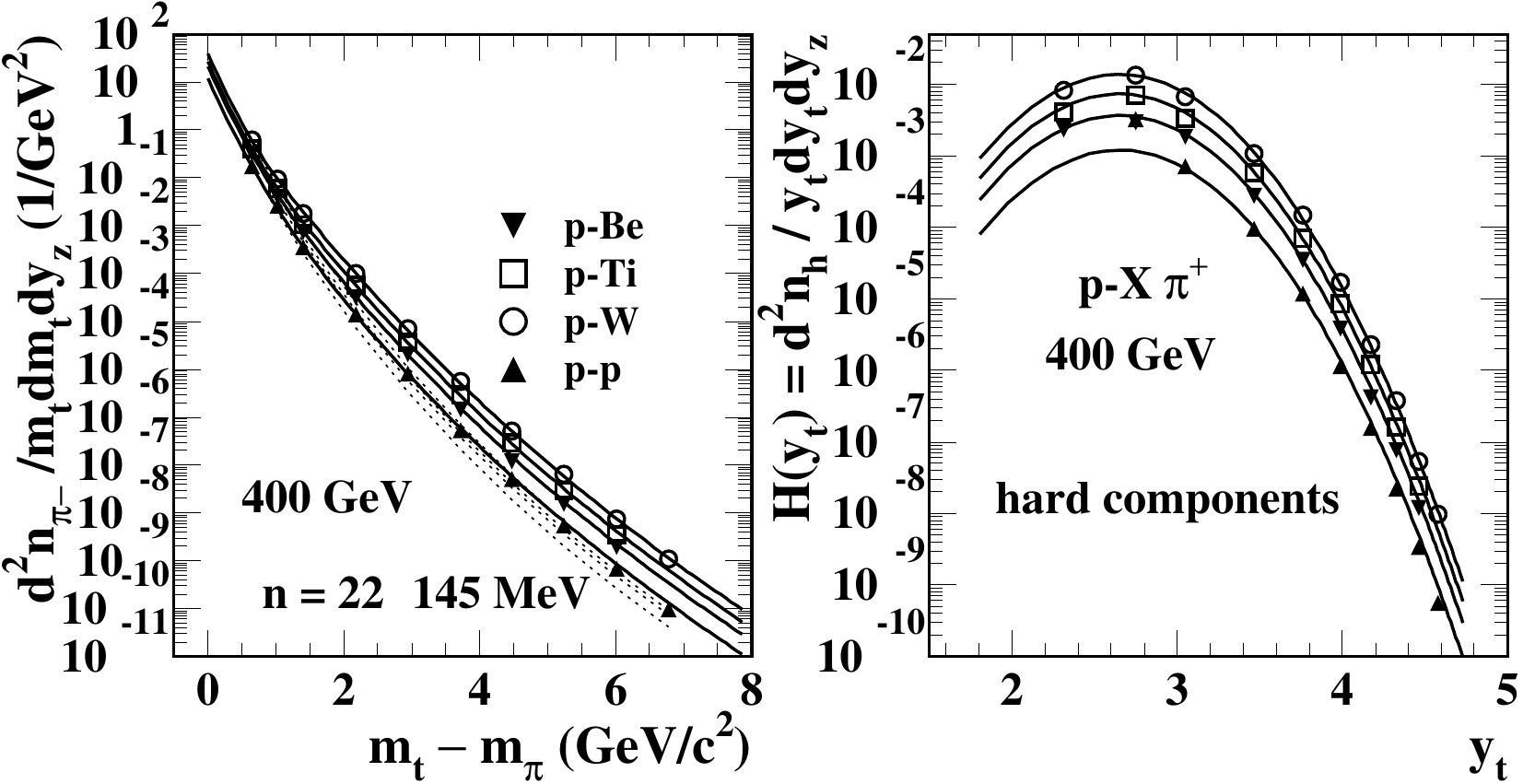}	
		\caption{\label{20d}
		Left: $\pi^+$ spectra from four collision systems (points) with varying target weight A for fixed-target beam energy 400 GeV~\cite{cronin0}. The solid and dotted curves are the corresponding TCM described in the text.
		Right: Data (points) and TCM (curves) hard components inferred as described in the text.
	} 
\end{figure}

Figure~\ref{20d} (left) shows $\pi^+$ spectra (hadron densities) for four collision systems (points) corresponding to variation of target atomic weight A~\cite{cronin0}. TCM soft components (dotted) use $n$ values determined for 400 GeV \pp\ and $p$-W collisions (the same for both systems) as described in the previous subsection, with coefficient $\bar \rho_s(A)$ values adjusted to accommodate spectrum data at low \pt. Of the available $p$-X data the $\pi^+$ spectra are preferred in this case as having more complete \pt\ coverage.

\begin{figure}[h]
	\includegraphics[width=3.3in]{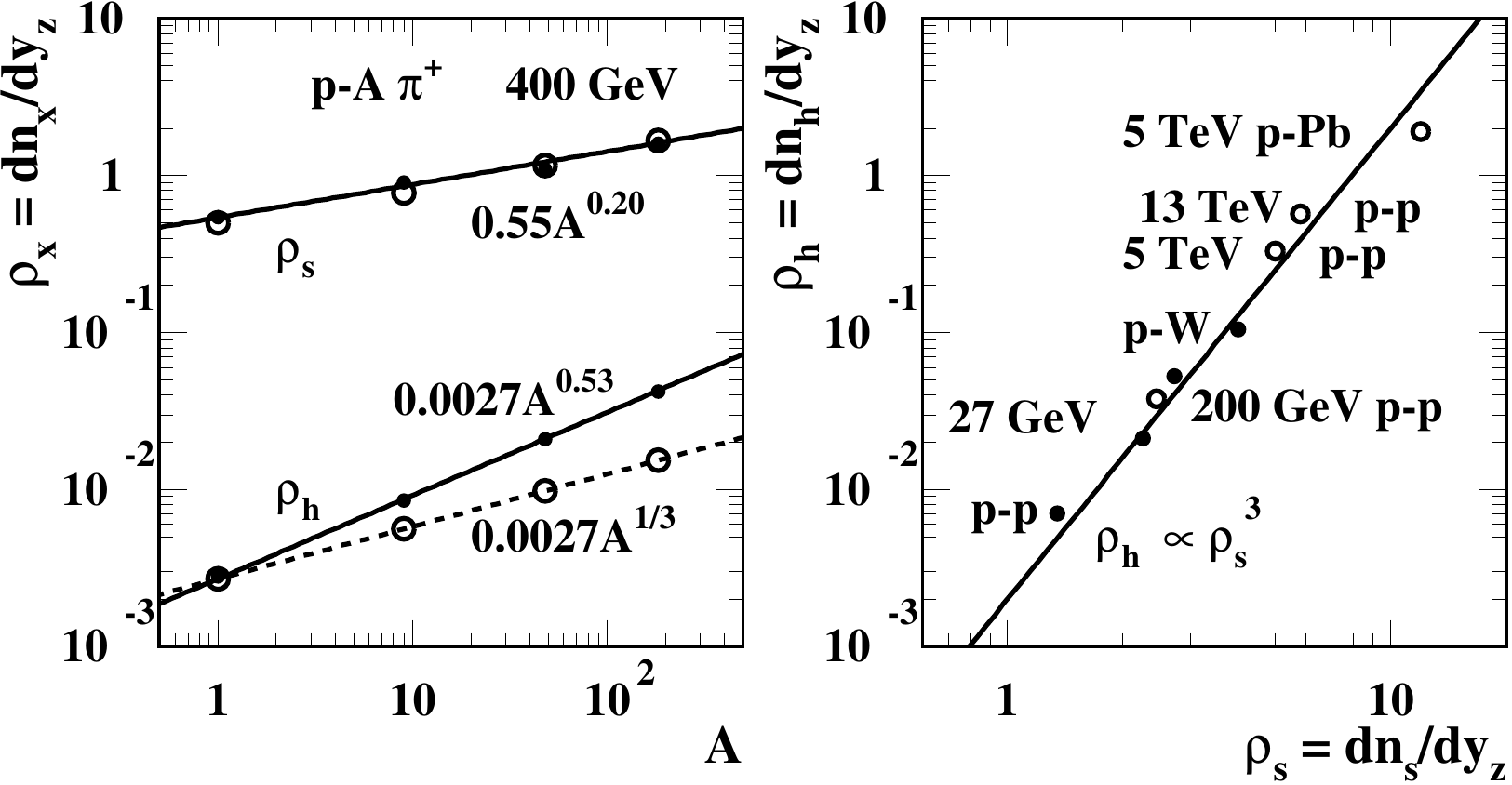}	
		\caption{\label{20g}
		Left: $\pi^+$ trends for charge densities $\bar \rho_s(A)$ and $\bar \rho_h(A)$ vs atomic weight A for four collision systems inferred from spectra in Fig.~\ref{20d}. The lines are described in the text.
		Right: 	Charge densities $\bar \rho_s(A)$ vs $\bar \rho_h(A)$ (points) from  the left panel. The inferred values follow a cubic relation $\bar \rho_h \propto \bar \rho_s^3$ (line).
		Data from other collision systems are added for comparison.
	}
\end{figure}

Figure~\ref{20d} (right) shows spectrum hard components (points) inferred as $H_\text{pX}(m_t) = \bar \rho_{0\text{pX}}(m_t) - S_\text{pX}(m_t)$ transformed to densities on \yt\ with an appropriate Jacobian. The curves are hard-component model $H_\text{pX}(m_t)$ with coefficient $\bar \rho_h(A)$ values adjusted to best accommodate data. The widths are discussed in connection with Fig.~\ref{20i}. Resulting parameter values, with soft and hard density trends at low and high \pt\ appearing in Fig.~\ref{20g} (left), are used in Eq.~(\ref{crontcm}) to generate the solid curves in the left panel. The resulting TCM describes the data well and may provide insight into the Cronin effect.

Table~\ref{aparams} shows $p$-X soft-component charge density $\bar \rho_{s\pi^+}$, hard-component charge density $\bar \rho_{h\pi^+}$ for $\pi^+$ and hard-component Gaussian width $\sigma_{y_t}$ vs target atomic weight A. Widths $\sigma_{y_t}$ are based on NMFs in Fig.~\ref{20i}. The small uncertainties for $\sigma_{y_t}$ reflect {\em relative} values determined differentially from Fig.~\ref{20i} as opposed to absolute uncertainties $\pm$0.005 in Table~\ref{rootsparams} from Fig.~\ref{softnx} (right).

\begin{table}[h]
	\caption{$p$-X soft-component charge density $\bar \rho_{s\pi^+}$, hard-component charge density $\bar \rho_{h\pi^+}$ for $\pi^+$ and hard-component Gaussian width $\sigma_{y_ti}$ vs target atomic weight A at 400 GeV.
	} \label{aparams}
	\begin{center}
		\begin{tabular}{|c|c|c|c|c|} \hline
			X	& A & $\bar \rho_{s\pi^+}$ & $100\bar \rho_{h\pi^+}$ & $100\sigma_{y_t}$   \\ \hline
			$p$     & 1 &  $0.55\pm0.02$ & $0.28\pm0.02$ & $35.1\pm0.1$    \\ \hline
			Be       & 9 & $0.90\pm0.04$ & $0.85\pm0.04$ & $35.6\pm0.1$   \\ \hline	
			Ti     & 48 &  $1.09\pm0.05$ & $2.10\pm0.10$ & --   \\ \hline
			W    & 183 & $1.60\pm0.05$  & $4.20\pm0.20$ & $36.0\pm0.1$   \\ \hline
		\end{tabular}
	\end{center}
\end{table}

Optimal accuracy of the fit procedure is achieved recursively as follows: The soft components (shapes) in Fig.~\ref{20d} (left) are identical with the 400 GeV soft component from Figs.~\ref{softn} and \ref{softnx}. The hard component shapes at right are {\em initially} identical with the 400 GeV hard component in Fig.~\ref{softnx} (right). After visually matching soft components via $\bar \rho_s(A)$ to the lowest \pt\ points in the left panel, $\bar \rho_h(A)$ values are adjusted to match the higher-\pt\ points in the right panel. The $\bar \rho_s(A)$ are then optimized to ensure that the lowest-\pt\ points at right coincide with the model curves. The higher-\pt\ points are insensitive to those adjustments. Because of the simplifying assumption that $\hat H_0(m_t)$ is independent of A, description of the lower-A data at higher \pt\ is not ideal. Related deviations appear in Fig.~\ref{20i} ratios (solid curves). Widths $\sigma_{y_t}$ for $p$-Be and \pp\ must be reduced to  the values in Table~\ref{aparams} to obtain the dashed and dotted curves through data.

The value $\bar \rho_{h\pi^+} \approx 0.042$ for 400 GeV $p$-W collisions in Table~\ref{aparams} is 55\% greater than the value $\bar \rho_{h\pi^-} \approx 0.027$ for the same collision system inferred from data in Fig.~\ref{softnx} (right). Of that total approximately 15\% arises from the systematic difference between $\pi^+$ and $\pi^-$ spectra at higher \pt\ (above 4 GeV/c), consistent with Table~IX of Ref.~\cite{cronin0} and Fig.~13 of Ref.~\cite{cronin10}. The other 40\% arises from systematic differences between spectra reported in two publications. Figure~\ref{softnx} includes data from  Ref.~\cite{cronin10} because of better \pt\ coverage across three collision energies. Figure~19 includes 400 GeV \pa\ data from Ref.~\cite{cronin0} because of its A coverage. Detailed inspection of 400 GeV $p$-W spectra from the two papers reveals that the spectra from Ref.~\cite{cronin0} are about 40\% higher than those from Ref.~\cite{cronin10} at higher \pt. This description broadly and approximately explains the discrepancy in $\bar \rho_{h\pi}$ values noted above. A detailed examination of PID spectra from the C-P collaboration will be presented in a follow-up study.

Figure~\ref{20g} (left) shows charge densities $\bar \rho_x \approx dn_x/dy_z$ (solid dots) plotted vs atomic weight A for four collision systems. Experimentally, soft-component density $\bar \rho_s$ varies $\propto A^{0.20}$ (upper solid line) and $\bar \rho_h$ varies $\propto A^{0.53}$ (lower solid line). As a simple collision model assume the number of \pn\ binary collisions goes as the nuclear radius, $N_{bin} \approx A^{1/3}$. The number of nucleon participants in \pa\ is then $N_{part} \approx A^{1/3} + 1$. Models $\bar \rho_s = 0.55 N_{part}/2$ and $\bar \rho_h = 0.0027 N_{bin}$ then appear as the open circles. {\em Predicted} hard component $\bar \rho_h$ assumes jet production is the same for all \pn\ collisions in a \pa\ collision (one version of linear superposition). According to the Glauber model jet production per participant pair should then increase with system size as $\bar \rho_h \propto \bar \rho_s^{4/3}$. Experimentally, these data indicate that the actual relation is $\bar \rho_h \propto \bar \rho_s^{8/3}$.

Figure~\ref{20g} (right) shows values $\bar \rho_h(A) = dn_h/dy_z$ plotted vs $\bar \rho_s(A) = dn_s/dy_z$. The 27 GeV $\pi^+$ charge densities (solid dots) are the basis for TCM curves in Fig.~\ref{20d} that describe the spectrum data well. Those $\pi^+$ densities have been multiplied by 2.5 (including factor 1.25 corresponding to pion  fraction 0.80) for proper comparison with {\em total} charge densities from other systems. The line represents a {\em cubic relation} $\bar \rho_h  \propto \bar \rho_s^3$ used here as a reference. Values (open circles) for 200 GeV \pp, 5 TeV \pp\ and 5 TeV \ppb\ (the last based on entries in Table~\ref{rppbdata} for event class 5) are included for comparison. The 200 GeV and 5 TeV \pp\ $\bar \rho_h$ values are obtained from the relation $\bar \rho_h = \alpha(\sqrt{s}) \bar \rho_s^2$ with NSD $\bar \rho_0 ~(= \bar \rho_s + \bar \rho_h)$ values 2.5 and 5 and with $\alpha = 0.006$ and 0.013 respectively. A 13 TeV \pp\ point is added with $\alpha = 0.017$ and $\bar \rho_0 \approx 6.3$. Note that the 27 GeV fixed-target data seem to agree with the \pa\ trend (line) observed at higher energies per Ref.~\cite{alicetomspec}. 

As noted, published spectra from Refs.~\cite{cronin10,cronin1,cronin0} have the form $\sigma_\text{$p$A} \times d^2n_i/2\pi m_t dm_t dy_z$ for hadron species $i$ and are typically rescaled by atomic weight A for comparison in ratios per Eq.~(\ref{cronratio}). Consequences are examined below.

\begin{figure}[h]
	\includegraphics[width=3.3in]{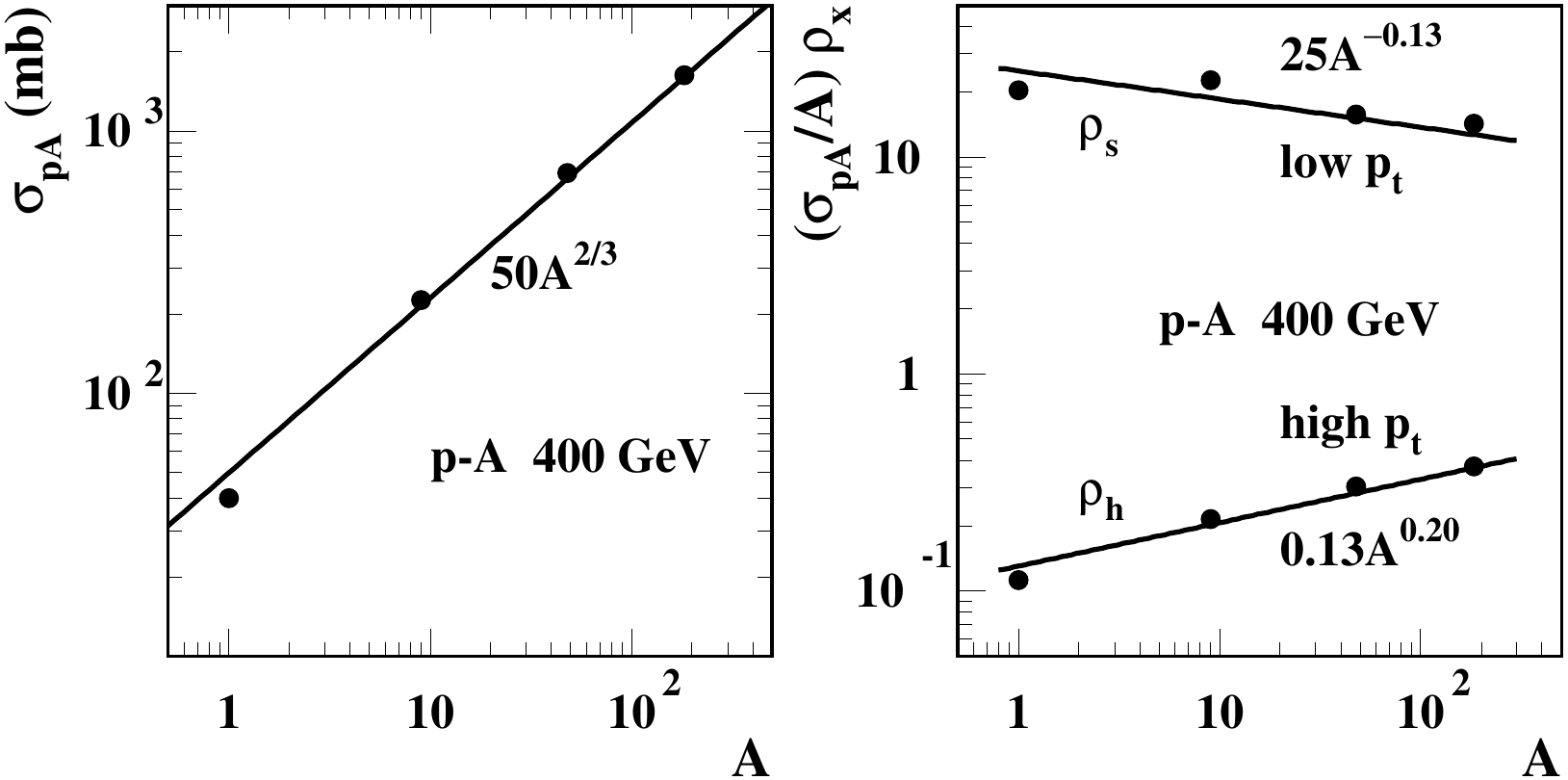}	
	\caption{\label{20h}
		Left: Cross sections $\sigma_\text{$p$A}$ (points) used in Refs.~\cite{cronin1,cronin10,cronin0} to generate differential cross-section spectra from measured particle densities. The cross-section trend is described by $\propto A^{2/3}$.
		Right: Charge-density trends in Fig.~\ref{20g} (left) multiplied by ``Cronin effect''  rescale factor $\sigma_\text{$p$A}/A \propto A^{-1/3}$.
	}  
\end{figure}

\begin{figure}[h]
	\includegraphics[width=3.3in]{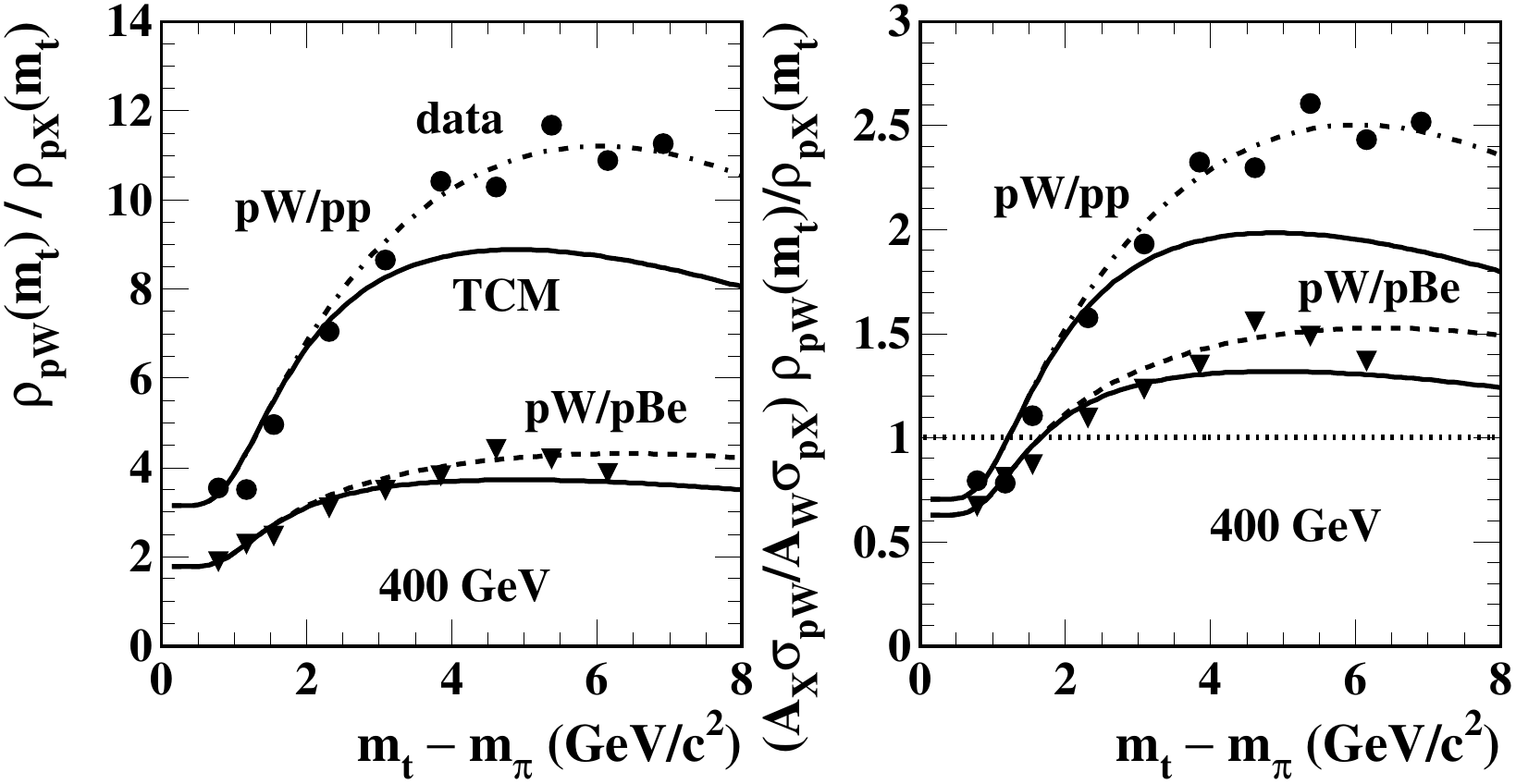}	
	\caption{\label{20i}
		Left: Spectrum (particle density) ratios denoted by pW/pp and pW/pBe (points) in the form of $R'_\text{pA}$ as defined by Eq.~(\ref{rpbp}) from fixed-target energy 400 GeV \pa\ collisions for data (points) and TCM (curves). Solid curves are the spectrum TCM with fixed hard component independent of A. Dashed and dash-dotted curves correspond to varying hard-component width $\sigma_{y_t}$ as described in the text.
		Right: Spectrum ratios in the left panel including factors $\sigma_\text{$p$A}/A$. 
	} 
\end{figure}

Figure~\ref{20h} (left) shows absorption cross section  $\sigma_{pX}$ (mb) vs atomic weight A for four $p$X collision systems. The line is the trend $\propto A^{2/3}$ for a nuclear area corresponding to the nuclear radius formula $R = R_0A^{1/3}$. If spectra are defined as differential cross sections, as noted above, rescaling the spectra by atomic weight A is equivalent to multiplying particle densities by a factor $\propto A^{-1/3}$. It is notable that for $p$-W or \ppb, $A^{1/3} \approx 6$ which is close to $N_{bin}$ values estimated by a classical Glauber Monte Carlo. See Secs.~\ref{geometry} and \ref{interpret} for responding comments.

Figure~\ref{20h} (right) shows charge densities $\bar \rho_s$ and $\bar \rho_h$ (points) rescaled by factor $\sigma_{pA}/A \propto A^{-1/3}$. Given the charge-density trends in Fig.~\ref{20g} (left) the result from rescaling is trends $\propto A^{-0.13}$ for the soft component and $\propto A^{0.20}$ for the hard component. That explains why spectrum ratios in the form of Eq.~(\ref{cronratio}) show a lesser {\em decrease} at low \pt\ where soft components dominate and a greater {\em increase} at high \pt\ where hard components dominate. Interpreting the low-\pt\ trend as ``suppression'' is misleading. The reduction is simply due to a questionable choice of prefactors for measured charge densities.

These relations explain Fig.~12 of Ref.~\cite{cronin0}. Plotted there is the differential cross-section spectrum (not rescaled) vs atomic weight A in a log-log format using spectrum values at $p_t = 4.6$ GeV/c that is dominated by the hard component. The trend expected from this study is $A^{2/3} \times A^{0.53} \approx A^{1.20}$, slightly higher than for the straight line $A^{1.17}$ in that figure, possibly due to a small contribution from the soft component even at 4.6 GeV/c.

Figure~\ref{20i} (left) shows  data particle-density spectrum ratios (points) for collision systems $p$-X, with {\em no rescaling} based on ratio $R_{pA}'$ as defined in Eq.~(\ref{rpbp}), derived from spectrum data reported by the C-P collaboration~\cite{cronin0}. The solid curves represent a TCM for spectrum data as described above which {\em assumes no A dependence} for model functions. Hard-component widths are fixed at 0.360 that best accommodates 400 GeV $p$-W data in Fig.~\ref{softnx} (right). Deviations above 3 GeV/$c^2$ correspond to width differences above the mode that are not discernible in Fig.~\ref{20d} (right). Dashed and dash-dotted curves correspond to reducing hard-component widths $\sigma_{y_t}$ from 0.360 for \mbox{$p$-W} to 0.356 for p-Be and 0.351 for \pp\ (width changes of 1-2\%). Ratio data are then described within point-to-point uncertainties {\em down to zero \pt}. Those results (especially for $p$-W/\pp) are comparable to right panels of figures in Sec.~\ref{nmfdata}, especially for event class $n = 5$ that approximates data for NSD \ppb\ collisions.

\begin{figure}[h]
	\includegraphics[width=1.65in]{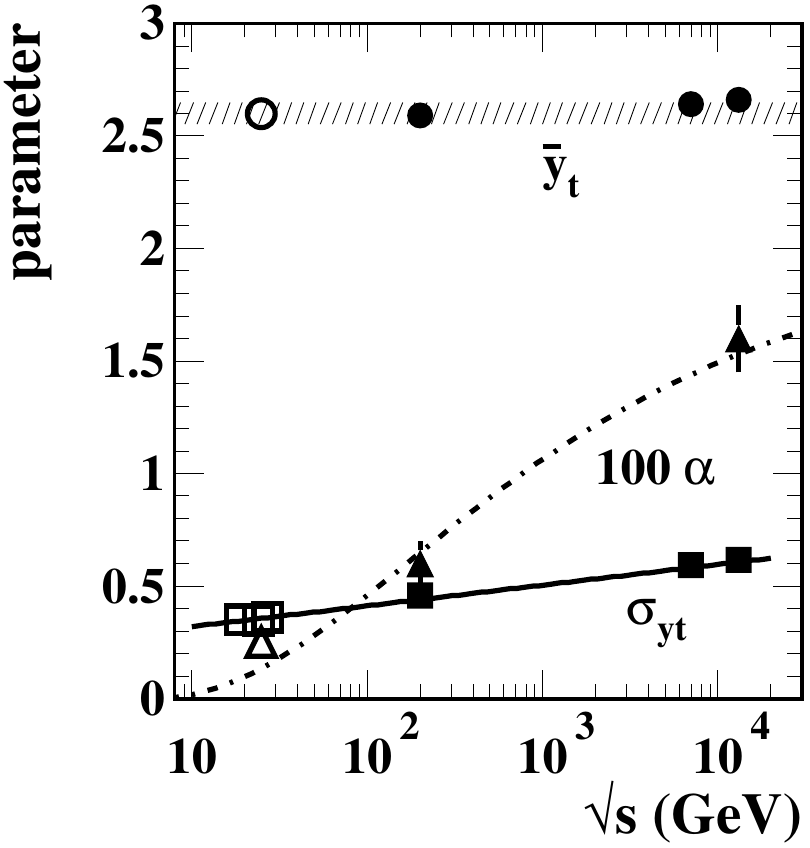}	
	\includegraphics[width=1.65in]{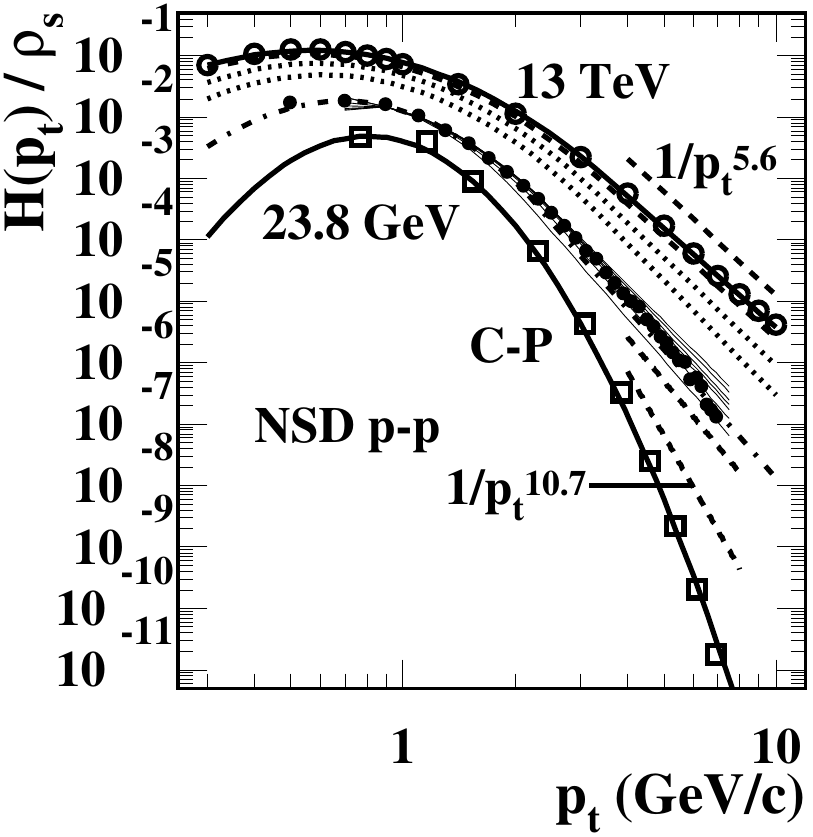}	
	\caption{\label{hardcom}
		Left: Figure~16 (left) from Ref.~\cite{alicetomspec} showing hard-component parameters $\bar y_t$ (mode) and $\sigma_{y_t}$ (Gaussian width) vs \pp\ collision energy $\sqrt{s}$. The figure has been updated with values (open squares) inferred from C-P spectra in the present study. Also updated is a value for hard/soft parameter $\alpha$ (open triangle) from this study.
		Right: Updated Fig.~15 (right) from Ref.~\cite{alicetomspec} showing \pp\ spectrum hard components for a variety of collision energies inferred from data (points) and corresponding TCM (curves).
		The solid dots are 200 GeV data. The open squares are inferred from C-P spectrum data in the present study. The corresponding solid curve is defined by 23.8 (300) GeV parameter values in the left panel.
	} 
\end{figure}

Figure~\ref{20i} (right) shows the same $\pi^+$ data and curves rescaled by conventional factors $\sigma_{pA}/A \propto A^{-1/3}$. Low-\pt\ ``suppression'' and {\em reduced} high-\pt\ enhancement are then apparent. The TCM curves indicate trends that are consistent with the data peaks (Gaussians) in Fig.~\ref{20d} (right) but not with theory predictions as for instance Fig.~9 of Ref.~\cite{wangcronin}.  As noted above, the trends above 4 GeV/c represent {\em less than two percent} of jet fragments and give a misleading impression of data-model disagreement. The great majority of jet fragments appears near 1 GeV/c ($y_t \approx 2.7$ in Fig.~\ref{20d} and see Fig.~\ref{hardcom}, right). In  this ratio format that contribution is not represented. The ratio format demands precision for a vanishing fraction of jet fragments that is {\em not statistically significant}. This figure also demonstrates that $R_\text{$p$A}$  trends at higher \pt\ have no requirement to converge to unity. Such a result would require an extremely precise correspondence between two hard components that is not encountered in nature.

Figure~\ref{hardcom} (left) summarizes TCM hard-component parameters $\bar y_t$ (mode) and $\sigma_{y_t}$ (Gaussian width) plus hard/soft parameter $\alpha$ appearing in the quadratic relation $\bar \rho_h \approx \alpha \bar \rho_s^2$ for NSD \pp\ collisions over three orders of magnitude collision energy. The solid points were obtained in previous TCM spectrum analyses as summarized in Ref.~\cite{alicetomspec} of which this is Fig.~16 (left). The open triangle is the $\alpha \approx 0.0025$ value inferred from C-P spectrum data in the present study. Open squares  are Gaussian widths inferred from C-P spectrum data for three collision energies also in this study that are consistent with the trend (solid line) established some years ago. The value $\alpha \approx 0.0025$ from C-P spectra agrees, within uncertainties, with a prediction (dash-dotted curve) from Ref.~\cite{alicetomspec} extending down to the limit of jet production in high-energy nuclear collisions near $\sqrt{s} = 10$ GeV.

Figure~\ref{hardcom} (right) shows spectrum hard components in the form $H(p_t) / \bar \rho_s \approx \alpha \bar \rho_s \hat H_0(p_t)$ plotted vs \pt. This plot is adapted from Fig.~15 (right) of Ref.~\cite{alicetomspec}. The points are inferred from spectrum data. The curves summarize TCM parametrizations corresponding to the left panel. The 23.8 GeV open squares and curve are inferred from the present study of C-P spectrum data as discussed below. The dashed lines represent power laws $1/p_t^n$ corresponding to exponential tails $\propto \exp(-qy_t)$ of $\hat H_0(y_t)$ models, where $n \approx q + 2.2$. The power law for 23.8 GeV is an extrapolation from Fig.~13 (left) of Ref.~\cite{alicetomspec} giving $q \approx 8.5$, but C-P data are inconsistent with any exponential tail on the hard component at that energy: $1/q$ for C-P energies is experimentally consistent with zero. 

The 23.8 GeV open squares in the right panel are derived from $p$-W data in Fig.~\ref{softnx} (right). The 300 GeV data from that figure are divided by $A^{1/2} \approx 13.6$ per the $\bar \rho_h$ trend in Fig.~\ref{20g} (left), multiplied by 2.5 to obtain nonPID \pp\ charge density $\bar \rho_h \approx 0.005$ and then divided by $\bar \rho_s \approx  1.4$ to obtain the required expression above for $H(p_t) / \bar \rho_s$. The solid curve, also per the expression above, is $H(p_t) / \bar \rho_s \approx 0.0025 \times 1.4 \hat H_0(p_t)$ with $\bar y_t = 2.65$ and $\sigma_{y_t} = 0.345$ defining 23.8 GeV $\hat H_0(p_t)$. Given the transformation as defined above the solid curve in this figure is equivalent to the dashed curve in Fig.~\ref{softnx} right.

In light of the present analysis multiple Cronin effects are enumerated (for pions) in the context of the TCM as follows: 
(a) For given target A and increasing collision energy the spectrum soft component becomes harder at higher \pt, possibly due to Gribov diffusion in the transverse momentum plane.
(b) For given target A and increasing collision energy the spectrum hard component becomes harder at higher \pt, probably due to extension of underlying scattered-parton or jet spectrum to higher energies. The dominant effect at these low energies is hardening of the soft component.
(c) For given collision energy and increasing target A the soft charge density increases as $\propto A^{0.20}$ consistent with centrality dependence intermediate between single peripheral \pn\ and central \pa\ collisions.
(d) For given collision energy and increasing target A the hard charge density increases as $\propto A^{0.53}$ indicating $\bar \rho_h \propto \bar \rho_s^{2.65}$ as inferred from data.
(e) Rescaling spectra as differential {\em cross sections} (including factors $\sigma_\text{pA}$) with factor $1/A$ introduces a factor $\propto A^{-1/3}$ that produces misleading trends in spectrum {\em ratios} wherein {\em the great majority of jet fragments is not represented}. 

Structures in Fig.~\ref{20i} that are conventionally identified with the ``Cronin effect'' arise primarily from the charge-density trends in Fig.~\ref{20g}, one from projectile nucleon dissociation via parton splitting cascade and the other from \pn\ binary collisions leading to jet production. Modifications of soft- and hard-component spectrum {\em shapes} appear as secondary effects such as data-TCM differences in Fig.~\ref{20i} and  soft-component hardening in Figs.~\ref{softn} and \ref{softnx}, the latter effect not being evident in spectrum ratios.

A two-component approach to the C-P collaboration spectrum data greatly simplifies what has been seen as a complex problem within the context of spectrum ratios. Rather than addressing {\em ad hoc} measure $A^{n(p_t)}$ one has separate empirical trends $A^{0.20}$ and $A^{0.53}$ with momentum-dependent factors $\hat S_0(m_t,\sqrt{s})$ and $\hat H_0(y_t,\sqrt{s})$ that are simply related to basic mechanisms.


\section{Discussion} \label{disc}

Four  topics are here pursued further: NMF interpretation, the background for C-P spectrum measurements and  Cronin effect, variations on two-component spectrum models and NMF-associated theoretical predictions.

\subsection{NMF Interpretation}

Small deviations of data hard components shown in Sec.~\ref{spechard} from a TCM reference model (e.g.\ $\hat H_0(y_t)$) reveal all the novel information conveyed by newly-acquired data spectra beyond the TCM reference. The TCM is in turn interpreted in terms of conventional QCD theory (e.g.\ factorization theorem~\cite{fragevo}) and related measurement (e.g.\ jet spectra~\cite{jetspec2} and fragmentation functions~\cite{eeprd}). Given that the TCM parametrization as summarized in Sec.~\ref{tcmparams} describes spectrum data within their point-to-point uncertainties (Sec.~\ref{fitqual}) one may question whether NMFs comprised of spectrum {\em ratios} carry more or less information than the TCM. In fact, such ratios {\em discard} information just as the difference 2 does not convey whether the inputted numbers are 9 and 7 or 3 and 1. In the specific instance of NMFs based on Ref.~\cite{alicenucmod} data the resulting NMF ratios include distortions ({\em dis}information) because of a biased reference (MB \pp\ spectra).

In Ref.~\cite{alicenucmod} the spectra compared in ratio are specifically NSD \ppb\ spectra vs MB \pp\ spectra, with an estimated $N_{bin}$ value for NSD \ppb\ to complete the conventional NMF formulation. That practice is presumably motivated by what was described in C-P publications where a minimum-bias average over an event ensemble and somewhat large \pt\ bins were possibly motivated by limited event numbers (see Sec.~\ref{cronedep}). Given that the nominal goal is detection of jet modification within a hypothetical dense medium in central \aa\ one should want to know how jet structure {\em evolves} with \ppb\ centrality. Compare evolution of $R'_\text{$p$Pb}$ spectrum ratios (solid dots) in Sec.~\ref{nmfdata} with single NSD/MB ratios (dash-dotted curves) there.

The conventional plotting format for NMF $R_{p\text{Pb}}$ is linear on \pt. That choice confines the low-\pt\ region within a small interval, especially for spectra that extend to tens of GeV/c. The ``suppression'' at low \pt\ is typically ignored. As demonstrated above, alternative ratio $R_{p\text{Pb}}'$ within that interval could be compared with conjectured $N_{part}$ that is just as important as $N_{bin}$ for \ppb\ collisions.

In Fig.~9 of Ref.~\cite{alicenucmod} the region of interest is above \pt\ $\approx 4$ GeV/c ($y_t \approx 4$). From the left panels of Figures in Sec.~\ref{spechard} that interval includes roughly {\em 2\% of total fragments} from MB dijets and thus grossly misrepresents the overall jet production in high energy nuclear collisions. In contrast, TCM isolation of {\em entire} hard components and accurate determination of their evolution with \ppb\ centrality provide a full picture of a critical issue: are jets modified by a dense medium in \ppb\ (or {\em any}) collisions.

The basic collision mechanism that produces the Cronin effect in conventional ratios $R_\text{AB}$, MB jet production, is present in \pp\ as well as \pa\ collisions and to varying degrees in both systems depending on \pp\ (\pn) event multiplicity and on \pa\ collision centrality.  Details are provided in Sec.~\ref{nmfdata}. That rescaled $R_{p\text{Pb}}$ might exceed 1 at higher \pt\ is inevitable if MB jet production {\em{and} hadron species abundances} vary with \pa\ centrality as they do~\cite{ppbpid, pidpart1,pidpart2,pppid}. One should then examine the high-\pt\ limit of Eq.~(\ref{rpbp}) that describes \ppb\ data within their point-to-point uncertainties. For higher \pt\ 
\bea
R_{p\text{Pb}}' &\rightarrow& \frac{z_{hi}(n_s) N_{bin} \bar \rho_{hNN} \hat H_{0ip\text{Pb}}(y_t,n_s)}{z_{hipp} \bar \rho_{hpp} \hat H_{0ipp}(y_t)}.
\eea
For rescaled $R_{p\text{Pb}} \rightarrow 1$ $N_{bin}$ must be estimated accurately {\em assuming} the constraint $\bar \rho_{hNN} \approx \bar \rho_{hpp}$ and $z_{hi}(n_s) \approx z_{hipp}$, {\em all within a few percent}, to test the hypothesis. But those assumptions are contradicted by \ppb\ data~\cite{ppbpid, pidpart1,pidpart2}. The spectrum ratio is very sensitive to any variation in hard-component properties (e.g.\ mode position, peak width above the mode, slope of exponential tail), exaggerating actual variation of hard-component properties.

\subsection{Background for C-P spectrum measurements}

The invariant differential cross section as defined in fixed-target C-P publications can be expressed as
\bea \label{croncross}
E\frac{d\sigma_{pAi}}{d^3p} &=& \sigma_{pA} \frac{1}{N_{evt}} \frac{\text{hadron species $i$ yield}}{p^2 \Delta \Omega \Delta p/p}
\\ \nonumber
&\rightarrow& \sigma_{pA} \frac{d^2 n_i}{2\pi m_t dm_t dy_z},
\eea
where $p$ is lab momentum and $i$ denotes hadron species. $\sigma_\text{$p$A}$ is reported in units of cm$^2$ and may be invoked as the \pa\ absorption cross section~\cite{cronin0} or alternatively effectively as the \pn\ inelastic cross section (e.g.\ either by replacing $ \sigma_\text{$p$A} \rightarrow  \sigma_{pp}$~\cite{cronin1} or by dividing  $p$-W data defined by Eq.~(\ref{croncross}) by cross-section ratio 40.9 \cite{croninconf}). That definition motivates the conversion factor $2\pi 10^{27}/\sigma_\text{$p$A}$ invoked for the present analysis, $\sigma_\text{$p$A}$ depending on data source. 

The initial measurements of Ref.~\cite{cronin1} indicated spectrum decrease slower than exponential on \pt\ and more so with increasing collision energy.  However, subsequent Ref.~\cite{cronin10} observes that ``For all energies and at large \pt\ the cross sections fall exponentially; they do not show a manifest power law dependence.'' That observation applies specifically to the C-P data shown in Fig.~\ref{softnx}. The present analysis demonstrates that such spectrum structure is composite: Soft component $\hat S_0(m_t)$ does have a power-law tail but $\hat H_0(y_t)$ at that energy falls off as a Gaussian and dominates the soft component at higher \pt\ for $p$-W. In contrast, for \pp\ data in Fig.~\ref{softn} the soft  power law is comparable to the hard component at higher \pt.

Interestingly, Ref.~\cite{cronin1} observes that based on additional spectra from $p$-Be and $p$-Ti ``...none of the important features observed in tungsten were dependent on atomic number [weight A].'' Despite that initial observation A dependence of differential cross sections became a central issue: how to compare \pa\ data to \pp\ data and interpret differences. Reference~\cite{cronin10} asserts that ``For the purpose of comparison with p-p cross sections it is necessary to convert the cross sections per nucleus [A] into equivalent p-{\underline {nucleon}} cross sections. The simplest way...is to divide the cross sections per nucleus by the atomic number [sic] A....'' Alternatively, per-nucleus cross sections are divided by ratios $\sigma_{pA}/\sigma_{pp}$ in which case for any spectrum ratios the common factor $\sigma_{pp}$ would cancel. From several alternatives ``The best method...is the one which yields cross sections independent of A.'' However, Ref.~\cite{cronin10} observes that ``Tungsten, for example, is much more efficient then Be in producing large transverse momentum hadrons'' which simply results from the measured hard vs soft particle-density trends in Fig.~\ref{20g} (left).

The intent of measuring spectra on various targets A was to extrapolate the \pa\ spectra to a \pp\ equivalent simply by rescaling with factor A: ``We felt, naively, that for the `hard' collisions only a single nucleon in the nucleus would be involved. It was therefore a surprise when we found that although the cross sections did extrapolate as $A^\alpha$, the power alpha is a function of \pt\ and for all particle types grows to be greater than 1.0 at large \pt\, implying that more than one nucleon is involved''~\cite{cronin0}.
After considerable attempts to express the resulting spectrum phenomenology Ref.~\cite{cronin0} comments that ``It now appears that with a theory [QCD] which actually predicts the single particle spectra to be rather complicated functions of $x_\perp$ [$= 2 p_\perp / \sqrt{s}$] and $p_\perp$, precise measurements over a wide range in these variables will again be important. ...it has been a field in which experiment leads theory.'' 

In light of the present analysis the C-P spectrum trends can be efficiently summarized per Eq.~(\ref{crontcm}): Each of two spectrum components has a simple collision energy-dependent {\em form} that is approximately independent of nuclear size A. The associated charge densities vary simply with nuclear size A per Fig.~\ref{20g} (left) and are approximately independent of collision energy (over the limited C-P energy range). Projectile protons do interact with multiple target nucleons on average. The resulting spectra reflect a combination of ``wounded nucleon'' soft components (with form independent of collision number) and jet-related hard components whose structure (at higher energies) does depend on the number of \pn\ collisions. 

\subsection{Cronin and two-component spectrum models}

From the fixed-target era of the seventies up to the present various two-component spectrum models have been proposed in response to new spectrum data. A range of examples appears to follow one or two of three categories: (a) functional forms intended to mimic certain data features, (b) functional forms based on {\em a priori} physical assumptions or models and (c) functional forms based on information-theoretic methods applied to spectrum data. The effectiveness of such models may depend in part on how broad a range of spectrum data is addressed and what data description accuracy is required.

Reference~\cite{kuhn} adopts a soft+hard model in which the soft component is assumed to be an exponential on \pt\ (corresponding to soft-scattering events) and the hard component is defined initially as complementary to the data soft component and as arising from ``hard-scattering events.'' Linear superposition is invoked: ``These processes [hard] occur independently of the first ones [soft]. ... Soft and hard components of the cross section can be added incoherently.'' There follow attempts to describe the hard component theoretically, especially in light of the C-P spectrum data and ratios in the form of Eq.~(\ref{cronratio}).

In the context of the early RHIC program twenty-five years later Ref.~\cite{accardi} refers to ``...so-called two component models of hadron spectra, and call[s] hard a scattering which is described by a power law...cross-section at large \pt, and soft a scattering whose cross-section is decreasing faster than any inverse power of [\pt] at large \pt.'' Again, a point of emphasis is modeling of spectrum ratios in the form of Eq.~(\ref{cronratio}) and the associated Cronin effect. Ironically, it is the soft component as defined by Eq.~(\ref{s000}) that can be described {\em asymptotically} as a power law $\propto p_t^{-n}$ whereas the hard component, at least within the \mbox{C-P} energy interval, appears as a Gaussian (on \yt) that descends at higher \pt\ faster than any inverse power law.

Some theoretical models emphasize multiple scattering (at the nucleon and/or parton level) as the source of spectrum ``hardening.'' Reference~\cite{wangcronin} comments that ``..semi-hard parton scattering is the dominant particle production mechanism underlying the hadron spectra at moderate \pt\ $> 1$ GeV/c.'' Both \pp\ and \pa\ spectra are modeled in that case by QCD factorization describing parton scattering as the only particle source above some low \pt\ value 1-2 GeV/c. That model is applied to C-P \pp\ and \pa\ spectrum data from Ref.~\cite{cronin0}. The result is, in effect, a {\em single-component} model: ``...we assume that the inclusive differential cross section for large \pt\ particle production is still given by a single hard parton-parton scattering. However, due to multiple parton scattering prior to the hard processes, we consider the initial transverse momentum $k_t$ of the beam partons is broadened.''

Analysis of the C-P spectrum data in the present study leads to the factorization in Eq.~(\ref{crontcm}) wherein model functions depend only on collision energy and charge densities depend only on system size A. In that case multiple scattering as invoked in the text above seems unlikely to contribute to model-function details. The soft component arises from isolated projectile nucleons if they have suffered at least one single inelastic collision -- the wounded- or participant-nucleon model. Additional encounters within target A {\em do not change that form}. Likewise, to good approximation the spectrum hard component remains invariant in shape no matter what the target A; again multiple scattering does not alter its shape.

The TCM utilized in the present study was also developed in the early days of the RHIC program. In contrast to categories (a) and (b) above the TCM  was derived from the {\em evolution} of spectrum shapes with  {\em event-multiplicity} \nch\ based on (c), information-theoretic techniques with no {\em a priori} assumptions about spectrum structure or physical models. Details are described in Ref.~\cite{ppprd} and include introduction of spectrum {\em running integrals} to enhance the signal(s) of interest compared to statistical noise  and  transverse rapidity \yt\ to improve visualization of (e.g.\ jet-related) details at lower \pt\ that turn out to be critical to physical interpretation. 

It was newly discovered in that analysis that two nearly-fixed spectrum structures are linearly superposed and that, for \pp\ collisions at least, the particle yield associated with one (hard) varies exactly quadratically relative to the other (soft). That relation permitted definition of the TCM soft-component model as the limiting case of data spectra rescaled by soft charge density $\bar \rho_s$ in the limit $\bar \rho_s \rightarrow 0$. The TCM hard component represents the data complement. Both components have been subsequently interpreted in terms of up-to-date QCD theory and jet-related measurements~\cite{eeprd,jetspec2,fragevo}. The TCM has been applied to an assortment of collision systems~\cite{hardspec,alicetomspec,ppbpid,pidpart1,pidpart2,pppid} where it provides insight as to physical mechanisms. The present study is an example where spectrum structure is not assumed. Differential analysis of C-P spectrum variation with some control feature (e.g.\ $\sqrt{s}$ or $A$) permits accurate description of spectrum structure and the discovery of new features as in Eq.~(\ref{crontcm}) and Fig.~\ref{20g}.

\subsection{Theoretical predictions and interpretations}

It is informative to review theoretical approaches to the Cronin effect as they vary over time and their implications for interpretation of NMF spectrum ratios.


Reference~\cite{kuhn} responds to C-P $p$-W pion spectrum data for three energies reported in Ref.~\cite{cronin10}. The basic finding is that $Ed\sigma/d^3p \propto A^{n(p_t)}$ with $n < 1$ for low \pt\ and $n > 1$ for higher \pt. Within the context of a multiple-scattering picture a two-component model is proposed: a soft component follows an exponential on \pt\ with $A$ dependence as for low \pt\ and a hard component (``hard scattering events'') represents deviations from the exponential at higher \pt\ with a different $A$ dependence. The hard component is attributed to ``high momentum fragments of the original hadrons.... These [hard] processes occur independently of the first [soft] ones.'' A formula for cross section hard component is then derived assuming multiple scattering within a \pa\ collision. Figure~3 of that reference, with two model components related to a C-P spectrum, may be compared with Fig.~\ref{softnx} above.


Referring to its own C-P data shown in Fig.~\ref{softn} (left) above, Ref.~\cite{cronin0} invokes a theoretical conjecture $B(1/p_t^n )(1 - x_t)^b$ with $x_t = 2 p_t / \sqrt{s}$. Fitted values $n \approx 8.5$ and $b \approx 9.5$ describe spectrum data well above 1.5 GeV/c but rise sharply above data below that point due to singular factor $1/p_t^n$. Those \pp\ spectra are actually dominated by nonjet soft components, whereas the theoretical expression is presumably directed toward jet production.


Some thirteen years later Ref.~\cite{straub} reports $p$-Be and $p$-W spectra at 800 GeV fixed target or $\sqrt{s} = 38.8$ GeV. The general results for ratio $R$ are similar to those from the C-P collaboration as reported in Ref.~\cite{cronin0} but extend to higher particle momentum. The theoretical context at the time is suggested by ``The species [$\pi$, K, p] dependence seen in Fig.~1 is partially explained by existing constituent multiple scattering models. ...the larger enhancement for protons compared to $\pi^+$'s indicates either that diquarks have a larger rescattering cross section than $u$ quarks or that the process of binding a diquark to a $u$ quark at high $z$ is enhanced in nuclei.'' The emphasis on quarks is interesting in that the phenomenon that is central to the Cronin effect may be gluon excitation and scattering emerging above a threshold near 10 GeV.


Reference~\cite{wangcronin} addresses an issue relevant just before RHIC startup: If high-\pt\ hadrons are to be employed as possible indicators of jet quenching in a dense QGP medium within \aa\ collisions their behavior in the absence of such a medium (e.g.\ within \pa\ collisions) must be well understood as a control or reference. But the Cronin effect already appears in \pa\ spectrum data as a poorly understood phenomenon, thus jeopardizing the value of the reference system. In this approach a {\em single}-component model is invoked in which all hadron production arises from parton scattering described by a factorization model -- e.g.\ its Eqs.~(2) and (7). Resulting predictions for \pp\ collisions are compared to spectrum data from Ref.~\cite{cronin0}. The curves with and without $Q$-dependent intrinsic $k_t$ differ by an order of magnitude. It has been established in several TCM analyses that pion spectra are dominated by the soft component up to 1 GeV/c, and the hard component, what might be related to Eq.~(2) of Ref.~\cite{wangcronin}, peaks near 1 GeV/c and then descends rapidly below that point. See Sec.~\ref{smallspec} above.

Concerning \pa\ collisions Ref.~\cite{wangcronin} states ``...single-particle spectra at high \pt\ in p+A collisions have been shown, both experimentally and theoretically, to be sensitive to multiple initial-state scattering, or Cronin effect. One can study the Cronin effect in a model of multiple parton scattering.'' Concerning the spectrum model ``...we assume that the inclusive differential cross section for large \pt\ particle production is still given by a single hard parton-parton scattering [QCD factorization]. However, {\em due to multiple parton scattering prior to the hard processes} [emphasis added], we consider the initial transverse momentum $k_t$ of the beam partons is broadened.'' Concerning spectrum ratios $R$ ``If there was no nuclear dependence due to multiple scattering, the ratios would have a flat value 1....'' Concerning measured \pt\ trends of ratios $R$ ``The ratios should become smaller than 1 at very small \pt\ because of the absorptive processes.... At larger \pt, the spectra are enhanced because of multiple parton scattering.'' Concerning rescaling of ratios $R$ ``$\langle N_\text{binary}\rangle = A$ for minimum-bias evens of \pa\ collisions'' implies that the number of \pn\ binary collisions in MB \mbox{$p$-W} is 183 whereas, per Table~\ref{rppbdata}, the number is close to 1.3 for equivalent 5 TeV \ppb\ event class $n = 5$. Emphasis on multiple scattering (of nucleons or partons?) seems misguided.  C-P spectra show no dependence on A for the soft component and only a percent-level width dependence for the hard component. Figure~9 of Ref.~\cite{wangcronin} shows theory curves with sharp structure (unjustified by the large data uncertainties) that is dramatically different from comparable TCM curves in Fig.~\ref{20i} (right).


Reference~\cite{boris} emphasizes that ``An adequate interpretation of the Cronin effect has become especially important...''\ in relation to claimed evidence emerging from RHIC at that time for ``jet quenching'' in more-central \auau\ collisions. ``However, in spite of the qualitative understanding of the underlying dynamics of this effect, no satisfactory quantitative explanation of existing \pa\ data has been suggested so far.'' It is proposed that a ``...QCD-dipole approach...allows to explain available data without fitting to them.... We point out that the mechanism causing Cronin effect drastically changes between the energies of fixed target experiments and RHIC-LHC.'' As to comparisons with data Fig.~1 of Ref.~\cite{boris} appears very similar to Fig.~9 of Ref.~\cite{wangcronin}.
And direct comparison of C-P hard components with those obtained from much higher collider energies in Fig.~\ref{hardcom} shows a remarkable consistency of trends over three orders of magnitude collision energy. Drastic change is not apparent.


Reference~\cite{accardi}, in defining Cronin ratio $R$ as Eq.~(\ref{cronratio}),  states that ``In the absence of nuclear effects one would expect R = 1, but for $A > B$ a suppression is observed experimentally at small \pt\ and an enhancement at moderate \pt\ with $R \rightarrow 1$ for large \pt.'' Such statements beg the question what are ``nuclear effects'' in the context of Sec.~\ref{nmfdata} above. Concerns similar to those of Ref.~\cite{boris} are expressed about \pa\ control experiments: ``...extrapolation to RHIC and LHC energies of the known Cronin effect [in \pa\ collisions] at lower energies is haunted by large theoretical uncertainties, which may make unreliable any interpretation of signals of this kind.'' [i.e.\ due to Cronin effect NMF ratios are {\em uninterpretable}.] 

``Multiple interactions'' is invoked but not well defined: ``...it was realized that the observed nuclear enhancement [Cronin effect]...could be explained in terms of multiple interactions.''  A scenario is introduced where projectile protons are sequentially excited by successive \pn\ interactions: ``Soft proton-nucleon interactions are assumed to excite the projectile proton's wavefunction, so that when the proton interacts with the next target nucleon its partons have a broadened intrinsic momentum [distribution].'' See Ref.~\cite{wangcronin} and $k_t$ broadening. The suggestion ``...the $A$  systematics, or the study of collision centrality cuts, would be interesting since would allow to change the opacity of the target -- then the sized of the Cronin effect -- in a controllable way'' invites Sec.~\ref{nucmod} above.


Reference~\cite{dima} considers Cronin effects within the context of the color-glass condensate (CGC)~\cite{cgc}. It considers gluon production in \pa\ collisions based on Eq.~(\ref{cronratio}) comparing \pa\ to \pp\ in ratio. The model addresses ``multiple rescatterings of the produced gluon and the proton in the target nucleus.'' As for other approaches a basic premise is  ``Multiple scatterings of partons inside the nucleus are believed to be the cause of Cronin effect. Phenomenologically these multiple rescatterings are usually modeled by introducing transverse momentum broadening in the nuclear structure functions.'' 

Referring to its Fig.~4 and in the context of the CGC model ``...we conclude...that in the quasi-classical approximation...the $k_t$-position and the height of the Cronin peak should increase with centrality of the \pa\ collision.'' That conclusion is consistent with the results in Sec.~\ref{nmfdata} above, but possibly not for the same reasons. It is notable that in that same Fig.~4 $R_{pA}$ increases {\em from zero} at low $k_t$. What is presented is a {\em one}-component model of particle (gluon) production. Reference~\cite{dima} extends the CGC application to higher collision energies ``We have also shown that at higher energies..., when quantum evolution [?] becomes important, it introduces suppression of gluons produced in \pa\ collisions {\em at all values of $k_t$} [emphasis added], as compared to the number of gluons produced in \pp\ collisions scaled up by the number of collisions $N_{coll}$ [A?].... The resulting $R_{pA}$ at high energy... is a {\em decreasing function of centrality} [emphasis added].'' Refer to Sec.~\ref{nmfdata} for data $R_{pA}$ centrality trends.

It is notable that especially in the RHIC/LHC era theoretical treatments of the Cronin effect have proceeded strictly in the context of spectrum {\em ratios}, as in Eq.~(\ref{cronratio}), where much of the information carried by intact individual spectra is discarded. Compare Sec.~\ref{nucmod} with Sec.~\ref{smallspec} above for example. And treatments are typically further restricted to comparing rescaled minimum-bias event averages rather than investigating centrality dependence as in Sec.~\ref{nucmod}. It seems ironic that a TCM as implemented in Sec.~\ref{smallspec}, introduced nearly twenty years ago~\cite{ppprd} and able to accurately separate jet and nonjet contributions to particle production that have been quantitatively related to QCD mechanisms, effectively plays no significant role in addressing what is broadly acknowledged to be a major problem for high-energy nuclear physics at this time: what is the Cronin effect and how does it affect our ability to interpret nucleus-nucleus collision data?

Theory should redirect its efforts to the findings presented in Sec.~\ref{cronineff}. Their implications are as follows: 
(a) Inelastically-scattered participant (wounded) nucleons proceed to dissociation by parton splitting cascade independently, without multiple scattering in the target. The resulting spectrum soft components are independent of target A or \pa\ centrality.
(b) Large-angle scattered partons (gluons) proceed to jet formation in \pa\ collisions without further rescattering. To good approximation the resulting fragment distributions are independent of target A or \pa\ centrality.
(c) Soft components are strongly energy dependent, consistent with Gribov diffusion as the mechanism for shape evolution.
(d) Hard components are strongly energy dependent consistent with development of the underlying scattered-parton energy spectrum.
(e) The soft and hard particle-density trends in Fig.~\ref{20g} remain to be completely understood and lie at the heart of the Cronin phenomenon.
(f) The quadratic relation between hard and soft charge densities appears to be universal across a broad range of collision systems and also remains to be understood. The theories summarized above do not address those issues in any substantive way.

\section{Summary}\label{summ}

This article reports a study of identified-hadron (PID) spectra for pions, charged kaons and protons from 5 TeV \ppb\ collisions extended up to $p_t = 20$ GeV/c.  A two-component (soft+hard) model (TCM) of hadron production is applied to the spectra and describes them within point-to-point uncertainties. In particular, evolution of spectrum hard components (jet fragment distributions) with \ppb\ centrality is accurately parametrized. The resulting TCM is then statistically equivalent to data. 

A primary motivation for this study is development of improved understanding of {\em nuclear modification factors} (NMFs) that have been invoked since first RHIC operation to claim formation of a quark-gluon plasma (QGP) in more-central \aa\ collisions but not small asymmetric systems such as \ppb. Such arguments are now questionable in that more-recent data from small collision systems has been interpreted also to indicate QGP formation.

The strategy adopted is to calculate not conventional spectrum ratio $R_\text{$p$Pb}(p_t)$ rescaled by estimated number of binary \nn\ collisions $N_{bin}$, but instead {\em unrescaled} spectrum ratios denoted here by $R'_\text{$p$Pb}(p_t)$, thus eliminating major uncertainty in $N_{bin}$ estimates via classical Glauber Monte Carlo. A section is devoted to discussion of accuracy issues for such Monte Carlos. Data NMFs are compared with corresponding TCM NMFs wherein the hadron production mechanisms are well understood.

Several observations emerge from this study: The structure of  $R'_\text{$p$Pb}(p_t)$ below 4 GeV/c is determined by the combination of soft and hard model functions $\hat S_0(p_t)$ and $\hat H_0(p_t)$ with (ironically) no sensitivity to jet production {\em details}. The ratios at low \pt\ correspond to participant number $N_{part}$ while the ratios at high \pt\ correspond to binary-collision number $N_{bin}$. However the ``correspondence'' is not simple proportionality because other centrality-dependent factors are also important. That is one reason that rescaled quantity $R_\text{$p$Pb}(p_t)$ is misleading.

Above 4 GeV/c evolution of $R'_\text{$p$Pb}(p_t)$ depends dramatically on evolution of the spectrum hard component, the jet fragment distribution. The direct connection between TCM and data spectra makes possible correspondence between the TCM hard-component parametrization and the $R'_\text{$p$Pb}(p_t)$ data trends. But the fraction of total jet fragments falling within that interval is of order 2\%. The great majority of jet fragments is {\em hidden} by the spectrum ratio format. That format amplifies small changes in hard-component widths and tail structure at the percent level to become 100\% changes in ratio characteristics.

Interpretation of NMFs has been further complicated by continuing uncertainty about the ``Cronin effect'' described as suppression of spectrum ratios at lower \pt\ accompanied by enhancement at higher \pt. The present study includes application of the TCM to the original Chicago-Princeton (C-P) fixed-target PID spectrum data (with CM energies near 25 GeV) that first manifested the Cronin effect. The result is a simple parametrization of the C-P data involving accurate separation of soft and hard components and factorization of collision energy and target-size A dependence for each component. The hard component at those energies is accurately predicted by previous TCM analysis of NSD \pp\ collisions at RHIC and LHC collider energies. The ``Cronin effect'' is in effect the first observation of minimum-bias jet production manifested in hadron spectra near mid-rapidity.

This study demonstrates that spectra combined as ratios in the form of rescaled NMFs $R_\text{$p$Pb}(p_t)$ are not interpretable. They represent a small fraction of total jet fragments and amplify small variations in jet spectrum contributions into large excursions at higher \pt. While the TCM representation of spectra can be used to interpret NMF trends, as in the present study, NMFs could never be used to anticipate spectrum structure and evolution. The only {\em interpretable} approach is differential analysis of {\em individual} spectra in the context of the TCM.


\end{document}